\documentclass[12pt,letterpaper]{article}
\usepackage[utf8]{inputenc}

\usepackage{graphicx,array}
\usepackage{url}
\usepackage{xcolor}
\usepackage{amsthm,amsmath,amssymb}
\usepackage{bbm}                
\usepackage{slashed}
\usepackage{mathrsfs}           
\usepackage{cancel}
\usepackage{esvect}             

\usepackage{bm}
\usepackage[numbers,sort&compress]{natbib}
\usepackage{siunitx}
\usepackage{setspace}
\usepackage[shortlabels]{enumitem}
\usepackage{booktabs,multirow,multicol}
\usepackage{subcaption}
\usepackage{tcolorbox}
\usepackage{tikz}
\usetikzlibrary{tikzmark, fit, calc, arrows.meta}
\usepackage{floatrow}
\newfloatcommand{capbtabbox}{table}[][\FBwidth]
\usepackage[normalem]{ulem}
\numberwithin{equation}{section}

\usepackage{hyperref}           
\hypersetup{colorlinks=true,linkcolor=red,citecolor=blue,filecolor=magenta,urlcolor=blue}
\usepackage[all]{hypcap}        

\setlist[description]{leftmargin=\parindent,labelindent=\parindent}
\newcolumntype{M}[1]{>{\centering\arraybackslash}p{#1}}

\newcommand{\nc}{\newcommand}

\nc{\beq}{\begin{equation}}   \nc{\eeq}{\end{equation}}
\nc{\beqa}{\begin{eqnarray}}  \nc{\eeqa}{\end{eqnarray}}
\nc{\bit}{\begin{itemize}}    \nc{\eit}{\end{itemize}}

\newcommand{\Z}{\mathbb{Z}}
\newcommand{\ev}[1]{\langle #1 \rangle}
\newcommand{\dd}[1][]{\mathop{}\!\mathrm{d}\ifx\\#1\\\else^{#1}\fi}
\DeclareMathOperator{\Tr}{Tr}

\DeclareMathOperator{\sech}{sech}

\newcommand{\eval}[1]{\left.#1\right|}

\allowdisplaybreaks

\title{
{\bf The Great Wall of the Standard Model
}
\author{\large Yang Bai$^{\,a,b}$, and Yiming Yang$^{\,a}$}
\date{\small \it 
$^a$Department of Physics, University of Wisconsin-Madison, Madison, WI 53706, USA \\
$^b$HEP Division, Argonne National Laboratory, 
Argonne, IL 60439, USA \\
}
}

\begin{document}

\maketitle

\begin{abstract}
The global form of the Standard Model (SM) gauge group, $[SU(3)_c\times SU(2)_L\times U(1)_Y]/\Gamma$, with $\Gamma\subseteq\mathbb{Z}_6$, remains undetermined. For $\Gamma=\mathbb{Z}_p$, the SM possesses a residual global electric one-form symmetry $\mathbb{Z}_{6/p}$. Compactification on the thermal circle yields an ordinary discrete zero-form symmetry whose spontaneous breaking can produce cosmological domain walls. We calculate the temperature-dependent tension and width of the $\mathbb{Z}_{6/p}$ domain walls below the electroweak scale, which exist for every quotient except $\Gamma=\mathbb{Z}_6$, finding that their thickness can reach the kilometer scale around reionization in our benchmarks. These macroscopic thermal interfaces, if produced, would constitute the ``Great Walls'' of the Standard Model. We develop a phenomenological thermal velocity-dependent one-scale model to follow their network evolution, incorporating temperature-dependent surface tension, inertia, and plasma drag. For sufficiently early formation, our benchmarks give an estimated wall energy fraction of order $10^{-14}$ near the onset of Big Bang nucleosynthesis, with further suppression at later times. Uncertainties arise from relating surface free energy to physical wall energy and from late-time plasma evolution and structure formation. Despite their small energy fraction, these walls could probe the global structure of the SM gauge group, motivating dedicated searches for their cosmological signatures.
\end{abstract}

\thispagestyle{empty}  
\newpage    
\setcounter{page}{1}  

\begingroup
\hypersetup{linkcolor=black,linktocpage}
\renewcommand{\baselinestretch}{0.98}\selectfont
\tableofcontents
\endgroup

\newpage


\section{Introduction}

Despite the success of the Standard Model (SM), the global structure of its gauge group remains undetermined. The known particles and their local interactions are compatible with
\begin{equation}
G_{\rm SM}=\frac{SU(3)_c\times SU(2)_L\times U(1)_Y}{\Gamma}\,,
\qquad
\Gamma=1,\ \mathbb{Z}_2,\ \mathbb{Z}_3,\ \mathbb{Z}_6\,,
\end{equation}
where the nontrivial quotients are subgroups of a diagonal $\mathbb{Z}_6$ that acts trivially on all SM fields. These choices share the same gauge algebra but differ in their allowed representations, line operators, and electric and magnetic charge lattices. The relation between the global gauge group and charge quantization has long been recognized; see, for example, Refs.~\cite{ORaifeartaigh:1986agb,Hucks:1990nw}. Determining this global structure is therefore another fundamental question about the SM, alongside understanding the Higgs sector and the origin and properties of neutrino masses and mixings.

Generalized global symmetries provide a useful framework for formulating this question~\cite{Gaiotto:2014kfa,Gomes:2023ahz,Bhardwaj:2023kri,Iqbal:2024pee}. In particular, an electric one-form symmetry acts on Wilson lines rather than local operators and, in four spacetime dimensions, is generated by topological surface operators. An invertible, group-like one-form symmetry is necessarily Abelian, even when the underlying gauge group is non-Abelian. With the known SM matter content, the direct-product gauge theory could have an electric $\mathbb{Z}_6$ one-form symmetry. Taking the quotient by $\Gamma=\mathbb{Z}_p$, with $p=1,2,3,6$, gauges this subgroup and leaves the electric symmetry $\mathbb{Z}_6/\Gamma\simeq\mathbb{Z}_{6/p}$. Thus, the quotient changes which extended operators are genuine observables, even though it does not change the perturbative interactions of the known particles~\cite{Aharony:2013hda,Tong:2017oea,Koren:2024xof}.

There are several complementary ways to probe the quotient. First, the allowed electric and magnetic line operators depend on $\Gamma$; measurements of the charges of suitable magnetic objects, potentially including magnetically charged black holes, could therefore constrain the global gauge group~\cite{Tong:2017oea}. Second, different quotients permit different representations for additional particles, including particles with fractional electric charge~\cite{Alonso:2024pmq,Koren:2024xof}. Their discovery would provide information about the quotient, although the group structure alone does not determine their masses or guarantee their existence. Such particles can also explicitly break part or all of the electric one-form symmetry, leaving an approximate symmetry at energies and temperatures well below their masses. Third, the thermal realization of the surviving global symmetry can support topological defects. The central aim of this work is to develop this defect-based perspective by studying thermal domain walls and their possible
cosmological evolution.

On the thermal spacetime $S^1_\beta\times\mathbb{R}^3$, with $\beta=1/T$, a four-dimensional one-form symmetry gives rise to both a zero-form symmetry acting on wrapped lines and a one-form symmetry acting on spatial lines. Wrapped Wilson lines are Polyakov loops, which serve as local order parameters in the three-dimensional equilibrium description~\cite{Polyakov:1978vu,Gaiotto:2014kfa}; related compactification methods are reviewed in Ref.~\cite{Poppitz:2021cxe}. At high temperature and weak coupling, thermal fluctuations generate a perturbative potential for the holonomy, with degenerate center-related minima in pure Yang--Mills theory~\cite{Gross:1980br,Weiss:1981ev}. A constrained effective potential for Polyakov loops provides a gauge-invariant formulation, including the contributions required beyond leading order~\cite{KorthalsAltes:1993ca}. When a discrete thermal zero-form symmetry is spontaneously broken, the gauge-inequivalent minima form a disconnected set $\mathcal{M}$, and the nontrivial zeroth homotopy  $\pi_0(\mathcal{M})$ allows domain-wall interface sectors. The physical interpretation of these interfaces has a long history. Smilga questioned the identification of Euclidean center interfaces with physical walls, while Hansson, Nielsen, and Zahed exhibited subtleties of symmetry breaking and domain coexistence in a two-dimensional unequal-charge QED model~\cite{Smilga:1994sn,Hansson:1994ep}. In pure Yang--Mills theory, a gauge-invariant interface diagnostic is supplied by the spatial 't Hooft loop. From the electric--magnetic order--disorder viewpoint, Wilson loops probe electric charges, while 't Hooft loops insert quantized center magnetic flux, producing a center phase in linked Wilson loops. In the thermal path integral, a spatial 't Hooft loop $V(C)$ imposes a center twist across a surface bounded by $C$. In the deconfined phase, its leading area law is $-\log\langle V(C)\rangle\simeq\sigma A(C)/T$, where $A(C)$ is the minimal spanning area and $\sigma$ is the interface free energy per unit area~\cite{Bhattacharya:1990hk,Korthals-Altes:1999cqo,Korthals-Altes:2000tia,Giovannangeli:2001bh}. For the SM application, the relevant symmetry and twist must respect the dynamical matter and the chosen quotient. The generalized symmetry viewpoint thus establishes the gauge-invariant basis of the equilibrium interface construction, while cosmological production and real-time network dynamics require additional analysis.

Although the full SM can possess such discrete symmetries, we concentrate on the color--electromagnetic sector below electroweak symmetry breaking. Below the electroweak scale the residual $\mathbb{Z}^{\mathfrak{e}(1)}_{6/p}$ symmetry acts on the $SU(3)_c$ and $U(1)_{\rm EM}$ holonomies. On the thermal circle it becomes a $\mathbb{Z}^{(0)}_{6/p}$ symmetry with $6/p$ degenerate thermal vacua. Domain walls therefore exist for $\Gamma=1,\mathbb{Z}_2,\mathbb{Z}_3$ and are absent only for $\Gamma=\mathbb{Z}_6$. The elementary wall is the same interface in all three cases; the quotients differ in the number of vacua (six, three, two) and hence in the topology of the wall network. All of our results concern the theory below the electroweak scale; formation above it requires matching to the full $SU(3)_c\times SU(2)_L\times U(1)_Y$ holonomy potential, which we do not compute. High-temperature SM holonomies have been studied previously~\cite{KorthalsAltes:1994be}, and a recent analysis classifies their vacua for the different quotients~\cite{Xu:2026frw}. Here we follow the wall profiles and their plasma environment toward much lower temperatures, where QCD and electromagnetism control the relevant dynamics. We use a phenomenological Polyakov-loop-extended Nambu--Jona-Lasinio (PNJL) description to connect the deconfined regime to the regime below the QCD crossover~\cite{Fukushima:2003fw,Ratti:2005jh,Fukushima:2017csk}.

Above the QCD crossover, the walls involve both color and electromagnetic thermal holonomies, together with a nontrivial chiral-field profile. Below $T_{\rm QCD}$, the color holonomy becomes approximately fixed and the wall is predominantly electromagnetic. In relativistic regimes, the leading scalings are $\sigma\propto T^3$ and $\ell\propto T^{-1}$, up to coupling dependence and changes in the active plasma species. Below the electron mass, the charged-particle density becomes essential: in the equilibrium, dilute electron--proton approximation, $\sigma\propto T^{3/2}(n_e^{\rm free})^{1/2}$ and $\ell\propto(T/n_e^{\rm free})^{1/2}$. Electron--positron annihilation and recombination consequently reduce the tension and increase the width. For illustration, a residual free-electron density $n_e^{\rm free}\simeq2\times10^{-7}~{\rm cm}^{-3}$ and a matter temperature $T_m\simeq8~{\rm K}$ near $z\simeq20$ give $\ell\simeq 1~{\rm km}$. The possibility of such macroscopic widths motivates the name ``The Great Wall of the Standard Model.''

Thermal walls could be seeded as the plasma thermalizes during reheating, if regions select different symmetry-related thermal phases. Causal ordering provides a possible basis for domain formation~\cite{Kibble:1976sj,Zurek:1985qw}; alternatively, nonthermal fluctuations produced during preheating could seed domains upon subsequent thermalization~\cite{Kofman:1994rk,Kofman:1997yn}. Neither possibility fixes a production rate here, and we adopt $L_{\rm form}=H_{\rm form}^{-1}$ and $v_{\rm form}=0$ as benchmark initial conditions, where $L_{\rm form}$ and $v_{\rm form}$ denote the characteristic length scale and root-mean-square velocity of the wall network at formation, respectively. Formation above the electroweak scale additionally requires matching to the full SM, which is not calculated in this work. We describe the subsequent evolution using a thermal velocity-dependent one-scale (TVOS) model, motivated by Refs.~\cite{Avelino:2005kn,Martins:2016ois}. Unlike conventional fixed-tension walls, these interfaces have a temperature-dependent free-energy tension; their surface inertia need not equal that tension, especially in a nonrelativistic plasma. Their drag also changes with the plasma composition, from quark scattering above the QCD crossover to electron and positron interactions below it, and to electromagnetic plasma response, including photon-induced drag, below $m_e$. The disappearance of thermal pairs suppresses the charged-particle response, whereas the surviving electron--ion population continues to contribute. Photons themselves are not Boltzmann suppressed. These distinctions require separate inputs for the static tension, inertia, and drag.

The resulting benchmark solutions exhibit successive periods of stretching and curvature-driven coarsening. The characteristic number of wall crossings along a Hubble-length line, $N_{\rm cross}\sim(HL)^{-1}$, can evolve nonmonotonically as the friction and other plasma-dependent inputs change. For representative choices of the late-time response and inertia, the network reaches $HL\sim10^{-3}$ and $v\sim10^{-5}$ by our chosen endpoint near $z_{\rm rei}=20$; these values are model-dependent benchmarks rather than universal predictions. We stop before modeling astrophysical reionization, after which gas heating, inhomogeneous ionization, and structure formation complicate the evolution. Local wall widths and tensions may then differ between intergalactic gas, galaxies, and clusters. A larger local wall abundance in denser environments is possible if the walls are entrained by the matter flow, but it does not follow from the electron-density dependence of an individual wall. Within the adopted energy-density prescription, the wall contribution is already tiny at the endpoint, and the illustrative present-day estimates are likewise negligible; neither estimate constitutes a prediction of the post-reionization network distribution.

The remainder of this paper is organized as follows. Section~\ref{sec:generalize_sym} reviews the generalized symmetries associated with the SM global structure and their thermal reduction. Section~\ref{sec:EFT} constructs the effective theory of the color and electromagnetic thermal holonomies, including the PNJL description of the QCD crossover. Section~\ref{sec:Z3_DW_property} determines the wall profiles, tensions, and widths across the relevant cosmological epochs. Section~\ref{sec:plasma_friction} studies the interactions with the thermal plasma and the inputs needed for real-time wall motion. Section~\ref{sec:s56_network_evolution} presents the formation assumptions, TVOS equations, and benchmark network evolution, together with a discussion of possible present-day wall properties. Section~\ref{sec:conclusions} contains our discussion and conclusions. Appendix~\ref{app:1Loop-selfE} gives the holonomy-dependent one-loop self-energies, Appendix~\ref{app:WKB_reflection} develops the fermion WKB reflection calculation, and Appendix~\ref{app:modified_VOS} derives the TVOS equations with a non-unity plasma thermal propagation speed.

\section{From generalized symmetries to thermal discrete symmetries}\label{sec:generalize_sym}

In this section, we first provide some pedagogical discussion for the generalized symmetries and the corresponding ones in the SM given the known gauge symmetries and matter fields. We will then discuss the remaining global discrete symmetries in the thermal field theory.

\subsection{Global 1-form symmetries in the Standard Model}

Ordinary global symmetries are also called $0$-form symmetries because they act on
local operators.  In four spacetime dimensions, a group element
$g\in G^{(0)}$ is represented by an invertible topological codimension-one
operator $U_g(\Sigma_3)$.  When $\Sigma_3$ is chosen to be a spatial
Cauchy surface, this reduces to the usual unitary symmetry operator acting
on the Hilbert space.  A $1$-form symmetry $G^{(1)}$, by contrast, is
generated by invertible topological codimension-two operators
$U_g(\Sigma_2)$ and acts on line operators.

For a closed line operator $\mathcal O_R(C)$ and a closed surface
$\Sigma_2$, the symmetry action is characterized by
\begin{align}
 \left\langle U_g(\Sigma_2)\,\mathcal O_R(C)\,\cdots\right\rangle
 =
 \chi_R(g)^{{\rm Link}(\Sigma_2,C)}
 \left\langle \mathcal O_R(C)\,\cdots\right\rangle ,
 \label{eq:one_form_action}
\end{align}
where $\chi_R:G^{(1)}\to U(1)$ is the character specifying the
$1$-form charge of $\mathcal O_R$.  Topological invariance means that
$\Sigma_2$ may be continuously deformed without changing the correlation
function, provided that it does not cross a charged line.  If
$\Sigma_2=\partial V_3$, then
${\rm Link}(\Sigma_2,C)=I(V_3,C)$ is the oriented intersection number of
$V_3$ with $C$.

For example, for $G^{(1)}=\mathbb Z_N$, writing
$g_k=e^{2\pi i k/N}$ and labeling the line electric charge, or called $N$-ality of the representation, by
$z_R\in\mathbb Z_N$, one has
\begin{align}
 \chi_R(g_k)^{{\rm Link}(\Sigma_2,C)}
 =
 \exp\left[
 \frac{2\pi i}{N}\,k\,z_R\,{\rm Link}(\Sigma_2,C)
 \right].
\end{align}
For $p\geq1$, a group-like $p$-form symmetry is necessarily Abelian.
In gauge theories, an electric $1$-form symmetry is often realized as
the subgroup of the gauge-group center that acts trivially on all
dynamical matter fields.

Consider the unquotiented direct-product Standard Model gauge group
\begin{align}
    \widetilde G_{\rm SM}
    = SU(3)_c\times SU(2)_L\times U(1)_Y .
\end{align}
In the absence of dynamical matter fields and dynamical magnetic
monopoles, the complete global $1$-form symmetry is
\begin{align}
    \mathbb Z_{3,c}^{\mathfrak e(1)}
    \times
    \mathbb Z_{2,L}^{\mathfrak e(1)}
    \times
    U(1)_{Y}^{\mathfrak e(1)}
    \times
    U(1)_{Y}^{\mathfrak m(1)} ,
    \label{eq:pure_SM_one_form_symmetry}
\end{align}
where $\mathfrak e$ and $\mathfrak m$ denote electric and magnetic
$1$-form symmetries, respectively. The first two factors arise from
the centers of $SU(3)_c$ and $SU(2)_L$. The electric
$U(1)_{Y}^{\mathfrak e(1)}$ symmetry is present because there are no
dynamical electrically charged particles, while
$U(1)_{Y}^{\mathfrak m(1)}$ follows from the Abelian Bianchi identity
and is exact in the absence of dynamical magnetic monopoles.

Including the SM matter fields listed in Table~\ref{tab:SMgauge_charge},
screening by dynamical electrically charged particles breaks the electric
$1$-form symmetry of the unquotiented gauge theory down to a diagonal
subgroup.  The surviving global $1$-form symmetry is
\beqa
    G^{(1)}
    =
    \mathbb{Z}_{6}^{\mathfrak e(1)}
    \times
    U(1)_{Y}^{\mathfrak m(1)} ~.
\eeqa

\begin{table}[hb!]
    \centering
    \renewcommand{\arraystretch}{1.2} 
    \begin{tabular}{c|c|c} \hline\hline
        particle & $SU(3)_c\times SU(2)_L\times U(1)_Y$ & $(z_3^{\mathfrak{e}},z_2^{\mathfrak{e}},6Y)$  \\ \hline\hline
        $Q_L$ & (\textbf{3,2},1/6)  & (1,1,1) \\
        $u_R$ & (\textbf{3,1},2/3)  & (1,0,4) \\
        $d_R$ & (\textbf{3,1},-1/3)  & (1,0,-2) \\
        $L_L$ & (\textbf{1,2},-1/2) & (0,1,-3)\\
        $e_R$ & (\textbf{1,1},-1)  & (0,0,-6) \\\hline
        $g$ & (\textbf{8,1},0)  & (0,0,0) \\
        $W$ & (\textbf{1,3},0)  & (0,0,0) \\ 
        $B$ & (\textbf{1,1},0)  & (0,0,0) \\ 
        $H$ & (\textbf{1,2},1/2)  & (0,1,3) \\ \hline\hline
    \end{tabular}
    \caption{SM matter content and charges under the gauge group and its center symmetries. Here, $z_3^{\mathfrak{e}}$ and $z_2^{\mathfrak{e}}$ are the center charges of $SU(3)_c$ and $SU(2)_L$, respectively. }
    \label{tab:SMgauge_charge}
\end{table}

To see this explicitly, consider a Wilson line operator
\begin{align}
    \mathcal{W}_{R}(C)
    \equiv
    {\rm Tr}_{R}\,
    {\cal P}\exp\left[
        i\oint_C dx^\mu
        \left(
            g_s G_\mu^a T^a_{R_c}
            +g_L W_\mu^i T^i_{R_L}
            +g_Y Y B_\mu
        \right)
    \right] ,
    \label{eqn:WilsonLine}
\end{align}
where $R=(R_c,R_L)_Y$ specifies the representations under
$SU(3)_c\times SU(2)_L$ and the hypercharge $Y$ and $C$ is a closed spacetime curve,  so that the traced Wilson line is gauge invariant. The Wilson line is characterized by the $SU(3)_c$ triality $z_3^{\mathfrak e}\in\mathbb{Z}_3$,
the $SU(2)_L$ $2$-ality $z_2^{\mathfrak e}\in\mathbb{Z}_2$, and the
hypercharge $Y$.  Its charge under the diagonal electric
$\mathbb{Z}_{6}^{\mathfrak e(1)}$ symmetry is
\begin{align}
    q_6(R)
    =
    2\,z_3^{\mathfrak e}
    +3\,z_2^{\mathfrak e}
    +6\,Y
    \pmod 6 ~.
    \label{eqn:Z6charge}
\end{align}
The action of its element $k\in\mathbb{Z}_6$ is
\begin{align}
    U_k(\Sigma_2)\,\mathcal{W}_{R}(C)
    =
    \exp\left[
        \frac{2\pi i k}{6}\,
        q_6(R)\,
        {\rm Link}(\Sigma_2,C)
    \right]
    \mathcal{W}_{R}(C)\,U_k(\Sigma_2) ~.
    \label{eqn:Z6transformation}
\end{align}
Equivalently, the generator corresponds to the simultaneous center
transformation
\begin{align}
    \left(
        e^{2\pi i/3}\,\mathbb{I}_3,\,
        -\mathbb{I}_2,\,
        e^{2\pi iY}
    \right) ,
\end{align}
whose action on a representation $R$ gives
$\exp[2\pi i\,q_6(R)/6]$.  It is straightforward to verify that every
SM matter multiplet satisfies
\begin{align}
    2\,z_3^{\mathfrak e}
    +3\,z_2^{\mathfrak e}
    +6\,Y
    \equiv 0 \pmod 6 ~.
    \label{eqn:Z6condition}
\end{align}
Consequently, all SM matter fields are neutral under this diagonal
transformation and cannot screen the corresponding Wilson-line charge.
Therefore, $\mathbb{Z}_{6}^{\mathfrak e(1)}$ remains an exact global
$1$-form symmetry of the unquotiented theory with the SM matter content. 

Another way to understand the global electric $1$-form symmetry of the unquotiented SM gauge theory is to consider the lattice of possible Wilson lines $\mathcal{W}_{z^{\mathfrak{e}}_3,z^{\mathfrak{e}}_2,6Y}$, labeled by $G_W= \mathbb{Z}_3 \times \mathbb{Z}_2\times \mathbb{Z}$. Dynamical charged matter fields can screen Wilson lines. For instance, since the left-handed quark doublet $Q_L$ carries the charge label $(1,1,1)$, the composite object
\begin{align}
\mathcal{W}_{z^{\mathfrak{e}}_3,z^{\mathfrak{e}}_2,6Y}(C)\, Q_L(x\in C)
\end{align}
transforms in the same way as the Wilson line $\mathcal{W}_{z^{\mathfrak{e}}_3+1,z^{\mathfrak{e}}_2+1,6Y+1}(C)$. These two Wilson lines therefore belong to the same screening class, implying that the group of inequivalent unscreened Wilson-line charges is smaller than $G_W$. More generally, two Wilson lines are identified if their charge labels differ by the charge vector of a combination of dynamical SM matter fields. After quotienting by all such screening relations, the group of unscreened Wilson-line charges is
\begin{align}
D_1 =
G_W/Q_{\rm matter} =
\frac{\mathbb{Z}_3\times\mathbb{Z}_2\times\mathbb{Z}}{
\left\langle \left(z_3^{\mathfrak e},z_2^{\mathfrak e},6Y\right)
:{\rm matter}\in{\rm SM}\right\rangle
} ~.
\end{align}
Using the Smith normal form, one obtains $D_1\simeq\mathbb{Z}_6$. The six distinct screening classes may be represented, for example, by $\mathcal{W}_{0,0,0}$, $\mathcal{W}_{0,0,1}$, $\mathcal{W}_{1,0,0}$, $\mathcal{W}_{0,1,0}$, $\mathcal{W}_{2,0,0}$, and $\mathcal{W}_{1,1,0}$. The global electric $1$-form symmetry is the Pontryagin dual of $D_1$. Since $\mathbb{Z}_6$ is self-dual, one obtains the global electric 1-form symmetry
\begin{align}
G^{\mathfrak e(1)}
=
\widehat{D}_1
\simeq
\mathbb{Z}_6^{\mathfrak e(1)} ~.
\end{align}

As noted long ago in Refs.~\cite{ORaifeartaigh:1986agb,Hucks:1990nw}, the global structure of the SM gauge group has a fourfold ambiguity, corresponding to the choices $\Gamma_p\simeq\mathbb{Z}_p$ with $p=1,2,3,6$. Including the SM matter content and assuming the absence of dynamical magnetic monopoles, the corresponding global $1$-form symmetry is
\beqa
G_{{\rm SM},p}
=
\frac{SU(3)_c\times SU(2)_L\times U(1)_Y}{\Gamma_p}
\quad\Rightarrow\quad
G_p^{(1)}
=
\mathbb{Z}_{6/p}^{\mathfrak e(1)}
\times
U(1)_{Y}^{\mathfrak m(1)} ~,
\eeqa
where the $\mathbb{Z}_{6/p}^{\mathfrak e(1)}$ symmetry is exact in the absence of additional matter charged under it and emerges only approximately at energies below the masses of any such additional matter fields.  

Without imposing a charge-completeness assumption, the spectrum of dynamical electric matter does not uniquely determine the global form of the gauge group; it only requires the quotient subgroup $\Gamma_p$ to act trivially on all dynamical electric matter fields. The quotient may also be probed by the allowed magnetic or dyonic charge lattice. Electric, magnetic, and dyonic objects can screen Wilson, 't Hooft, and Wilson--'t Hooft lines, respectively, and therefore affect the corresponding electric, magnetic, or dyonic $1$-form symmetries in different ways.

Different choices of the quotient admit different representations for additional heavy states beyond the SM. For $\Gamma_6=\mathbb{Z}_6$, every allowed electric matter representation must satisfy
\begin{align}
q_6
=
2\,z_3^{\mathfrak e}
+3\,z_2^{\mathfrak e}
+6\,Y
\equiv 0 \pmod 6 ~,
\label{eqn:Z6quotient_condition}
\end{align}
as do all existing SM particles. Consequently, there is no nontrivial discrete electric $1$-form symmetry, because the entire diagonal $\mathbb{Z}_6^{\mathfrak e(1)}$ symmetry has been gauged.

For $\Gamma_3=\mathbb{Z}_3$, an allowed matter representation must be invariant under the gauged $\mathbb{Z}_3$ subgroup and therefore satisfy
\begin{align}
2\,z_3^{\mathfrak e}
+6\,Y
\equiv 0 \pmod 3 ~,
\end{align}
but it may be charged under the residual $\mathbb{Z}_2^{\mathfrak e(1)}$ symmetry. For example, a field in the $(\mathbf{3},\mathbf{2},2/3)$ representation satisfies the $\mathbb{Z}_3$ quotient condition but carries a nontrivial $\mathbb{Z}_2$ charge. If all additional states carrying this charge are very heavy and decouple, an approximate emergent
$\mathbb{Z}_{6/p}^{\mathfrak e(1)}
=\mathbb{Z}_2^{\mathfrak e(1)}$
$1$-form symmetry is present at energies well below their masses. The inequivalent unscreened Wilson-line charges are $\mathcal{W}_{0,0,0}$ and $\mathcal{W}_{0,1,0}$. Alternatively, the $\mathbb{Z}_3$ component of the quotient may be probed through the magnetic sector. In particular, the $\Gamma_3=\mathbb{Z}_3$ theory admits a purely magnetic line with $(z_3^{\mathfrak m},z_2^{\mathfrak m},g_Y^{\mathfrak m})=(1,0,1/3)$ for zero $\theta$ angles~\cite{Tong:2017oea}. This line is not allowed for $\Gamma_1=1$ or $\Gamma_2=\mathbb{Z}_2$, although it is also allowed for $\Gamma_6=\mathbb{Z}_6$. It therefore probes the presence of the $\mathbb{Z}_3$ quotient subgroup without uniquely determining the full quotient. If an object carrying this magnetic charge is dynamical and electrically neutral, it screens the corresponding magnetic lines but leaves the residual electric $\mathbb{Z}_2^{\mathfrak e(1)}$ symmetry unbroken. The simultaneous presence of this magnetic charge and an unbroken residual global $\mathbb{Z}_2^{\mathfrak e(1)}$ symmetry therefore distinguishes $\Gamma=\mathbb{Z}_3$ from $\Gamma=\mathbb{Z}_6$.   

Similarly, for $\Gamma_2=\mathbb{Z}_2$, an allowed matter representation must be invariant under the gauged $\mathbb{Z}_2$ subgroup and therefore satisfy
\begin{align}
z_2^{\mathfrak e}
+6\,Y
\equiv 0 \pmod 2 ~, 
\end{align}
but it may be charged under the residual $\mathbb{Z}_3^{\mathfrak e(1)}$ symmetry. For example, a field in the $(\mathbf{3},\mathbf{2},1/2)$ representation satisfies the $\mathbb{Z}_2$ quotient condition but carries a nontrivial $\mathbb{Z}_3$ charge. If all additional states carrying this charge are very heavy and decouple, an approximate emergent
$\mathbb{Z}_{6/p}^{\mathfrak e(1)}
=\mathbb{Z}_3^{\mathfrak e(1)}$
$1$-form symmetry is present at energies well below their masses. The inequivalent unscreened Wilson-line charges are $\mathcal{W}_{0,0,0}$, $\mathcal{W}_{1,0,0}$ and $\mathcal{W}_{2,0,0}$. Alternatively, the $\mathbb{Z}_2$ component of the quotient may be probed through the magnetic sector. In particular, the $\Gamma_2=\mathbb{Z}_2$ theory admits a purely magnetic line with $(z_3^{\mathfrak m},z_2^{\mathfrak m},g_Y^{\mathfrak m})=(0,1,1/2)$~\cite{Tong:2017oea}. This line is not allowed for $\Gamma_1=1$ or $\Gamma_3=\mathbb{Z}_3$, although it is also allowed for $\Gamma_6=\mathbb{Z}_6$. It therefore probes the presence of the $\mathbb{Z}_2$ quotient subgroup without uniquely determining the full quotient. If an object carrying this magnetic charge is dynamical and electrically neutral, it screens the corresponding magnetic lines but leaves the residual electric $\mathbb{Z}_3^{\mathfrak e(1)}$ symmetry unbroken. The simultaneous presence of this magnetic charge and an unbroken residual global $\mathbb{Z}_3^{\mathfrak e(1)}$ symmetry therefore distinguishes $\Gamma=\mathbb{Z}_2$ from $\Gamma=\mathbb{Z}_6$.

For $\Gamma_1=1$, there is no quotient constraint on additional electric matter representations. Heavy matter may therefore carry nontrivial $\mathbb{Z}_2$ and/or $\mathbb{Z}_3$ charges and break part or all of the diagonal $\mathbb{Z}_6^{\mathfrak e(1)}$ symmetry. In particular, heavy states whose charges generate the full $\mathbb{Z}_6$ charge group break this symmetry completely. The magnetic sector provides complementary information about the quotient. For example, the fractional combined magnetic charges $(z_3^{\mathfrak m},z_2^{\mathfrak m},g_Y^{\mathfrak m})=(1,0,1/3)$ and $(0,1,1/2)$ are not compatible with the full Wilson-line spectrum of the $\Gamma_1=1$ theory; their existence would instead probe the presence of a $\mathbb{Z}_3$ or $\mathbb{Z}_2$ quotient subgroup, respectively~\cite{Tong:2017oea}. A dynamical, electrically neutral magnetic object screens the corresponding 't Hooft lines and may break or reduce the magnetic $1$-form symmetry, but it does not break the electric $\mathbb{Z}_6^{\mathfrak e(1)}$ symmetry. By contrast, a dyonic object may also break the electric symmetry, depending on its electric $\mathbb{Z}_6$ charge. Thus, if all electrically or dyonically charged states that transform nontrivially under $\mathbb{Z}_6^{\mathfrak e(1)}$ are very heavy and decouple, an approximate emergent
$\mathbb{Z}_{6/p}^{\mathfrak e(1)}
=\mathbb{Z}_6^{\mathfrak e(1)}$
$1$-form symmetry is present at low energies.\footnote{An analogous distinction arises between $SU(2)$ and $SU(2)/\mathbb{Z}_2\simeq SO(3)$ gauge theories. The $SU(2)$ gauge group admits half-integer-isospin representations, whereas $SO(3)$ admits only integer-isospin representations. Pure $SU(2)$ gauge theory has an exact electric $\mathbb{Z}_2^{\mathfrak e(1)}$ symmetry. If half-integer-isospin matter is present but very heavy, this symmetry is broken in the full theory but emerges approximately at energies well below the matter mass scale. By contrast, in the $SO(3)$ theory the electric $\mathbb{Z}_2^{\mathfrak e(1)}$ symmetry has been gauged and is therefore absent. The $SO(3)$ theory instead possesses a dual magnetic $\mathbb{Z}_2^{\mathfrak m(1)}$ symmetry, or a dyonic $\mathbb{Z}_2^{(1)}$ symmetry for the alternative choice of discrete theta angle~\cite{Aharony:2013hda}.}

\subsection{Thermal compactification of one-form symmetries}

Finite-temperature quantum field theory is formulated in Euclidean time $\tau$, with the identification of $\tau \sim \tau + \beta$ with $\beta = 1/T$. Bosonic fields are periodic around $S^1_\beta$, while fermionic fields are antiperiodic. At spatial distances $r \gg \beta$, nonzero Matsubara modes are heavy and the static sector may be regarded as a three-dimensional effective theory on $\mathbb{R}^3$. A $p$-form global symmetry is characterized by topological symmetry operators of codimension
$p+ 1$ and by charged operators of dimension $p$. Compactifying one direction gives an unwrapped and a wrapped descendant. Consequently, a four-dimensional 1-form symmetry gives both a three-dimensional 1-form symmetry and a three-dimensional 0-form symmetry. This is the $p=1$ instance of the general reduction rule~\cite{Gaiotto:2017yup}
\beqa
G^{(p)}_{d} \rightarrow G^{(p)}_{d-1} \times G^{(p-1)}_{d-1} ~. 
\eeqa

In four-dimensional Euclidean thermal spacetime, there are two basic ways to orient a Wilson-line contour relative to the thermal circle. The first is a spatial Wilson loop $\mathcal{W}(C)$ with $C\subset\mathbb{R}^3$. From the three-dimensional perspective, it remains a line operator and is acted upon by the wrapped symmetry operator $U(\Sigma_2)$ with $\Sigma_2=\gamma_1\times S_\beta^1$. After dimensional reduction, $\Sigma_2$ becomes the codimension-two line $\gamma_1\subset\mathbb{R}^3$, which generates the three-dimensional $1$-form symmetry $G_{d=3}^{(1)}$. The wrapped symmetry surface can link with the spatial Wilson loop and therefore act nontrivially on it.

The second possibility is a Wilson line wrapping once around the Euclidean thermal circle while remaining at a fixed spatial position $\mathbf{x}$, $\mathcal{W}\!\left(\{\mathbf{x}\}\times S_\beta^1\right)$. From the three-dimensional perspective, this wrapped Wilson line becomes a local Polyakov-loop operator. It is acted upon by an unwrapped symmetry operator $U(\Sigma_2)$ with $\Sigma_2\subset\mathbb{R}^3$. In three dimensions, $\Sigma_2$ is a codimension-one surface that generates the ordinary $0$-form symmetry $G_{d=3}^{(0)}$ of the thermal effective theory. The unwrapped symmetry surface can intersect the wrapped Wilson line at a point and therefore act nontrivially on the local Polyakov-loop operator.

Explicitly, the thermal Wilson line, or Polyakov loop, in a representation
$r$ is
\begin{align}
    L_r(\mathbf{x})
    =
    \frac{1}{\dim(r)}
    \operatorname{Tr}_r
    \left[
        \textsc{P}
        \exp\left(
            i g\int_0^\beta \dd \tau\,
            A_0(\mathbf{x},\tau)
        \right)
    \right] ~,
    \qquad
    \beta=\frac{1}{T} ~.
\end{align}
Here $A_0$, equivalently denoted by $A_\tau$ in Euclidean notation, is
the component of the gauge field along the thermal circle. Indeed, for
a Wilson line evaluated on the contour
$C_{\mathbf{x}}=\{\mathbf{x}\}\times S_\beta^1$, one has
$\dd x^i=0$ and hence $\oint_{C_{\mathbf{x}}} A_\mu\,dx^\mu = \int_0^\beta \dd\tau\,A_0(\mathbf{x},\tau)$. Thus, the Polyakov loop measures the gauge holonomy around the Euclidean
thermal circle. The path-ordering symbol $\textsc{P}$ is required
because $A_0(\mathbf{x},\tau)$ at different values of $\tau$ need not
commute. 

\subsection{$SU(3)_c \times U(1)_{\rm EM}$ and its thermal $\mathbb{Z}^{(0)}_{6/p}$ symmetry}

In this study, we focus on the thermal Universe after electroweak symmetry breaking. The relevant SM particle content is listed in Table~\ref{tab:QCDgauge_charge}. The unbroken $U(1)_{\rm EM}\subset SU(2)_L\times U(1)_Y$ is the one-parameter subgroup $\big({\rm diag}(e^{i\theta/2},e^{-i\theta/2}),\,e^{i\theta}\big)$, which acts on a field of charge $Q=T_3+Y$ as $e^{i\theta Q}$. In the direct-product theory it closes only at $\theta=12\pi$: the $SU(2)_L$ factor requires $\theta\in4\pi\Z$ and, with $6Y\in\Z$, the $U(1)_Y$ factor requires $\theta\in12\pi\Z$. The element at $\theta=6\pi$ is $g^3=(\mathbb{I}_3,-\mathbb{I}_2,e^{6\pi iY})$, and $g^2$ times a color-center element appears at $\theta=4\pi$. Writing $t=\theta/2\pi$, the diagonal $\Z_6$ generator acts on the unbroken group as $g=(e^{2\pi i/3}\mathbb{I}_3,\,t\to t+1)$, and genuine Wilson lines carry $6Q\in\Z$ for $\Gamma=1,\Z_3$ (e.g.\ $Q=\pm1/2$ from $(\mathbf 1,\mathbf 2,0)$, $Q=1/6$ from $(\mathbf 1,\mathbf 1,1/6)$) and $3Q\in\Z$ for $\Gamma=\Z_2,\Z_6$. Since the Higgs and the $W^\pm$ are $\Z_6$-neutral, $\Gamma_p$ lies inside the unbroken group and the infrared gauge group is $H_p=[SU(3)_c\times U(1)_{\rm EM}]/\Gamma_p$.

\begin{table}[hb!]
    \centering
    \renewcommand{\arraystretch}{1.2}
    \begin{tabular}{c|c|c} \hline\hline
        particle & $SU(3)_c\times U(1)_{\rm EM}$ & $(z_3^e,6\,Q)$  \\ \hline\hline
        $\ell$ & (\textbf{1},-1) & (0,-6)\\
        $u$ & (\textbf{3},2/3)  & (1,4) \\
        $d$ & (\textbf{3},-1/3)  & (1,-2) \\ \hline
        $g$ & (\textbf{8},0)  & (0,0) \\
        $A$ & (\textbf{1},0)  & (0,0) \\ \hline\hline
    \end{tabular}
    \caption{The $SU(3)_c\times U(1)_{\rm EM}$ gauge group and the electric charges under its center. All fields have $0=2z^{\mathfrak{e}}_3+6Q \pmod 6$. }
    \label{tab:QCDgauge_charge}
\end{table}

In the infrared the Wilson lines are labelled by $(z^{\mathfrak{e}}_3,6Q)\in\Z_3\times\Z$ and the dynamical charges are those of Table~\ref{tab:QCDgauge_charge}; quotienting by the screening relations, $\Z_3\times\Z/\langle(1,-2),(0,6)\rangle\simeq\Z_6$, generated by $(e^{2\pi i/3})^{z^{\mathfrak{e}}_3}e^{2\pi iQ}$. The electric one-form symmetry of the unquotiented theory below the electroweak scale is therefore again $\Z_6^{\mathfrak{e}(1)}$, as it must be, since Higgsing by a $\Z_6$-neutral field does not change the screening classes.

After electroweak symmetry breaking, the possible global forms of the gauge group are
\begin{align}
H_p=\frac{SU(3)_c\times U(1)_{\rm EM}}{\Gamma_p} \quad\mbox{for} \quad p=1,2,3,6\,,
\end{align}
with $U(1)_{\rm EM}$ of period $t\sim t+6$. For the SM matter content and in the absence of dynamical magnetically charged objects, the corresponding global $1$-form symmetry is
\beqa
H_p=\frac{SU(3)_c\times U(1)_{\rm EM}}{\Gamma_p}
\quad\Rightarrow\quad
H_p^{(1)}
=
\mathbb{Z}_{6/p}^{\mathfrak e(1)}
\times
U(1)_{\rm EM}^{\mathfrak m(1)} ~.
\eeqa

After compactification on the thermal circle, and for the quotient $\Gamma_p$, the four-dimensional electric discrete $1$-form symmetry decomposes into a three-dimensional $1$-form symmetry and a three-dimensional $0$-form symmetry:
\beqa
\mathbb{Z}_{6/p,d=4}^{\mathfrak e(1)}
\longrightarrow
\mathbb{Z}_{6/p,d=3}^{\mathfrak e(1)}
\times
\mathbb{Z}_{6/p,d=3}^{\mathfrak e(0)} ~.
\eeqa
The thermal Wilson line, or Polyakov loop, is charged under the $0$-form symmetry $\mathbb{Z}_{6/p,d=3}^{\mathfrak e(0)}$.

Similarly, the four-dimensional magnetic $1$-form symmetry decomposes as
\beqa
U(1)_{{\rm EM},d=4}^{\mathfrak m(1)}
\longrightarrow
U(1)_{{\rm EM},d=3}^{\mathfrak m(1)}
\times
U(1)_{{\rm EM},d=3}^{\mathfrak m(0)} ~.
\eeqa
An 't Hooft line wrapping the thermal circle becomes a local monopole operator from the three-dimensional perspective and is charged under the magnetic $0$-form symmetry $U(1)_{{\rm EM},d=3}^{\mathfrak m(0)}$.


\section{Effective theory of thermal Wilson lines in
$SU(3)_c \times U(1)_{\rm EM}$}
\label{sec:EFT}

\subsection{Parametrization of thermal Wilson lines}
We first introduce the following parametrization of the thermal Wilson line in the fundamental representation in terms of the temporal components of the gauge fields~\cite{Gross:1980br, Bhattacharya:1990hk}:
\beqa
G_0(\mathbf{x})
&=&
\frac{2\pi T}{g_s}
\left[
q_1(\mathbf{x})H_1+q_2(\mathbf{x})H_2
\right],
\qquad
H_1=\mbox{diag}\left(\frac{1}{2},-\frac{1}{2},0\right),
\qquad
H_2=\mbox{diag}\left(\frac{1}{3},\frac{1}{3},-\frac{2}{3}\right),
\nonumber\\
A_0(\mathbf{x})
&=&
\frac{2\pi T}{e}\,t(\mathbf{x})~.
\eeqa
Here, $q_1$ and $q_2$ are dimensionless coordinates on the Cartan subalgebra of the color holonomy, while $t$ is a dimensionless electromagnetic holonomy coordinate. These variables should not be regarded as three independent real scalar fields with independently chosen ranges. Large gauge transformations and Weyl transformations identify many apparently distinct coordinate values.
With $\beta=1/T$, the untraced color and electromagnetic holonomies are
\beqa
\mathcal{W}_c
&=&
e^{ig_s\beta G_0}
=
e^{2\pi i(q_1H_1+q_2H_2)}
=
\mbox{diag}
\left(
e^{2\pi i\phi_1},
e^{2\pi i\phi_2},
e^{2\pi i\phi_3}
\right),
\nonumber\\
&&
\qquad\qquad
\mbox{where}
\quad
\phi_1=\frac{q_1}{2}+\frac{q_2}{3},
\qquad
\phi_2=-\frac{q_1}{2}+\frac{q_2}{3},
\qquad
\phi_3=-\frac{2q_2}{3},
\\
\mathcal{W}_{\rm EM}^{(Q)}
&=&
e^{ieQ\beta A_0}
=
e^{2\pi iQt}~.
\eeqa

For the color coordinates, the corresponding periodic identifications are
\beqa
(q_1,q_2)
&\sim&
(q_1+2,q_2) \,,
\qquad
(q_1,q_2)
\sim
\left(q_1-1,q_2+\frac{3}{2}\right) \,.
\eeqa
Permutations of the diagonal entries of $\mathcal{W}_c$ are gauge
transformations. We may therefore impose the ordering convention  $\phi_1\geq\phi_2\geq\phi_3$ 
and $\phi_1-\phi_3\leq 1$.
In terms of $(q_1,q_2)$, these conditions define the fundamental
$SU(3)$ Weyl alcove:
\beqa
0\leq q_1\leq1 \,,
\qquad
\frac{q_1}{2}\leq q_2\leq1-\frac{q_1}{2} \,.
\eeqa
Its vertices are 
$A=(0,0)$, $B=(0,1)$, 
$C=\left(1,\frac{1}{2}\right)$.
The corresponding color holonomies are
$\mathcal{W}_c(A)=\mathbb{I}_3$,  
$\mathcal{W}_c(B)=z\,\mathbb{I}_3$ and
$\mathcal{W}_c(C)=z^2\,\mathbb{I}_3$, with
$z=e^{2\pi i/3}$.

\begin{figure}[th!]
    \centering
    \includegraphics[width=0.6\linewidth]{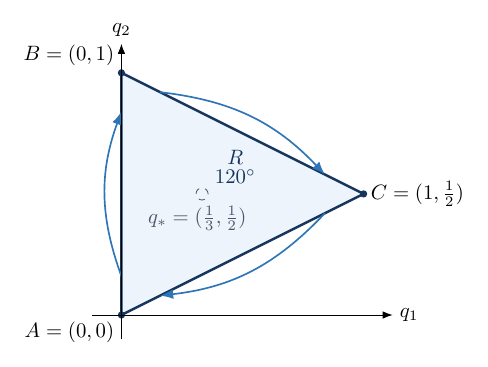}
    \caption{The fundamental $SU(3)$ Weyl alcove. Multiplication of the
    color holonomy by the center element $z$, followed by the Weyl
    permutation required to return the transformed configuration to the
    chosen alcove, acts as the affine rotation $R$ and cycles the
    vertices as $A\rightarrow B\rightarrow C\rightarrow A$.}
    \label{fig:su3_weyl_alcove}
\end{figure}

Multiplication of the color holonomy by the center element $z$ maps the
$SU(3)$ Weyl alcove into itself, up to a Weyl transformation. Since 
$e^{2\pi iH_2}=z\,\mathbb{I}_3$, 
the center action is represented in the unrestricted coordinates by the
shift
\beqa
(q_1,q_2)\longrightarrow(q_1,q_2+1) ~.
\eeqa
The shifted point generally lies outside the chosen Weyl alcove and must
therefore be returned to it by an appropriate Weyl permutation. Within
the alcove, the resulting transformation can be represented by the
affine map
\beqa
R(q_1,q_2)
=
\left(
q_2-\frac{q_1}{2},
\,
1-\frac{3q_1}{4}-\frac{q_2}{2}
\right).
\eeqa
This map is a $120^\circ$ rotation with respect to the natural metric on
the Cartan subalgebra and satisfies $R(A)=B$, $R(B)=C$, $R(C)=A$, $R^3=1$.
More explicitly, one finds
\beqa
\mathcal{W}_c\bigl(R(\bm q)\bigr)
=
U_{\rm cyc}\times
\left[
z\,\mathcal{W}_c(\bm q)
\right] \times
U_{\rm cyc}^{-1},
\eeqa
where $U_{\rm cyc}\in SU(3)$ is a cyclic permutation matrix and $\bm q$ is the notation for $(q_1,q_2)$. Equivalently,
\beqa
\mathcal{W}_c\bigl(R(\bm q)\bigr)
\simeq
z\,\mathcal{W}_c(\bm q),
\eeqa
where $\simeq$ denotes gauge equivalence under a Weyl transformation.
This construction is illustrated schematically in
Fig.~\ref{fig:su3_weyl_alcove}.

Since the electromagnetic charges of Standard Model particles lie in
$(1/3)\mathbb{Z}$, $\mathcal W^{(Q)}_{\rm EM}=e^{2\pi iQt}$. Since genuine lines carry $6Q\in\Z$, the large-gauge periodicity of $t$ alone is $t\sim t+6$. The remaining identifications are the elements of $\Gamma_p$, which act as $g^k:(\bm{q},t)\to(R^k(\bm{q}),t+k)$:
\begin{align}
\Gamma=1:&\ \text{none};&\Gamma=\Z_2:&\ (\bm{q},t)\sim(\bm{q},t+3); \nonumber \\
\Gamma=\Z_3:&\ (\bm{q},t)\sim \big(R^2(\bm{q}),t+2\big);&\Gamma=\Z_6:&\ \big(\bm{q},t)\sim(R(\bm{q}),t+1\big).
\end{align}

For $\Gamma=1$ a convenient fundamental domain is the Weyl alcove together with $0\le t<6$. The global thermal $\Z^{\mathfrak{e}(0)}_{6,d=3}$ transformation acts as $(\bm{q},t)\to(R(\bm{q}),t+1)$ with $t$ understood modulo 6, and the six configurations
\beqa
(q_1,q_2,t)=(0,0,0),\ (0,1,1),\ (1,\tfrac12,2),\ (0,0,3),\ (0,1,4),\ (1,\tfrac12,5)
\eeqa
are distinct physical configurations related by successive actions of the symmetry. For $\Gamma=\Z_2$ the domain is $0\le t<3$, the symmetry is $\Z^{\mathfrak{e}(0)}_{3,d=3}$, and the first three configurations are the distinct ones.

For $\Gamma=\Z_6$ a convenient fundamental domain is the alcove with $0\le t\le1$ and the twisted identification $(\bm{q},0)\sim(R(\bm{q}),1)$; the combined action is a gauge identification and there is no global thermal symmetry. For $\Gamma=\Z_3$ the domain is the alcove with $0\le t\le2$ and the twisted identification $(\bm{q},0)\sim(R^2(\bm{q}),2)$; the residual $\Z^{\mathfrak{e}(0)}_2$ acts as $(\bm{q},t)\to(R(\bm{q}),t+1)$, which squares to a gauge identification, and its two vacua are $(0,0,0)$ and $(0,1,1)\sim(0,0,3)$. In the remainder of this work we consider $\Gamma\neq\Z_6$, for which $(R(\bm{q}),t+1)$ is a genuine global symmetry and the elementary wall is the same interface; where a definite number of vacua is needed we quote it as $6/p$.

\subsection{PNJL description of the QCD crossover and thermal vacua}\label{subsec:PNJL}

At physical quark masses, finite-temperature QCD undergoes a smooth crossover rather than a genuine confinement or chiral phase transition. To describe the continuous evolution of the chiral condensate and color holonomy, we employ a Polyakov-loop-extended Nambu–Jona-Lasinio (PNJL) model~\cite{Meisinger:1995ih,Ratti:2005jh}. Our purpose is not to construct a precision equation of state, but to obtain a symmetry-preserving effective potential for the color and electromagnetic holonomies throughout the QCD crossover. In particular, the PNJL potential allows us to follow how the $6/p$ thermal vacua related by the diagonal $\mathbb{Z}^{\mathfrak e(0)}_{6/p}$ symmetry evolve with temperature.

Defining the Polyakov loop as the trace of the thermal Wilson line or 
\beqa
L_c(\bm{q};\bm{x}) \equiv \frac{1}{3}\,\mbox{Tr}\left(\mathcal{W}_c(\bm{q};\bm{x}) \right) \,, ~~ L(\bm{q}, t;\bm{x}) \equiv \frac{1}{3}\,\mbox{Tr}\left(\mathcal{W}_c(\bm{q};\bm{x})\,\mathcal{W}_{\rm EM}^{(Q_f)} \right) = e^{2\pi i Q_f\,t(x)}\,L_c(\bm{q};\bm{x}) ~,
\eeqa
the mean-field thermodynamic potential density, equivalently the finite-temperature effective potential, is~\cite{Ratti:2005jh} 
\begin{align}
    &\Omega(\bm{q}, t,\Sigma; T) \nonumber\\
=~& U_{\rm glue}\big(L_c(\bm{q}), L_c^\dagger(\bm{q}); T\big) + \frac{\Sigma^2}{2\,G} + \Omega_{\rm matter}(\bm{q}, t, \Sigma; T) - \sum_{f}\int^{\Lambda}_0\frac{6\times 4\pi\,p^2 \dd p}{(2\pi)^3}E_{p,f}(\Sigma)\,. 
\label{eq:potential_PNJL_3fields}
\end{align}
Here, $U_{\rm glue}$ denotes the effective gluonic potential for the color Polyakov loop $L_c$, which serves as an indicator of confinement. The auxiliary scalar field $\Sigma$ is associated with chiral symmetry breaking, and $G$ denotes the coupling of the effective four-fermion interaction in the NJL model. The term $\Omega_{\rm matter}$ is the finite-temperature fermionic contribution in the background color and electromagnetic holonomies, including contributions from both quarks and charged leptons. The final term is the zero-temperature quark contribution, regularized by the three-momentum cutoff $\Lambda$; the factor of $6=2\times3$ accounts for the spin and color degrees of freedom. The corresponding single-particle energy is $E_{p,f}^2(\Sigma) = |\bm{p}|^2 + m^2_f(\Sigma)$ with $m_f = m_{0, f} - \langle \Sigma \rangle$, where $m_{0,f}$ is the bare quark mass and $- \langle \Sigma \rangle$ is the constituent quark mass.

\begin{table}[hb!]
    \centering
    \renewcommand{\arraystretch}{1.2}
    \begin{tabular}{c|c|c|c|c} \hline\hline
        $T_0$ & $a_0$ & $a_1$ & $a_2$ & $a_3$ \\ \hline
        270~MeV & 6.75 & -1.95 & 2.625 & -7.44\\ \hline\hline
        $b_3$ & $b_4$ & $\Lambda$ & $G$ & $m_0$ \\ \hline
        0.75 & 7.5 & 651~MeV & 10.08~GeV$^{-2}$ & 5.5~MeV \\ \hline\hline
    \end{tabular}
    \caption{The fitted values of parameters in PNJL potential~\cite{Ratti:2005jh}. }
    \label{tab:PNJL_parameters}
\end{table}

In the presence of a $U(1)_{\rm EM}$ holonomy background, we assume that the purely gluonic effective potential $U_{\rm glue}$ remains unchanged, since gluons carry no electromagnetic charge. Its parameters are fixed by fits to pure-gauge lattice-QCD results, while the dependence on the electromagnetic holonomy induced by charged matter is included separately in $\Omega_{\rm matter}$. We adopt the parametrization
\begin{align}
\label{eq:U_eff_gluon}
    &\frac{U_{\rm glue}\big(L_c(\bm{q};\bm{x}), L_c^\dagger(\bm{q};\bm{x}); T\big)}{T^4}=-\frac{b_2(T)}{2}L_c^\dagger(\bm{x})L_c(\bm{x})-\frac{b_3}{6}\big(L_c^3(\bm{x})+L_c^{\dagger3}(\bm{x})\big)+\frac{b_4}{4}\big(L_c^\dagger(\bm{x})L_c(\bm{x})\big)^2,\\
    &\qquad \qquad \qquad \qquad \mbox{with}\quad b_2(T)=a_0+a_1\left(\frac{T_0}{T}\right)+a_2\left(\frac{T_0}{T}\right)^2+a_3\left(\frac{T_0}{T}\right)^3\,. \nonumber 
\end{align}
After fitting the experimental data and the numerical Lattice QCD results~\cite{Kaczmarek:2002mc,Ratti:2005jh}, the numerical values of $a_i$ and $b_i$ are shown in Table~\ref{tab:PNJL_parameters}.

Both quarks and charged leptons contribute to the effective potential $\Omega_{\rm matter}(\bm{q}, t, \Sigma; T)$ with~\cite{Weiss:1981ev, KorthalsAltes:1999cp} 
\beqa
&&\Omega_{\rm matter}(\bm{q}, t, \Sigma; T) = \Omega_{\rm quark}(\bm{q}, t, \Sigma; T) + \Omega_{\rm lepton}(t; T) \,, \\
&&\Omega_{\rm quark}(\bm{q}, t, \Sigma; T) = -2 \sum_{f = u, d} T\int\frac{\dd^3 p}{(2\pi)^3}\Tr[\log(1+ \mathcal{W}_c(\bm{q}) \,e^{2\pi i Q_f t} e^{-E_{p, f}/T})]+ {\rm c.c.}~, \nonumber \\
&& \hspace{0.5cm}=-2\sum_{f=u,d}T\int\frac{\dd^3 p}{(2\pi)^3}\log\Big(1+3\,L_c\,e^{2\pi i Q_f t} e^{-E_{p,f}/T}+3\,L_c^\dagger\, e^{4\pi i Q_f t}e^{-2E_{p,f}/T}  \nonumber \\
 && \hspace{5.5cm} +\,e^{6\pi i Q_f t}e^{-3E_{p,f}/T}\Big)+{\rm c.c.} ~, \nonumber \\
&&  \Omega_{\rm lepton}(t; T) = -2 \sum_{f = e, \mu} T\int\frac{\dd^3 p}{(2\pi)^3} \,\log(1+ e^{2\pi i Q_f t} e^{-E_{p, f}/T}) + {\rm c.c.}~.
\eeqa
We include all charged leptons with masses below approximately $1\,\mathrm{GeV}$, namely the electron and muon. For $T<100\,\mathrm{MeV}$, we retain only the electron. This scale is sufficiently high to cover the deconfined and chirally symmetric regime of QCD considered in this work. We also neglect the strange-quark contribution, as we expect it to induce only small quantitative changes in the numerical results presented below. It is straightforward to verify that $\Omega_{\rm matter}(\bm q,t,\Sigma;T)$ is invariant under the $\mathbb{Z}_{6/p}^{{\mathfrak e}(0)}$ transformation $(\bm q,t)\longrightarrow \bigl(R(\bm q),t+1\bigr)$. Indeed, under this transformation, $\mathcal{W}_c\to z\,\mathcal{W}_c$, and the combined quark holonomy transforms by the factor $z\,e^{2\pi iQ_f}=1$ for both $Q_u=2/3$ and $Q_d=-1/3$. The charged-lepton contributions are also invariant because their electric charges are integers. Because every SM field has $3Q\in\Z$, $\Omega_{\rm matter}$ is also invariant under $t\to t+3$ alone. For $\Gamma=1$ the vacua $t$ and $t+3$ are therefore exactly degenerate and indistinguishable by any local SM observable; they are distinguished only by Polyakov loops of half-integer-charge lines.

Besides, the effective potential is also invariant under charge conjugation, $\mathcal{C}$. Within the fundamental Weyl alcove, the variables $q_1$, $q_2$, and $t$ transform as
\beqa
(q_1,q_2,t)\xrightarrow{\quad\mathcal{C}\quad}\big(q_2-\frac{q_1}{2},\ \frac{q_2}{2}+\frac{3q_1}{4},\ -t \bmod 6\big) \quad (\operatorname{mod} 3 ~~\mbox{for}~~ \Gamma=\Z_2)~.
\eeqa
The $\mathcal{C}$-invariant configurations satisfy $q_2=3q_1/2$ and $t=0,3$ (respectively \ $t=0,3/2$); $\mathcal{C}$ fixes the vacuum $(A,0)$ and exchanges $(B,1)\leftrightarrow(C,5)$ and $(C,2)\leftrightarrow(B,4)$, since it maps $B\leftrightarrow C$.

\subsubsection{High-temperature vacuum: $T \gg T_{\rm QCD}$}

At temperatures well above the QCD scale, $T\gg T_{\rm QCD}\sim T_0$, we expect the theory to be in a chirally symmetric phase, with $\langle\Sigma\rangle \to 0$, and a deconfined phase, characterized by a nonzero Polyakov-loop expectation value, $\langle L\rangle\neq 0$~\cite{Gross:1980br}, as in pure Yang–Mills theory~\cite{Polyakov:1975rs}. In this high-temperature limit, we also neglect the bare quark masses. Rather than using the phenomenological potential for $L_c$ in Eq.~\eqref{eq:U_eff_gluon}, which is motivated by the pure-gauge theory in the transition region, we employ the analytically calculated gluon contribution derived in Refs.~\cite{Gross:1980br,KorthalsAltes:1994be,Fukushima:2017csk}. The effective perturbative potential is 
\begin{align}\label{eqn:perturbativeSU3U1_ud}
\frac{\Omega_{\rm per}(\bm{q},t,\langle \Sigma\rangle =0; T)}{\pi^2\,T^4} \approx &\frac{4}{3}\, \left[\mathcal{V}(q_1)+\mathcal{V}\left(\frac{q_1}{2}+q_2\right)+\mathcal{V}\left(-\frac{q_1}{2}+q_2\right)\right] -\sum_{\substack{f\in \mathrm{e, \mu, \tau}\\ m_f<T}}\frac{4}{3}\, \mathcal{V}\left(Q_f\,t+\frac{1}{2}\right)\nonumber \\
    &\hspace{-3.5cm}-\sum_{\substack{f\in \mathrm{quarks}\\ m_f<T}}\frac{4}{3}\left[ \mathcal{V}\left(\frac{q_1}{2}+\frac{q_2}{3}+Q_f \,t+\frac{1}{2}\right)+\mathcal{V}\left(-\frac{q_1}{2}+\frac{q_2}{3}+Q_f \,t+\frac{1}{2}\right)+\mathcal{V}\left(-\frac{2q_2}{3}+Q_f\,t+\frac{1}{2}\right)\right] ~.
\end{align}
These terms arise from the gluon, charged-lepton, and quark contributions, respectively. Here, we have omitted field-independent terms, as well as threshold effects that become important when $T$ is comparable to a fermion mass. The function $\mathcal{V}(x) \equiv \{x\}^2 (1 - \{x\})^2$ with $\{x\}\equiv x - \lfloor x\rfloor \in [0, 1)$, arises from the momentum integration and can be expressed in terms of the real part of the polylogarithm $\mbox{Li}_4(z)$, where $z$ is complex.

Minimizing the effective potential in terms of $q_1, q_2, t$ and noticing that each individual contribution prefers to have the minimum located at the origin of the dependent parameters, the global vacuum of the potential sits at 
\beqa
\mbox{Vac}(T\rightarrow \infty):&&\langle \Sigma \rangle = 0 \,,\quad \langle |L| \rangle  = 1\,, \nonumber \\
&&(\langle q_1 \rangle, \langle q_2 \rangle, \langle t \rangle)  = (0,0,0),\;(0,1,1), \; (1, 1/2, 2) ~,
\label{eq:vacuum-high-T}
\eeqa
with the three degenerate global minima related by the $\mathbb{Z}^{\mathfrak e(0)}_3$ discrete symmetry for $\Gamma = \Z_2$. It is also interesting to note that there could also exist a local metastable vacuum. For instance including $e$ and $\mu$ charged leptons and $u$ and $d$ quarks, one has $(q_1, q_2, t) = \big(1,1/2, (3\sqrt{409}-29)/284\big)$ and $(q_1, q_2, t) = \big(1,1/2, (313-3\sqrt{409})/284\big)$ as the local minima. Notice that the $t$ in the two local minima sum to 1 and are related by $\mathcal{C}$ composed with the $\mathbb{Z}_3$ generator.

\subsubsection{Low-temperature vacuum: $m_e \ll T \ll T_{\rm QCD}$}

In the limit of $T$ much below the QCD transition temperature, the chiral symmetry is spontaneously broken with the $\langle \Sigma \rangle\rightarrow \langle \Sigma \rangle_{T=0} \equiv \langle \Sigma \rangle_0 \approx -0.32$~GeV using the numerical numbers in Table~\ref{tab:PNJL_parameters}. The quarks have their effective mass dominated by the constituent mass or $m_f \approx - \langle \Sigma \rangle_0$. 
\begin{align}\label{eqn:low-T-Omega}
\frac{\Omega(\bm{q},t,\langle \Sigma\rangle_0; T)}{\pi^2\,T^4} \approx & \frac{U_{\rm glue}\big(L_c(\bm{q}), L_c^\dagger(\bm{q}); T\big)}{\pi^2\,T^4} -\frac{4}{3}\, \mathcal{V}\left(- t+\frac{1}{2}\right)\nonumber \\
    &\hspace{-2.5cm}-\sum_{\substack{f\in u, d}} \frac{\sqrt{2}}{\pi^{7/2}}\left(\frac{|\langle \Sigma\rangle_0|}{T}\right)^{3/2}\,e^{-|\langle \Sigma\rangle_0|/T}\left\{ \cos\left[2\pi\left(\frac{q_1}{2}+\frac{q_2}{3}+Q_f \,t\right)\right] +\cos\left[2\pi\left(-\frac{q_1}{2}+\frac{q_2}{3}+Q_f \,t\right)\right] \right.\nonumber \\
& \left.  \hspace{4cm} +\cos\left[2\pi\left(-\frac{2q_2}{3}+Q_f\,t\right)\right]\right\} ~.
  \end{align}
At the leading order of $\mathcal{O}(e^{-|\langle \Sigma\rangle_0|/T})$, the VEVs of $q_1$ and $q_2$ are determined by the phenomenological potential $U_{\rm glue}$, while the VEV of $\langle t\rangle = 0$ is determined by the electron contribution of the second term of the above equation. The minimum has $|L_c(\bm{q})| = 0$ at the zero temperature limit or $q_1 = 1/3$ and $q_2 = 1/2$ (the center point of the the fundamental $SU(3)$ Weyl alcove in Fig.~\ref{fig:su3_weyl_alcove}). 

At the next leading order of $\mathcal{O}(e^{-|\langle \Sigma\rangle_0|/T})$, one has 
\beqa\label{Eq:VEV_exp_closeto center}
\mbox{Vac}(T\rightarrow 0):&&\langle \Sigma \rangle \approx \langle \Sigma \rangle_0 \,,\quad \langle |L| \rangle  \approx \frac{6\sqrt{2}}{|b_2(T)|}\left(\frac{|\ev{\Sigma}_0|}{\pi T}\right)^{3/2}e^{-|\ev{\Sigma}_0|/T}\,, \\
&&\hspace{-2cm}(\langle q_1 \rangle, \langle q_2 \rangle, \langle t \rangle)  \approx \left(\frac{1}{3} - \epsilon(T), \frac{1}{2} - \frac{3}{2}\epsilon(T),0\right),\;
\left(\frac{1}{3} - \epsilon(T), \frac{1}{2} + \frac{3}{2}\epsilon(T),1\right), \nonumber \\ 
&& \hspace{1cm} \left(\frac{1}{3} + 2\,\epsilon(T), \frac{1}{2} , 2\right) ~, 
\hspace{1.2cm}\mbox{with} \quad \epsilon(T) \equiv \frac{3\sqrt{6}}{\pi\,|b_2(T)|}\left(\frac{|\ev{\Sigma}_0|}{\pi T}\right)^{3/2}e^{-|\ev{\Sigma}_0|/T} \,, \nonumber 
\eeqa
for $\Gamma = \Z_2$; For $\Gamma =1$ the three further vacua are obtained by $t\rightarrow t+ 3$.
The VEVs of $q_1$ and $q_2$ in the first vacuum obey the same leading-order relation, $\langle q_2\rangle/\langle q_1\rangle=3/2$. Consequently, the effective potential is an even function of $t$, yielding $\langle t\rangle=0$ at this order. These VEV relations follow from charge-conjugation symmetry, $\mathcal{C}$, which is not spontaneously broken in QCD at zero chemical potential.

In pure $SU(3)$ Yang–Mills theory, the Polyakov loop has a vanishing expectation value at low temperatures, and the theory undergoes a first-order deconfinement transition. The discrete center symmetry is therefore fully restored in the low-temperature phase. In QCD with physical quark masses, as realized in the SM, the transition is instead a crossover. Nevertheless, for $T\ll |\Sigma_0|$, the Polyakov loop retains a small but nonzero expectation value, implying that the discrete $\mathbb{Z}^{\mathfrak e(0)}_{6/p}$ symmetry remains broken even at low temperatures. Because $\epsilon(T)$ is exponentially small, the three degenerate vacua have nearly identical values of $q_1$ and $q_2$ but remain widely separated along the $t$ direction.

\subsubsection{Vacuum structure in the QCD crossover region}

\begin{figure}[th!]
    \centering
    {
    \includegraphics[width=0.48\linewidth]{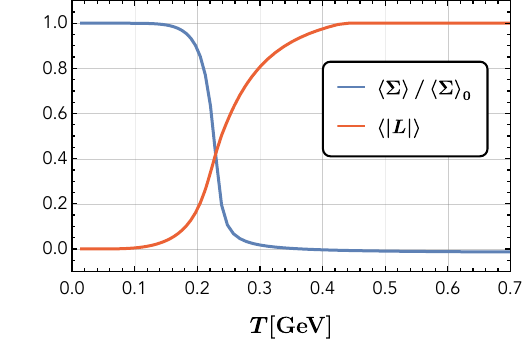}
    \includegraphics[width=0.48\linewidth]{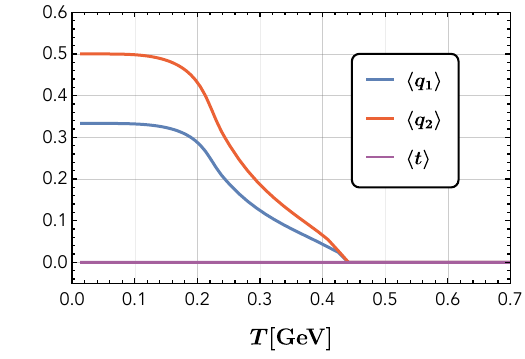}
    }
    \caption{{\it Left}: The VEVs $\ev{\Sigma}$ and $\ev{|L|}$ as functions of temperature in the PNJL $SU(3)_c\times U(1)_{\text{EM}}$ model, obtained from the full potential in Eq.~\eqref{eq:potential_PNJL_3fields}. {\it Right}: The temperature dependence of one of the three $\mathbb{Z}_{6/p}$-degenerate vacua.}\label{fig:vev_sigmaL_T_PNJL_full_3fields}
\end{figure}

As the temperature decreases through the QCD crossover region, the two indicators of the QCD transition, $\langle\Sigma\rangle$ and $\langle|L|\rangle$, evolve smoothly from their high-temperature values to their low-temperature values. The pseudocritical temperature is close to the phenomenological temperature scale $T_0=0.27$~GeV. By minimizing the effective potential with respect to all fields, we obtain the VEVs of $\Sigma$ and $|L|$ as functions of $T$, shown in the left panel of Fig.~\ref{fig:vev_sigmaL_T_PNJL_full_3fields}, and the VEVs of the individual fields $q_1$, $q_2$, and $t$, shown in the right panel.

\section{Thermal $\mathbb{Z}_{6/p}$ domain walls across cosmological epochs}
\label{sec:Z3_DW_property}

Having determined the temperature dependence of the vacuum structure, we now study the thermal domain walls interpolating between neighbouring vacua, i.e.\ between $(\bm{q},t)$ and $(R(\bm{q}),t+1)$. All such interfaces are equivalent by symmetry; the interface between vacua $k$ and $k+2$ (or $k+3$ for $\Gamma=1$) is a stack of two (three) elementary walls rather than a new type. In contrast to the homogeneous vacuum analysis, the domain-wall solution depends on both the effective potential and the gradient terms of the background fields. We characterize the wall by its field profile, tension $\sigma(T)$, and width $\ell(T)$.

The physical origin of the wall potential changes substantially as the Universe cools. At $T\gg T_{\rm QCD}$, the effective potential is dominated by perturbative gluon, quark, and charged-lepton contributions, allowing the wall properties to be studied analytically using the high-temperature potential. Near $T\sim T_{\rm QCD}$, the rapid evolution of the chiral condensate and the Polyakov loop requires the full PNJL effective potential and a numerical solution of the coupled wall equations. Below the QCD scale, the color holonomy and chiral condensate approach their low-temperature values, while the electromagnetic holonomy $t$ continues to be affected by the charged particles in the thermal plasma.

The charged-particle contribution itself passes through several qualitatively different regimes. For $m_e\ll T\ll T_{\rm QCD}$, electrons and positrons are relativistic and generate an unsuppressed thermal potential for $t$. For $T\ll m_e$, their pair density becomes Boltzmann suppressed. We define $T_{\rm asy}$ as the temperature at which the thermally produced electron-positron pair density becomes comparable to the electron density associated with the cosmological charge asymmetry. Below $T_{\rm asy}$, the charge-asymmetric electron and ion densities dominate, and the usual zero-chemical-potential grand-canonical potential must be replaced by a charge-constrained Helmholtz free-energy functional.

Recombination does not lead to an abrupt disappearance of the wall. Although the free-electron fraction decreases significantly, it approaches a nonzero residual value, $x_e^{\rm res}\sim10^{-4}$. These residual charged particles can continue to support a small holonomy barrier and hence a very weak and broad wall, provided that the plasma remains sufficiently responsive on the wall length and time scales. Nevertheless, even when $x_e$ becomes approximately constant, the physical free-electron density $n_e^{\rm free}=x_e n_{\rm H}$ continues to decrease as the Universe expands. The wall tension therefore decreases and its width increases toward late times.

In the following, we combine analytic approximations in the well-separated temperature regimes with numerical solutions near the QCD crossover and the relevant particle thresholds.

\subsection{Effective action and definitions of the wall observables}

To determine the domain-wall profiles, we must supplement the effective potential with the kinetic terms of the background fields. The kinetic terms of $q_1$, $q_2$, and $t$ follow from those of the temporal color and electromagnetic gauge fields. For the parametrization adopted here, they are
\begin{align}
\mathcal{L}_{\rm kin}
=&
\frac{1}{2}K_1(T)(\partial_i q_1)^2
+\frac{1}{2}K_2(T)(\partial_i q_2)^2 + \frac{1}{2}K_t(T)(\partial_i t)^2,
\label{eq:wall_kinetic_terms}
\end{align}
where
\begin{align}
K_1=\frac{4\pi^2T^2}{g_s^2}~,
\qquad
K_2=\frac{4\pi^2T^2}{g_s^2}\frac{4}{3} ~,
\qquad
K_t=\frac{4\pi^2T^2}{e^2} ~.
\end{align}
The numerical factors in $K_{1,2}$ follow from the normalization of the $SU(3)$ Cartan generators. The gauge couplings $e=\sqrt{4\pi \alpha}$ and $g_s=\sqrt{4\pi \alpha_s}$ are fixed values in the calculation, with $\alpha=1/137$, $\alpha_s=0.115$.

In the NJL model, $\Sigma$ is introduced as an auxiliary field representing the fermion bilinear~\cite{Klevansky:1992qe}. Consequently, the tree-level bosonized action does not contain an independent canonical kinetic term for $\Sigma$. Nevertheless, derivative terms of the form $\frac{1}{2}Z_\Sigma(\Sigma,T)(\partial_i\Sigma)^2$ can be generated by the quark determinant. In our leading-order analysis, we neglect these derivative corrections and determine $\Sigma$ locally by minimizing the grand-potential density:
\begin{align}
\left.
\frac{\partial\Omega(\bm{q},t,\Sigma;T)}
{\partial\Sigma}
\right|_{\Sigma=\Sigma_\star(\bm{q},t;T)}
=0 ~.
\end{align}
We then define the reduced effective potential
\begin{align}
\Omega_{\rm red}(\bm{q},t;T)
\equiv
\Omega\bigl(\bm{q},t,\Sigma_\star(\bm{q},t;T);T\bigr)~.
\label{eq:reduced_wall_potential}
\end{align}
This local-potential treatment is expected to be reliable when the wall varies slowly compared with the relevant fermionic correlation length.

Consider a static planar wall perpendicular to the $z$ direction. We denote the dynamical wall fields by
\begin{align}
\phi^a(z)=\bigl(q_1(z),q_2(z),t(z)\bigr),
\end{align}
with the boundary conditions
\begin{align}
\phi^a(z\rightarrow-\infty)=\phi^a_- ~,
\qquad
\phi^a(z\rightarrow+\infty)=\phi^a_+ ~,
\label{eq:wall-boundary-condition}
\end{align}
where $\phi_-$ and $\phi_+$ denote two adjacent degenerate vacua, related by $(\bm{q},t) \rightarrow
(R(\bm{q}),t+ 1)$. Because the holonomies are compact variables, a continuous lift to their covering space should be used when solving the wall equations to avoid artificial discontinuities when applying a modulo operation to the fields.

At a fixed temperature, the free energy per unit area of the equilibrium wall, or equivalently the wall tension, is
\beqa
\sigma(T) = \frac{F_{\rm wall}}{A}
&=&
\int_{-\infty}^{+\infty}dz\,
\left[
\frac{1}{2}K_{ab}(T)\partial_z\phi^a\partial_z\phi^b
+\Delta\Omega_{\rm red}(\phi;T)
\right]~,
\label{eq:wall_free_energy}
\eeqa
where $K_{ab}=\operatorname{diag}(K_1,K_2,K_t)$ and
\beqa
\Delta\Omega_{\rm red}(\phi;T)
\equiv
\Omega_{\rm red}(\phi;T)
-\Omega_{\rm red}(\phi_{\rm vac};T)~.
\eeqa
The wall profiles satisfy the coupled equations
\begin{align}
K_{ab}(T)\frac{d^2\phi^b}{dz^2} =\frac{\partial\Omega_{\rm red}}{\partial\phi^a}~,
\label{eq:wall_eom}
\end{align}
subject to the boundary conditions in Eq.~\eqref{eq:wall-boundary-condition}. The numerical calculation of the domain-wall profiles depends on the ratio $\alpha_s/\alpha$, while the tension can be written as an overall factor $1/g_s$ multiplying a function of $\alpha_s/\alpha$ and $T$.

For the numerical solutions, we define the wall width as the interval centered on the wall that contains $90\%$ of its total energy. Choosing the wall center to be at $z=0$, we define
\beqa
\ell(T)&\equiv&\ell_{90}(T)\,,~ \mbox{with}~
0.9 = \frac{1}{\sigma(T)}
\int_{-\ell_{90}/2}^{+\ell_{90}/2}dz\,
\left[
\frac{1}{2}K_{ab}(T)\partial_z\phi^a\partial_z\phi^b
+\Delta\Omega_{\rm red}(\phi;T)
\right]\,.
\label{eq:wall-width-definition}
\eeqa

\subsection{Domain walls in the deconfined QCD phase}
\subsubsection{Asymptotically high temperatures: $T\gg T_{\rm QCD}$}

At temperatures well above the QCD crossover temperature, the effective potential is given by Eq.~\eqref{eqn:perturbativeSU3U1_ud}, with three degenerate vacua specified in Eq.~\eqref{eq:vacuum-high-T}. To obtain basic domain-wall solutions, one must impose boundary conditions for which the vacuum label $t$ changes by one in the covering space before modulo operation, namely $(0,0,0)\to(0,1,1)$ and $(0,1,1)\to(1,1/2,2)$. Due to the periodicity of the fields, the algebraic solution connecting $(1,1/2,2)\to(0,0,0)$ is not the minimum-energy configuration. Instead, one should solve for the configuration $(1,1/2,2)\to(0,0,3)$. One can verify that these domain walls have identical properties, in agreement with the symmetry argument. Solving the equations of motion for the domain-wall field profiles, one finds that the wall tension scales as $\sigma \propto T^3$.

\begin{table}[bh!]
    \centering
    \renewcommand{\arraystretch}{1.2}
    \begin{tabular}{c||c c c c c c c}\hline\hline
        additional flavors & - & $s$ & $\mu$ & $c$ & $\tau$ & $b$ & $t$ \\ \hline
        $\sigma/T^3$ & 36.4 & 38.9 & 43.1 & 48.9 & 52.1 & 53.9 & 55.8 \\ \hline
        $\ell\, T$ & 9.67 & 9.02 & 8.08 & 7.02 & 6.56 & 6.35 & 6.15 \\\hline\hline
    \end{tabular}
    \caption{Numerical coefficients of $\sigma/T^3$ as heavier fermion flavors are added sequentially. The first column, ``-'', corresponds to the matter content $u,d,e$ used in the main text. No pronounced step-like change in $\sigma/T^3$ is observed, except when the top quark is included. For flavor contents up to and including the $b$ quark, we take the value at $T=10~{\rm GeV}$ as the asymptotic value, whereas when the top quark is included, we use the value at $T=450~{\rm GeV}$.}
    \label{tab:sigmaT3_moreflavors}
\end{table}

The PNJL-like model discussed in Section~\ref{subsec:PNJL} is based on two quark flavors. We therefore use, in the following discussion, the numerical results for the tension coefficient $\sigma/T^3$ obtained with two quark flavors and the electron. In $T\gg T_{\rm QCD}$ regime, the difference of $\sigma/T^3$ and $\ell\, T$ between lattice potential Eq.~\eqref{eq:U_eff_gluon} in PNJL and the perturbative one in Eq.~\eqref{eqn:perturbativeSU3U1_ud} is only percent level. Additional quark and lepton flavors can also be included at temperatures above $T_{\rm QCD}$. The high-temperature potential depends only on $\Omega_{\rm matter}$, allowing us to avoid the additional complications associated with chiral symmetry breaking. The masses of the second- and third-generation fermions, except for the top quark, are all at or below the GeV scale. Their contributions therefore remain in the slowly convergent regime in which $\sigma/T^3$ gradually approaches a constant. Consequently, no sharp step-function behavior in $\sigma/T^3$ is observed when these additional flavors are included. The resulting domain-wall tensions and widths are shown in Table~\ref{tab:sigmaT3_moreflavors}. By contrast, once the top quark is included, a clear step-like behavior appears between temperatures of $\mathcal{O}(10~{\rm GeV})$ and $T>m_t$. The electroweak interactions, $W,Z,h$ bosons are ignored in this simple estimation. For the pure gauge theory with the $q_1,q_2$ potential in Eq.~\eqref{eq:vacuum-high-T}, the result of Ref.~\cite{Bhattacharya:1990hk}, $\sigma/T^3=8\pi^2/(9g_s)$, can be recovered. The difference between this value and those in Table~\ref{tab:sigmaT3_moreflavors} arises from the large factor $\alpha_s/\alpha$ in the kinetic term of $t$, which appears only when matter fields are included. The fermion masses and gauge couplings are taken to be constant, without renormalization-group running. Including renormalization-group running causes $\sigma/T^3$ to increase with temperature rather than remain constant when $T\gg T_{\rm QCD}$. At one-loop, the numerical solutions show an $\mathcal{O}(10\%)$ variation in $\sigma/T^3$ in the GeV–TeV range.

\subsubsection{Evolution through the QCD crossover: $T\sim T_{\rm QCD}$}

\begin{figure}[th!]
    \centering
    \includegraphics[width=0.6\linewidth]{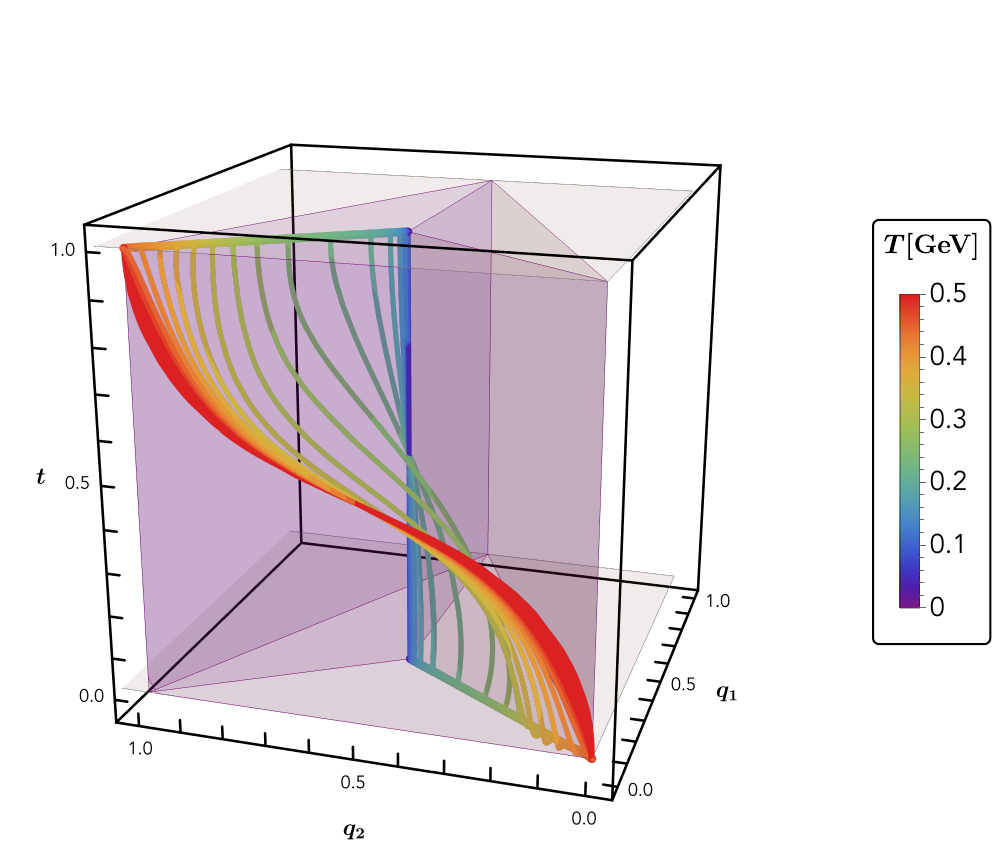}
    \caption{The local minima of the potential (the points on the top and bottom planes) and the domain wall paths in the field space $(q_1,q_2,t)$. The temperature range is $0.05~{\rm GeV}\le T\le 0.5~{\rm GeV}$. The domain wall paths and the minima are in rainbow colors, from red to blue corresponding to  dropping temperature.}
    \label{fig:potential_q1q2t_illus}
\end{figure}

As stated in Section~\ref{subsec:PNJL}, the QCD phase transition in our model follows that of the original PNJL model. During the crossover, the domain-wall trajectory moves from the boundary of the unit cell toward its interior and approaches the central line $\{(q_1,q_2,t)\mid q_1=1/3,\ q_2=1/2\}$. The domain-wall trajectories at various temperatures are shown in Fig.~\ref{fig:potential_q1q2t_illus}, where the color changes from red to blue as the temperature decreases. At temperatures well below $0.15~{\rm GeV}$, the trajectory is nearly a straight line along the $t$ direction because the VEVs of $(q_1,q_2)$ are exponentially close to $(1/3,1/2)$, as shown in Eq.~\eqref{Eq:VEV_exp_closeto center}. We therefore take $T_{\rm QCD}\equiv0.15~{\rm GeV}$.

\begin{figure}[th!]
    \centering
    \includegraphics[width=0.45\linewidth]{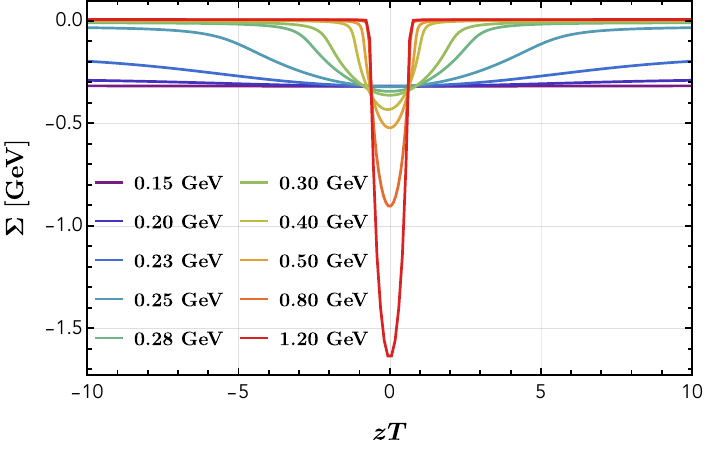} \hspace{3mm}
    \includegraphics[width=0.45\linewidth]{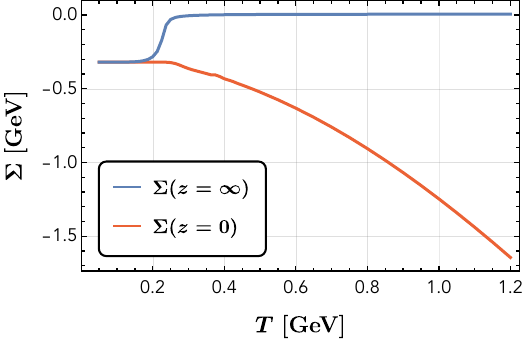}
    \caption{\textit{Left}: The $\Sigma(z)$ profiles at several temperatures, plotted as functions of the dimensionless coordinate $zT$. \textit{Right}: The values of $\Sigma$ at the boundary (blue) and the center (red) of the domain wall.}
    \label{fig:sigma_profile}
\end{figure}

\begin{figure}[th!]
    \centering
    \includegraphics[width=0.45\linewidth]{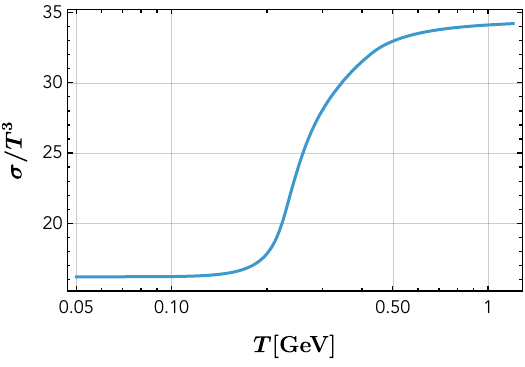} \hspace{3mm}
    \includegraphics[width=0.45\linewidth]{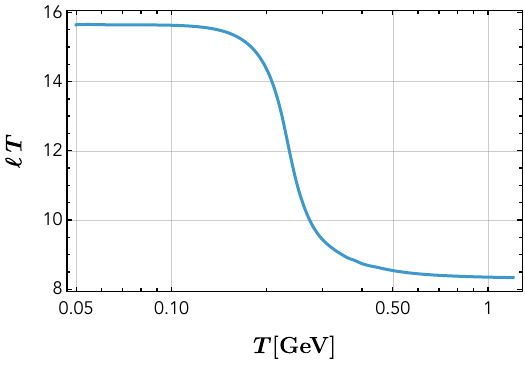}
    \caption{The domain-wall tension (\textit{left}) and width (\textit{right}) as functions of $T$ across the QCD crossover. At each temperature, the wall profiles are obtained by solving the coupled equations for the three dynamical fields $q_1$, $q_2$, and $t$, while the auxiliary field $\Sigma$ is determined from its local stationary condition.}
    \label{fig:dwtension_T3_3fields}
\end{figure}

In addition, we plot the $\Sigma$-field profiles in Fig.~\ref{fig:sigma_profile} at several representative temperatures by solving the coupled equations of motion. At high temperatures, the $\Sigma$ field develops a nontrivial spatial profile. When the domain-wall trajectory lies along the boundary of the unit cell, corresponding to $T>0.45~{\rm GeV}$, the minimum of $\Sigma$ occurs at the midpoint $(q_1,q_2,t)=(0,1/2,1/2)$ of the trajectory and is determined by the stationary condition $\partial\Omega/\partial\Sigma=0$.  The $\Sigma$ well gives equivalently the quark-mass barrier, and we use this observation to study the reflection in Section~\ref{sec:plasma_friction}. We only trust this PNJL up to 1.2 GeV, twice of the cutoff $\Lambda$. As the temperature decreases, the magnitude of the bulk VEV, $|\ev{\Sigma}|$, approaches its zero-temperature value $\ev{\Sigma}_0=0.32~{\rm GeV}$, while the contrast between the bulk and central values decreases. The $\Sigma$ profile therefore becomes progressively flatter and can be treated as approximately spatially constant for $T<T_{\rm QCD}$. The domain-wall tensions and widths across the QCD crossover are shown in Fig.~\ref{fig:dwtension_T3_3fields}; both exhibit smooth crossover behavior.

\subsection{Lepton-supported walls below the QCD scale}

Well below the QCD scale, the wall profile is dominated by the $U(1)_{\rm EM}$ holonomy, parametrized by $t$. Before analyzing this regime, we address historical questions about the physical interpretation of thermal center walls~\cite{Belyaev:1991np,Chen:1992sa,Borisenko:1994jn,Smilga:1994sn}. Throughout, a Euclidean $A_0$ profile should not be identified pointwise with a real electrostatic potential or a zero-temperature Minkowski soliton. Its equilibrium meaning is encoded in an interface free energy, while its response to probes and its motion require a real-time description of the surrounding plasma. Smilga emphasized that multiple minima of a Euclidean effective potential do not alone establish distinct physical phases. In particular, for pure Yang--Mills theory with gauge group $SU(N)/\mathbb{Z}_N$, center-related holonomies are gauge equivalent, and the fundamental Polyakov loop is not a genuine gauge-invariant line operator. Interpolating configurations can then describe Euclidean flux sectors without separating distinct symmetry-breaking phases. Hansson, Nielsen, and Zahed examined a different subtlety in two-dimensional QED with dynamical charge-$Ne$ particles and infinitely heavy charge-$e$ probes~\cite{Hansson:1994ep}. In the massless limit, they found spontaneous $\mathbb{Z}_N$ breaking, but integration over the constant $A_0$ mode enforcing the global Gauss law made their spatially varying source select a single global sector, preventing a domain wall in that construction. These results motivate careful specification of the global gauge group, admissible probes, and interface boundary conditions. The modern one-form symmetry framework resolves the gauge-equivalence issue by distinguishing gauged center transformations from global symmetries acting on genuine line operators~\cite{Gaiotto:2014kfa}. 
For $\Gamma\neq\Z_6$ the diagonal $\Z^{\mathfrak{e}(0)}_{6/p,d=3}$ transformation is a genuine thermal global symmetry, and its action on wrapped genuine Wilson lines distinguishes the symmetry-related sectors. In the broken phase, selecting different sectors at opposite spatial boundaries defines a gauge-invariant interface free energy, whose leading area contribution determines the thermodynamic wall tension. Averaging over all sectors in finite volume can make a charged Polyakov-loop expectation value vanish, but does not identify those sectors as gauge copies or invalidate this interface construction. Although the low-temperature profile lies predominantly along $t$, the sector distinction remains inherited from the full diagonal symmetry rather than from an independent periodicity of QED. As the charged plasma becomes dilute and the holonomy potential weakens, the tension decreases and the profile can broaden, eventually delocalizing if the potential vanishes. The resulting object is a plasma-supported thermal interface whose properties follow the evolving thermal state.

\subsubsection{Relativistic charged plasma: $m_e\ll T\ll T_{\rm QCD}$}

For $m_e\ll T\ll T_{\rm QCD}$, we can ignore all quark/hadron contributions to the effective potential of $t$. With the dominant contribution from electron/positron, the effective Lagrangian is~\cite{Smilga:1994sn} 
\beqa
\mathcal{L} = \frac{1}{2}\,\frac{4\pi^2 T^2}{e^2}\,(\partial_i t)^2 - \pi^2\,T^4\,\frac{4}{3}\,\mathcal{V}\left(-t+\frac{1}{2}\right) ~,
\eeqa
for $0\leq t \leq 1$. The domain wall profile has the analytical solution of
\beqa
t(z) = \begin{cases}
    \frac{1}{\sqrt{2}} \sech\left[ \sqrt{\dfrac{1}{3}}\,e\,T\,z - \ln(1+\sqrt{2}) \right] & \mbox{for} \; z \leq 0~, \\
    1- \frac{1}{\sqrt{2}} \sech\left[ \sqrt{\dfrac{1}{3}}\,e\,T\,z - \ln(1+\sqrt{2}) \right] & \mbox{for} \; z \geq 0  ~,
\end{cases}
\eeqa
with the wall tension of
\beqa
\sigma(T) = \frac{\pi^{2}(4-\sqrt{2})}{3\sqrt{3}}\,\frac{T^3}{e} ~,
\eeqa
and the wall width of $\ell(T) \approx 5/(e\,T)$.

\subsubsection{Nonrelativistic pair-dominated plasma:
$T_{\rm asy}\ll T\ll m_e$}

We define $T_{\rm asy}$ as the electron-asymmetry crossover temperature at which the positron density becomes comparable to the conserved net electron density,
\beqa
n_{e^+}(T_{\rm asy})
\simeq n_{e^-}(T_{\rm asy})-n_{e^+}(T_{\rm asy})
\simeq n_Q^{\rm ion}
\sim \eta_B n_\gamma ~.
\eeqa
For $T\gtrsim T_{\rm asy}$, the plasma is dominated by thermally produced, nearly symmetric $e^-e^+$ pairs, whereas for $T\lesssim T_{\rm asy}$, the positron abundance is exponentially suppressed and the electron density is fixed primarily by charge neutrality with the baryonic component. Using the nonrelativistic equilibrium density gives $T_{\rm asy}\sim20~{\rm keV}$~\cite{Kawasaki:2004qu,Mukhanov:2005sc}.

For $T_{\rm asy}\ll T\ll m_e$, the effective potential of the $t$ field is provided by non-relativistic electrons and positrons and is
\beqa
\Omega(t, T) = T\,(n^{\rm th}_{e^-} + n^{\rm th}_{e^+})\,\left[ 1 - \cos{2\pi\,t} \right]\,,~~ \mbox{with}~~ n^{\rm th}_{e^-} = n^{\rm th}_{e^+}\equiv n^{\rm th}_{e} = 2\,\left(\frac{m_e\,T}{2\pi}\right)^{3/2}\,e^{-m_e/T}. 
\eeqa
The domain wall profile for $t$ from $0$ to $1$ is the sine-Gordon kink with
\beqa
t(z) = \frac{2}{\pi}\,\tan^{-1}\,\left( e^{z/d_{\rm DW}}\right) \,, \qquad \mbox{with} \quad d_{\rm DW}(T) = \sqrt{\frac{T}{2\,e^2\,n^{\rm th}_{e}(T)}} = m_D^{-1}(T) ~,\label{eqn:sineGordon_tprofile_Tasy_me}
\eeqa
with $m_D(T)$ as the photon plasma mass. 

The wall tension is 
\beqa
\sigma(T) = \frac{8\,T}{e}\,\sqrt{2\,T\,n^{\rm th}_{e}(T)} = 8\, \,T^2\,m_D(T)/e^2 ~.
\eeqa
The wall width has $\ell \approx 3\,d_{\rm DW}$.

\subsubsection{Charge-asymmetry-dominated plasma:
$T_{\rm rec}\ll T\ll T_{\rm asy}$}

For $T_{\rm rec}\ll T\ll T_{\rm asy}$ with $T_{\rm rec}$ as the recombination temperature, the thermal positron abundance is exponentially suppressed, and the free-electron density is determined by the (comoving) conserved charge asymmetry,
\beqa
n_e^{\rm asy}(T)
\simeq n_Q^{\rm ion}(T)
\equiv \eta_Q n_\gamma(T)\, ,
\qquad
n_\gamma(T)=\frac{2\zeta(3)}{\pi^2}T^3 ~,
\eeqa
where $\eta_Q\simeq(1-Y_p/2)\eta_B = 0.878\,\eta_B \approx 5.4\times 10^{-10}$~\cite{Aver:2021rwi} accounts for the fraction of baryons bound in helium. In a local grand-canonical description, the electron and ion contributions to the vacuum-subtracted holonomy potential are
\beqa
\Omega_{\rm loc}(t,T) = Tn_e^{\rm asy}\left[1-\cos(2\pi t)\right]
+
T\sum_i n_i\left[1-\cos(2\pi Z_i t)\right] ~,
\eeqa
where $Z_i$ and $n_i$ are the charge and number density of the ion species $i$. Approximating the positive component as a gas of singly charged ions, such that $\sum_i n_i\simeq n_e^{\rm asy}$, gives the sine-Gordon potential
\beqa
\Omega_{\rm loc}(t,T)
\simeq
2\,T\,n_e^{\rm asy}(T)
\left[1-\cos(2\pi t)\right] ~.
\eeqa
Together with the kinetic term $2\pi^2T^2(\partial_i t)^2/e^2$, this potential gives the domain-wall profile interpolating from $t=0$ to $t=1$,
\beqa
t(z) = \frac{2}{\pi}
\tan^{-1}\left(e^{z/d^{\rm asy}}\right)\,,
\qquad
\mbox{with}
\qquad
d^{\rm asy}(T) = \sqrt{\frac{T}{2\,e^2\,n_e^{\rm asy}(T)}} = m_D^{-1}(T) ~.
\eeqa
The corresponding wall tension is
\beqa
\sigma^{\rm asy}(T) = \frac{8\,T}{e}
\sqrt{2\,T\,n_e^{\rm asy}(T)} = 8\,T^2\,m_D(T)/e^2 ~.
\eeqa
Using $n_e^{\rm asy}=\eta_Qn_\gamma$, these expressions become
\beqa
d^{\rm asy}=\frac{\pi}
{2\,e\,T\sqrt{\zeta(3)\,\eta_Q}}\,,
\qquad
\sigma^{\rm asy}(T)=\frac{16}{\pi\,e}
\sqrt{\zeta(3)\,\eta_Q}\,T^3 ~.
\eeqa
The domain wall width is approximately $\ell \approx 3\,d^{\rm asy}$.

At first sight, one may worry that fixing the conserved electron and proton numbers eliminates the holonomy potential. Indeed, for a spatially uniform holonomy, the phase acquired by the fixed number of electrons is exactly canceled by that acquired by the fixed number of protons in a globally neutral plasma. This cancellation, however, concerns only the spatial zero mode of $t$ and does not eliminate the surface free energy associated with a localized interpolation.

To see this explicitly, we define $\theta(z)=2\pi t(z)$ and consider a box of transverse area $A$ and length $L$ in the $z$ direction, with fixed particle numbers $N_{e}=N_p=N$. In the nonrelativistic ideal-gas limit, the holonomy-dependent part of the canonical partition function is
\beqa
\frac{Z_{N,N}[\theta]}{Z_{N,N}[0]}=\left[
\frac{1}{L}
\int_{-L/2}^{L/2}dz\,e^{i\theta(z)}
\right]^N
\left[
\frac{1}{L}
\int_{-L/2}^{L/2}dz\,e^{-i\theta(z)}
\right]^N ~.
\eeqa
The momentum-space factors, including their dependence on $m_e$ and $m_p$, cancel in this ratio. Therefore, at fixed densities, the leading holonomy dependence is determined by the particle densities and electric charges rather than by the particle masses.

For a uniform $\theta$, the two factors in the above expression are pure phases whose product is unity, and the canonical free energy is independent of $\theta$. For a localized wall satisfying $\theta(-\infty)=0$ and $\theta(+\infty)=2\pi$, however, $e^{i\theta(z)}$ differs from unity only within a region of thickness $d_{\rm DW}$. Expanding the exact canonical result in $d_{\rm DW}/L$ gives
\beqa
\frac{\Delta F_{\rm can}}{A}=2\,n_{e}\,T
\int dz\, \left[1-\cos\theta(z)\right] + {\cal O}\left(
n_{e}\,T\,\frac{d_{\rm DW}^2}{L}
\right) ~,
\eeqa
where charge neutrality, $n_{e}=n_p$, has been used. Thus, although the exact canonical functional is spatially nonlocal, its leading surface contribution in the thermodynamic limit is local and coincides with the sine-Gordon potential obtained in the grand-canonical calculation,
\beqa
\Omega_{\rm wall}(t,T)= T\left(n_{e}+n_p\right)
\left[1-\cos(2\pi t)\right]=2\,T\,n_{e}
\left[1-\cos(2\pi t)\right] ~.
\eeqa
The canonical constraint removes only the strictly uniform mode and does not cancel the finite interface tension.

\subsubsection{Recombination and residual ionization:
$T\ll T_{\rm rec}$}\label{subsubsec:tensionTlTrec}

Around $T_{\rm rec}\sim0.3~{\rm eV}$, free electrons and protons combine into neutral hydrogen through $e^-+p\leftrightarrow H+\gamma$. Although our numerical calculation follows the relevant hydrogen populations and radiative transitions, including the $1S$ and $2S$ states, the qualitative suppression of the free-electron density can be understood from the Saha equation. Defining
\beqa
x_e(T)
\equiv
\frac{n_{e}^{\rm free}(T)}{n_{\rm H}(T)}\,,
\eeqa
where $n_{\rm H}$ denotes the total number density of hydrogen nuclei, the Saha equation gives
\beqa
\frac{x_e^2}{1-x_e}=\frac{1}{n_{\rm H}}
\left(\frac{m_eT}{2\pi}\right)^{3/2}
e^{-13.6~{\rm eV}/T} ~.
\eeqa
Thus, during equilibrium recombination, $x_e$ develops an exponential suppression. The Saha approximation eventually breaks down because the recombination rate becomes slower than the Hubble expansion rate. The free-electron fraction then freezes out at a residual value of approximately $x_e\sim10^{-4}$, with its precise evolution determined by the nonequilibrium recombination calculation~\cite{Seager:1999bc,Ali-Haimoud:2010hou}.

After helium has recombined, the residual free-electron and proton densities are approximately
\beqa
n_{e}^{\rm res}(T)= n_p^{\rm res}(T) =
x_e(T)\,n_{\rm H}(T)\,,
\qquad
n_{\rm H}(T)=(1-Y_p)\,\eta_B\,n_\gamma(T) ~.
\eeqa
Neutral hydrogen and helium atoms do not generate the leading periodic potential for $t$, because the bound states carry zero electric charge. Their electric polarizabilities can generate derivative corrections, but these do not produce the leading local potential proportional to $1-\cos(2\pi t)$. The holonomy potential is therefore dominated by the residual free electrons and protons.

As long as Compton scattering maintains a common matter and photon temperature, $T_m\simeq T_\gamma\equiv T$, the canonical argument of the previous subsection remains applicable after replacing the fully ionized electron density by $n_{e}^{\rm res}$. The leading surface contribution is
\beqa
\Omega_{\rm res}(t,T) =2\,T\,n_{e}^{\rm res}(T)
\left[1-\cos(2\pi t)\right] ~.
\eeqa
The corresponding domain-wall profile is
\beqa
t(z)=\frac{2}{\pi}
\tan^{-1}\left(
e^{z/d^{\rm res}}
\right)\,,
\qquad
d^{\rm res}=\sqrt{
\frac{T}
{2e^2n_{e}^{\rm res}(T)}
} ~.
\eeqa
The wall tension is
\beqa
\sigma^{\rm res}(T)
&=&
\frac{8\,T}{e}
\sqrt{
2\,T\,n_{e}^{\rm res}(T)
}\,,
\eeqa
Using $n_{e}^{\rm res}=x_e\,(1-Y_p)\,\eta_B\,n_\gamma$, it becomes
\beqa
\sigma^{\rm res}(T)=
\frac{16}{\pi\,e}
\sqrt{
\zeta(3)\,x_e(T)\,(1-Y_p)\,\eta_B
}\,T^3 ~.
\eeqa
The wall width is $\ell^{\rm res} \approx 3 \, d^{\rm res}$.

At redshifts below $z\sim150$, Compton scattering can no longer maintain $T_m=T_\gamma$. The matter temperature then evolves approximately as $T_m\propto(1+z)^2$, whereas $T_\gamma\propto(1+z)$. Once the two temperatures differ, there is no single equilibrium temperature with which to define the Euclidean thermal holonomy potential. The equilibrium wall tension above should therefore be regarded as a benchmark rather than extrapolated using the photon temperature alone. The residual plasma still has the quasistatic Debye screening length $\lambda_{\rm D}^{-2} \simeq 2e^2n_{e}^{\rm res}/T_m$,
but a physical wall profile and tension in this nonequilibrium regime require a real-time plasma-response calculation. Finally, the free-electron fraction becomes spatially dependent during astrophysical reionization. We therefore terminate our late-time evolution at $z_{\rm rei}=20$ and do not attempt to describe the subsequent reionization epoch.

\subsection{Summary of the temperature dependence}
\label{subsec:temperature_summary}

\begin{figure}[th!]
    \centering
    \includegraphics[width=0.45\linewidth]{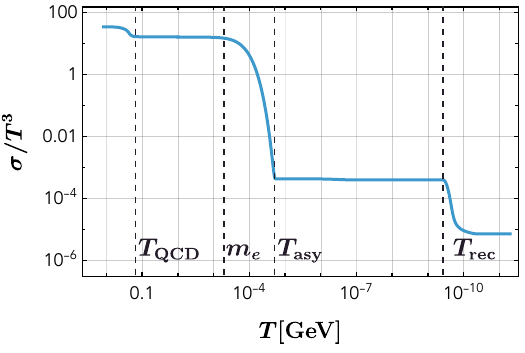}
    \hspace{3mm} \includegraphics[width=0.45\linewidth]{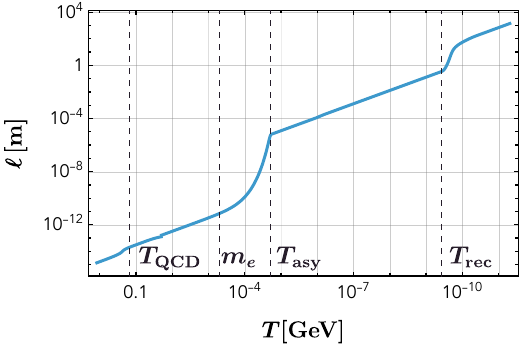}
    \caption{Temperature dependence of the domain-wall properties across all regimes considered. \textit{Left}: The normalized domain-wall tension $\sigma/T^3$. \textit{Right}: The wall width $\ell$ in meters. For both panels, the changes in behavior correspond to the QCD crossover, electron-positron annihilation/the transition to the charge-asymmetry-dominated plasma, and recombination.}
    \label{fig:dwtension_T3_3fields_all}
\end{figure}

The temperature dependence of the domain-wall tension and width is summarized in Fig.~\ref{fig:dwtension_T3_3fields_all}. The left panel shows the dimensionless ratio $\sigma/T^3$, while the right panel shows the physical wall width $\ell$ in meters. Plotting $\sigma/T^3$ makes it particularly easy to identify the temperature intervals in which $\sigma\propto T^3$, whereas the physical width illustrates how the wall evolves from a microscopic object at high temperatures to a macroscopically extended configuration at late times.

For $T\gtrsim T_{\rm QCD}$, the wall contains nontrivial profiles of the QCD holonomies $q_1$ and $q_2$, the electromagnetic holonomy $t$, and the auxiliary field $\Sigma$. The normalized tension $\sigma/T^3$ is of order $10$ and exhibits a smooth feature across the QCD crossover, reflecting the evolution of the QCD and chiral components of the wall. Below $T_{\rm QCD}$, the VEVs of $(q_1,q_2)$ become exponentially close to $(1/3,1/2)$, the $\Sigma$ profile becomes approximately spatially constant, and the wall is predominantly described by the electromagnetic holonomy $t$.

In the relativistic-electron regime, $m_e\lesssim T\lesssim T_{\rm QCD}$, the temperature is the only relevant scale in the electromagnetic wall action. Consequently,
\beqa
\sigma(T)\propto T^3\,,
\qquad
\ell(T)\propto T^{-1} ~,
\eeqa
and $\sigma/T^3$ remains approximately constant. The corresponding wall width decreases approximately as $T^{-1}$ toward higher temperatures.

For $T_{\rm asy}\ll T\ll m_e$, electrons and positrons become nonrelativistic, and their equilibrium densities are exponentially suppressed. Using $n_{e}^{\rm th}\propto (m_eT)^{3/2}e^{-m_e/T}$ gives the approximate scalings
\beqa
\frac{\sigma(T)}{T^3}
\propto
\left(\frac{m_e}{T}\right)^{3/4}
e^{-m_e/(2T)}\,,
\qquad
\ell(T)
\propto
m_e^{-3/4}T^{-1/4}e^{m_e/(2T)} ~.
\eeqa
This Boltzmann suppression produces the steep decrease of $\sigma/T^3$ and the corresponding rapid increase of the wall width visible in the two panels.

For $T_{\rm rec}\ll T\ll T_{\rm asy}$, the thermally produced positron density is negligible and the free-electron density is fixed by charge neutrality,
\beqa
n_e\simeq n_p\simeq \eta_B n_\gamma\propto T^3 ~.
\eeqa
The simple power-law behavior is therefore restored:
\beqa
\sigma(T)\propto T^3 \,,
\qquad
\ell(T)\propto T^{-1} ~.
\eeqa
This explains the intermediate low-temperature plateau, $\sigma/T^3\sim{\cal O}(10^{-4})$, in the left panel and the approximately straight power-law segment in the right panel.

Around recombination, the free-electron density becomes $n_e^{\rm free}=x_e n_H$, where the ionization fraction $x_e$ decreases rapidly. The wall properties consequently scale approximately as
\beqa
\frac{\sigma(T)}{T^3}\propto\sqrt{x_e}\,,
\qquad
\ell(T)\propto\frac{1}{T\sqrt{x_e}} ~.
\eeqa
Recombination therefore produces a second decrease in $\sigma/T^3$ and a corresponding increase in the wall width. Once the residual ionization fraction becomes approximately constant, $\sigma/T^3$ approaches another plateau of order $10^{-5}$, while the width again scales approximately as $T^{-1}$. The extrapolation into the later reionization era is more model dependent because $x_e$ becomes spatially inhomogeneous.

Overall, the wall tension decreases rapidly as the Universe cools, while the wall width grows from approximately $10^{-15}~{\rm m}$ near the GeV scale to macroscopic, potentially kilometer, scales at the lowest temperatures displayed. The curves remain continuous across the QCD crossover, electron-positron annihilation, the charge-asymmetry crossover, and recombination. These changes reflect the successive disappearance or redistribution of the charged plasma species that generate the wall potential. The resulting $\sigma(T)$ and $\ell(T)$ provide the microscopic inputs for the friction and domain-wall network evolution discussed in the following sections.

\section{Interactions with the thermal plasma}
\label{sec:plasma_friction}

The interactions of a thermal domain wall with the surrounding plasma change as the Universe cools. At high temperatures, the constituent-quark mass profile generated by the chiral field provides a direct scattering barrier. After confinement, the electromagnetic holonomy becomes the relevant wall background, and charged-particle and photon interactions are mediated by thermal response functions. Below the charge-asymmetry crossover, $T_{\rm asy}\simeq20\,\mathrm{keV}$, these functions must be evaluated with the surviving electron--ion population rather than a charge-symmetric pair plasma. Residual ionization remains relevant after recombination.

We distinguish the momentum transferred to the wall from its response to that momentum. The former determines a drag coefficient $\eta_X$, whereas the latter also depends on a surface inertia $\mu_X$. The static calculation of Section~\ref{sec:Z3_DW_property} determines a surface free-energy tension, denoted here by $\sigma$ to distinguish it from a physical surface energy density. These quantities need not coincide for an interface supported by a thermal medium. Throughout this section, $z$ denotes the physical coordinate normal to a locally planar wall and $\omega$ is a real frequency. The electromagnetic holonomy is denoted by $t(z)$; a time derivative is always indicated by a dot or written explicitly.

We define the normal force per physical wall area exerted by a homogeneous plasma as
\begin{equation}
 f_{\rm plasma}^{n}(v)=-P_{\rm drag}(v)\,,\qquad
 P_{\rm drag}(v)=\eta_X\,v+\mathcal O(v^3),\qquad \eta_X\geq0 ~,
 \label{eq:s56_drag_definition}
\end{equation}
where $v>0$ is the wall velocity relative to the plasma. We denote by $P_{\rm drag}(v)$ the magnitude of the plasma drag force per unit wall area, which opposes the wall velocity relative to the plasma. This convention assumes degenerate bulk phases and no static pressure bias. Scattering is
most conveniently evaluated in the wall rest frame, where the incident thermal distribution is boosted. At linear order in $v$, the coefficient in Eq.~\eqref{eq:s56_drag_definition} is also the coefficient entering the plasma-frame equation of motion.

\begin{figure}[ht!]
    \centering
    
    \includegraphics[width=0.6\linewidth]{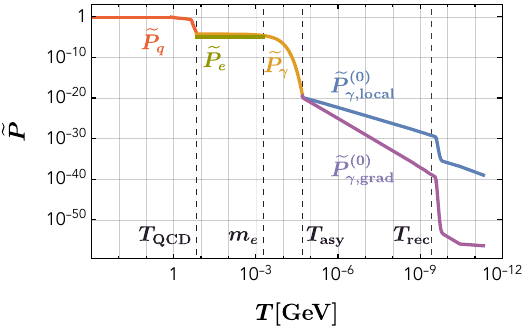}
    \caption{Reduced plasma drag pressure, $\widetilde{P}(T)=P_{\rm drag}(v,T)/(T^4v\gamma_v)$, showing the leading contributions in each temperature regime. Quarks (red) dominate for $T>T_{\rm QCD}$. Electrons and positrons (green) and photons (orange) provide comparable contributions for $m_e<T<T_{\rm QCD}$, while photons dominate for $T<m_e$. For $T<T_{\rm asy}$, the local $(c_0,C_e,C_p)=(1,0,0)$ (blue) and gradient $(c_0,C_e,C_p)=(0,1.9\times10^{12},0)$ (purple) response benchmarks are shown separately; in both, the coefficients are chosen so that $\widetilde{P}$ is continuous
at $T_{\rm asy}$.}
    \label{fig:tildeP_allT_summary}
\end{figure}

For elastic reflection with the same probability from either side and the same asymptotic dispersion on both sides, the momentum-transfer expression is~\cite{Blasi:2022ayo}
\begin{align}
 P_{{\rm drag},a}(v)
 &=2g_a\int\frac{\mathrm d^2p_\parallel}{(2\pi)^3}
      \int_0^\infty\!\mathrm dp_z\,
      \frac{p_z^2}{E_a}\,R_a(E_a,\bm p_\parallel)
      \left[f_a\bigl(\gamma_v(E_a-v\,p_z)\bigr)
            -f_a\bigl(\gamma_v(E_a+v\,p_z)\bigr)\right],
 \label{eq:s56_pressure}\\
 \eta_a &\equiv \frac{\partial P_{{\rm drag},a}}{\partial v}\Big|_{v=0} =\frac{4g_a}{T_a}
      \int\frac{\mathrm d^2p_\parallel}{(2\pi)^3}
      \int_0^\infty\!\mathrm dp_z\,
      \frac{p_z^3}{E_a}R_a(E_a,\bm p_\parallel)
      f_a(E_a)\bigl[1+\epsilon_af_a(E_a)\bigr] ~.
 \label{eq:s56_eta_integral}
\end{align}
Here $E_a^2=p_z^2+p_\parallel^2+m_{a,\infty}^2$, $\gamma_v=(1-v^2)^{-1/2}$, and $f_a(E)=[\exp((E-\mu_a^{\rm chem})/T_a)-\epsilon_a]^{-1}$, with $\epsilon_a=-1$ for fermions and $+1$ for bosons. The degeneracy $g_a$ counts the states included in the reflection probability $R_a$ for an incident particle of species $a$ scattering from the wall; particles and antiparticles are summed separately when their distributions differ. The factor $2p_z$ transferred in a reflection accounts for the factor two in Eq.~\eqref{eq:s56_pressure}. The dominant reduced drag coefficients $\widetilde{P}_a=\eta_a/T^4$ in each temperature regime are shown in
Fig.~\ref{fig:tildeP_allT_summary} and derived in the remainder of this section.

\subsection{Quark-induced friction for $T\gtrsim T_{\rm QCD}$}
\label{subsec:quark_friction}

In the deconfined bulk the chiral field $\Sigma$ is suppressed, whereas the wall solutions contain a localized peak in $\Sigma$ (see Fig.~\ref{fig:sigma_profile}). The associated position-dependent constituent mass $m_q(z)$ is a tree-level barrier for quarks. This contribution is distinct from the loop-induced scattering discussed below. To fix the self-energy convention, write the leading local isotropic symbol of the retarded inverse propagator as
\begin{align}
 \Xi_a^R(z;\omega,\bm p)
 &=A_a\gamma^0+B_a\bm\gamma\cdot\widehat{\bm p}+C_a,
 \label{eq:s56_selfenergy}\\
 \mathcal D_a^R
 &=\gamma^0(\omega-A_a)
   -\bm\gamma\cdot\widehat{\bm p}\,(p+B_a)-(m_a+C_a).
 \label{eq:s56_dirac}
\end{align}
All form factors depend on $(z;\omega,\bm p)$. Thus $\Omega_a=\omega-A_a$, $P_a=p+B_a$, and $M_a=m_a+C_a$. This notation avoids confusing the fermion self-energy with the chiral field. Wall-gradient corrections can generate further structures and are not included in the local isotropic symbol.

For the constituent-mass contribution we set $A_q=B_q=C_q=0$ and solve
\begin{equation}
 \left[-i\alpha_z\partial_z+\bm\alpha_\parallel\cdot\bm p_\parallel
                   +\beta_Dm_q(z)\right]\psi_q=\omega\,\psi_q \,,
 \qquad \alpha_i=\gamma^0\gamma^i,\quad \beta_D=\gamma^0~.
 \label{eq:s56_quark_hamiltonian}
\end{equation}
For this local scalar mass background, a boost parallel to the wall reduces the scattering problem to $\bm{p}_\parallel=0$ with energy $E_\perp=\sqrt{\omega^2-p_\parallel^2}$. The transverse momentum is restored in Eq.~\eqref{eq:s56_eta_integral}. This reduction applies to the tree-level mass-barrier problem, not to a general thermal self-energy.

Let $u_s(\pm p_z)$ be asymptotic spinors normalized to unit normal flux, where $p_z=\sqrt{E_\perp^2-m_{q,\infty}^2}$. The boundary conditions can be written
\begin{align}
 \psi_{q,s}(z)&\longrightarrow
 u_s(p_z)e^{ip_zz}+r_{q,s}u_s(-p_z)e^{-ip_zz}\,,&&z\to-\infty \,,
 \nonumber\\
 \psi_{q,s}(z)&\longrightarrow
 \tau_{q,s}u_s(p_z)e^{ip_zz}\,,&&z\to+\infty ~.
 \label{eq:s56_quark_bc}
\end{align}
The physical reflection probability is
\begin{equation}
 R_{q,s}=\frac{|j^z_{\rm reflected}|}{j^z_{\rm incident}}
        =|r_{q,s}|^2 \,,\qquad j^z=\bar\psi_q\gamma^z\psi_q\,,
 \label{eq:s56_quark_reflection}
\end{equation}
where the last equality uses the flux normalization above. Coefficients of a single spinor component do not in general have this normalization, see Ref.~\cite{Blasi:2022ayo} for details. For a Hermitian mass barrier with identical asymptotic phases, $R_{q,s}+|\tau_{q,s}|^2=1$ provides a useful check.

The pressure can then be parametrized as
\begin{equation}
 P_{{\rm drag},q}(v,T)\simeq T^4\widetilde P_q(T)\,v\,\gamma_v\,,
 \qquad \eta_q(T)=T^4\widetilde P_q(T) \,,
 \label{eq:s56_quark_eta}
\end{equation}
with the finite-velocity factorization understood as an approximation. The transfer from a local mass barrier to a friction force follows the standard scattering treatment of domain-wall friction~\cite{Blasi:2022ayo}. As the bulk condensate grows toward confinement, the mass contrast decreases and this contribution is suppressed. The numerical value is $\widetilde P_q(T)\approx\mathcal{O}(1)$ when $T>T_{\rm QCD}$ shown in Fig.~\ref{fig:tildeP_allT_summary}. For the temperature above 1.2~GeV, we assume $\widetilde P_q(T)=\widetilde P_q(T=1.2~{\rm GeV})$ because the PNJL model may be no longer valid when temperature is much higher than the cutoff scale $\Lambda$. After $T<T_{\rm QCD}$, the reduced drag pressure $\widetilde{P}_q(T)$ exhibits a sharp suppression toward the confinement region, as shown in Fig.~\ref{fig:tildeP_allT_summary}.

\subsection{Electron-induced friction for $m_e\lesssim T\lesssim T_{\rm QCD}$}
\label{subsec:s56_electron_friction}

Electrons do not acquire a constituent-mass barrier from $\Sigma$. Their interaction with the electromagnetic holonomy is instead described by $\Xi_e^R(z;\omega,\bm p)$. The imaginary-time background changes the charged thermal boundary conditions. The continuation to real frequency acts on fluctuations about that background; it does not turn the holonomy into a prescribed real electrostatic potential. This distinction is explicit in the real-time prescription of Hidaka and Pisarski~\cite{Hidaka:2009hs}.

The retarded form factors are generally momentum dependent. In position space they therefore define a nonlocal operator, rather than an ordinary Dirac equation obtained by replacing every momentum by a derivative. We express the nonlocal electron self-energy in a Wigner representation, using the local holonomy to determine the leading quasiparticle dispersion relation. Spatial variations of the wall are then incorporated through a gradient expansion with Weyl ordering. This approximation requires the background to vary slowly over the relevant microscopic propagation scales, while reflection near turning points requires separate matching. In particular, thermal particles have momenta of order $T$, so a soft-external-momentum approximation must not be used throughout their phase space without further justification.

With electron and positron reflection probabilities obtained from this scattering problem, their contribution follows from Eq.~\eqref{eq:s56_eta_integral}. In the relativistic regime the dimensional form is
\begin{equation}
 \eta_{e^-e^+}(T)=T^4\widetilde{P}_e(T)~. 
 \label{eq:s56_electron_eta}
\end{equation}
If the wall profile and the massless-electron form factors scale with $T$, $\widetilde{P}_e(T)$ is approximately temperature independent, apart from coupling evolution and departures from the scaling profile. The dimensionless coefficient $\widetilde{P}_e(T)$ is obtained by integrating the electron and positron reflection probabilities over their thermal momentum distributions. Its approximate temperature independence in the relativistic regime follows from the scaling of the wall profile and the electron self-energy with $T$, whereas its normalization depends on the momentum-dependent scattering problem. We present the self-energy and evaluation of the thermal integral in Appendix~\ref{app:1Loop-selfE}, where the existing one-loop scattering estimate corresponds to the benchmark $\widetilde{P}_e(T)\sim10^{-5}$ from the scattering prescription in Weyl expansion shown in Appendix~\ref{app:WKB_reflection}. We retain this as an approximate input, whose normalization and error depend on the gradient treatment, flux normalization, and treatment of quasiparticle damping. It is not extrapolated unchanged into the asymmetric nonrelativistic plasma. The momentum-dependent form factors and the scattering calculation are given in Appendices~\ref{app:1Loop-selfE} and \ref{app:WKB_reflection}.

\subsection{Electromagnetic response and friction for $T\lesssim m_e$}
\label{subsec:s56_photon_friction}

Photons interact with the electromagnetic holonomy through the charged-plasma response. For a planar wall, this response is described by the retarded kernel
\begin{equation}
 \Pi_{\mu\nu}^R(z,z';\omega,\bm k_\parallel)\,.
 \label{eq:photon_kernel}
\end{equation}
A local polarization is a gradient approximation to this kernel. Its physical modes must satisfy the electromagnetic constraints and Ward identities. Homogeneous plasma results provide useful reference limits~\cite{Braaten:1993jw, Scherer:2024uui}, but do not determine the wall-dependent response. In particular, the thermal twist must not be identified with an imposed real electrostatic potential~\cite{Hidaka:2009hs}. Nor do we assume that the wall carries a localized excess of electrons: the response of their thermal state can be inhomogeneous even when the physical number density is approximately uniform.

We first evaluate a local elastic-scattering benchmark. This assumes well-defined transverse photon quasiparticles, coherent propagation across the wall, reciprocal reflection, and negligible absorption and mode conversion. We approximate the two transverse polarizations by the same scalar equation, giving $g_T=2$. This last approximation requires a check against the full tensor kernel, especially at oblique
incidence.

\paragraph{Nonrelativistic pair plasma.}
For $T_{\rm asy}\ll T\ll m_e$, the electrons, positrons, and photons share a common temperature $T$. A formal local polarization suggested by the conjugate dilute thermal twist weights is
\begin{equation}
 \Pi_T(z)\simeq m_{\gamma,\infty}^2\cos\theta_0(z)\,,\qquad
 m_{\gamma,\infty}^2=\frac{2e^2n_{e}^{\rm th}}{m_e}\,,\qquad
 \theta_0(z)=2\pi t(z)\,.
 \label{eq:s56_cosine_polarization}
\end{equation}
The factor of two counts the two charge species. For the sine-Gordon profile in Eq.~\eqref{eqn:sineGordon_tprofile_Tasy_me}, the transverse equation becomes
\begin{equation}
 A_T''+\left[p_{z,\infty}^2+
 2m_{\gamma,\infty}^2\operatorname{sech}^2(z/d)\right]A_T=0\,,
 \qquad
 p_{z,\infty}^2=\omega^2-k_\parallel^2-m_{\gamma,\infty}^2\,.
 \label{eq:s56_pt_equation}
\end{equation}
The plus sign follows from $p_z^2=\omega^2-k_\parallel^2-\Pi_T(z)$. For this attractive P\"oschl--Teller problem,
\begin{equation}
 R_T(s)=\frac{\sin^2(\pi\lambda)}
 {\sinh^2(\pi s)+\sin^2(\pi\lambda)}\,,\qquad
 s=d\,p_{z,\infty}\,,\qquad
 \lambda(\lambda+1)=2m_{\gamma,\infty}^2d^2=\frac{2T}{m_e}\,.
 \label{eq:s56_pt_reflection}
\end{equation}
Here $\lambda=(\sqrt{1+8T/m_e}-1)/2$; it equals the potential strength $2T/m_e$ at leading order in $T/m_e$.

The corresponding pressure is $P_{{\rm drag},\gamma}=\eta_\gamma v+\mathcal O(v^3)$. In Eq.~\eqref{eq:s56_eta_integral}, take $g_T=2$, bosonic statistics, and $\omega_{\bm p}=\sqrt{p_z^2+p_\parallel^2+m_{\gamma,\infty}^2}$, where $p_z\equiv p_{z,\infty}>0$. Since the scalar reflection probability depends only on $p_z$, the parallel-momentum integral can be evaluated using
$p_\parallel\mathrm dp_\parallel=\omega\mathrm d\omega$ and $f_B(1+f_B)=-T\partial_\omega f_B$. This gives
\begin{equation}
 \eta_\gamma(T)=\frac{2}{\pi^2}
 \int_0^\infty\!\mathrm dp_z\,p_z^3\,R_T(dp_z)\,
 f_B\!\left(\sqrt{p_z^2+m_{\gamma,\infty}^2};T\right)\,,
 \qquad
 f_B(E;T)=\frac{1}{e^{E/T}-1}\,.
 \label{eq:s56_pair_photon_eta_exact}
\end{equation}
We retain this expression for numerical estimates within the local benchmark, including the asymptotic plasma gap.

For $Td\gg1$ and $T/m_e\ll1$, the dominant normal momenta satisfy $p_z\lesssim d^{-1}\ll T$, and $m_{\gamma,\infty}d=\sqrt{T/m_e}\ll1$. The leading weak-potential and soft-photon limits give
\begin{equation}
 R_T(s)\simeq\frac{4\pi^2(T/m_e)^2}{\sinh^2(\pi s)}\,,\qquad
 \eta_\gamma\simeq\frac{2T}{\pi^2d^3}
 \int_0^\infty\!\mathrm ds\,s^2R_T(s)\,.
\end{equation}
Although the reflection expansion fails at $s\lesssim2T/m_e$, it gives the leading integrated asymptote. Using $\int_0^\infty\mathrm ds\,s^2/\sinh^2(\pi s)=1/(6\pi)$ yields
\begin{equation}
 \eta_{\gamma,\rm pair}^{(0)}
 =\frac{4\,T^3}{3\pi\,m_e^2\,d^3}
 =\frac{8\sqrt{2}\,e^3}{3\pi}
 \frac{(n_{e}^{\rm th})^{3/2}\,T^{3/2}}{m_e^2}\,,
 \qquad e^2=4\pi\alpha\,.
 \label{eq:s56_pair_photon_eta_leading}
\end{equation}
The $(n_{e}^{\rm th})^{3/2}$ dependence comes from $d^{-3}$, since $2m_{\gamma,\infty}^2d^2$ is independent of density. In dilute pair equilibrium,
\begin{equation}
 n_{e}^{\rm th}\simeq
 2\left(\frac{m_eT}{2\pi}\right)^{3/2}e^{-m_e/T}\,,\qquad
 \eta_{\gamma,\rm pair}^{(0)}\propto
 T^{15/4}e^{-3m_e/(2T)}\,.
 \label{eq:s56_pair_photon_boltzmann}
\end{equation}
The asymptotic normalization need not be accurate at finite $T/m_e$: Eq.~\eqref{eq:s56_pair_photon_eta_exact} gives $\eta_\gamma/\eta_{\gamma,\rm pair}^{(0)}\simeq0.463$ and $0.331$
at $T=30$ and $50\,\mathrm{keV}$, respectively, using the dilute
pair density and $\alpha^{-1}=137$.

The same calculation can also be applied when $T$ is close to or above $m_e$. The procedure is similar, but there are no analytical expressions for reflection rate or friction force. The numerical results show the photon contribution $\eta_{\gamma}$ is of the same order as the electron one $\eta_{e^-e^+}$ when $T>m_e$. We add these two contributions together to present the total friction pressure for $m_e<T<T_{\rm QCD}$ as shown in Fig.~\ref{fig:tildeP_allT_summary}. Numerical results are used to ensure that  $\eta_{\gamma}$ remains smooth over the interval $T_{\rm asy}<T<T_{\rm QCD}$.

\paragraph{Charge-asymmetry-dominated plasma.}
For $T_{\rm rec}\ll T\ll T_{\rm asy}$, we use a hydrogen-only approximation, neglecting helium and heavier nuclei. Positrons are negligible, and charge neutrality gives $n_e=n_p= n_e^{\rm asy}$. The surviving carriers are fixed by the baryon asymmetry and are not Boltzmann suppressed by $m_e/T$. Their homogeneous nonrelativistic transverse plasma frequency is
\begin{equation}
 \omega_p^2=e^2n_e^{\rm asy}\left(\frac{1}{m_e}+\frac{1}{m_p}\right)\,.
 \label{eq:s56_plasma_frequency}
\end{equation}
At equal electron and proton densities, the proton contribution to the high-frequency transverse response is suppressed by $m_e/m_p$, reflecting its larger inertia. Static Debye screening instead depends on the equilibrium density susceptibility, $m_D^2=e^2(n_e/T_e+n_p/T_p)$. The two species therefore contribute equally at equal temperatures, provided both have time to establish their screening distributions.

A canonical formulation fixes the particle numbers while defining the response through a physical wall state. Let $\rho_W$ be the density operator of a specified quasistationary wall state and set
\begin{equation}
 \rho_{C,W}=\frac{\mathcal P\rho_W\mathcal P}
 {\operatorname{Tr}(\mathcal P\rho_W)}\,,\qquad
 \mathcal P=\mathcal P_{N_e}\mathcal P_{N_p}\,,\qquad N_e=N_p\,.
 \label{eq:s56_canonical_wall_state}
\end{equation}
The retarded electromagnetic response is then obtained from
\begin{equation}
 \Pi_{R,C}^{\mu\nu}(x,y)=
 -i\Theta(x^0-y^0)
 \operatorname{Tr}\!\left(\rho_{C,W}
 [J^\mu(x),J^\nu(y)]\right)
 +\Pi_{\rm contact}^{\mu\nu}(x,y)\,.
 \label{eq:s56_canonical_response}
\end{equation}
Here $J^\mu$ is the electromagnetic current; in a nonrelativistic description the contact term includes the diamagnetic contribution required by gauge invariance. The wall-dependent part is obtained by subtracting the homogeneous response at the same carrier densities. A spatially constant twist inserted solely as $e^{i\theta(N_p-N_e)}$ cancels in normalized neutral-sector correlators. 

To obtain explicit evolution inputs, we parameterize a real local reduction of the wall-dependent response as
\begin{equation}
 \delta\Pi_{T,H}(z)
 =-\frac{\mathcal A_\gamma}{d^2}\operatorname{sech}^2(z/d)\,.
 \label{eq:s56_asymmetric_response}
\end{equation}
The coefficients below are taken to be approximately independent of frequency and momentum over the range relevant for reflection. For the illustrative width $d^{-2}=2e^2n_e^{\rm asy}/T_e$, with $T_p=T_e$,
we write
\begin{equation}
 \mathcal A_\gamma
 =c_0T_e\left(\frac{1}{m_e}+\frac{1}{m_p}\right)
 +\frac{e^2n_e^{\rm asy}}{T_e}
 \left(\frac{C_e}{m_e^2}+\frac{C_p}{m_p^2}\right)\,.
 \label{eq:s56_asymmetric_strength}
\end{equation}
The first term corresponds to a local response $\delta\Pi_{T,H}=-2c_0\omega_p^2\operatorname{sech}^2(z/d)$. The second is an assumed second-gradient contribution, suppressed at the response level by the microscopic parameter $(m_sT_ed^2)^{-1}$ for species $s$. We have chosen the same profile for both terms and absorbed their signed amplitudes into the real matching coefficients $c_0,C_e,C_p$. The width relation and the gradient term are additional model assumptions; they do not follow from number projection.

For this scalar model, the reflection formula \eqref{eq:s56_pt_reflection} applies with $\lambda(\lambda+1)=\mathcal A_\gamma$ in the weak-potential regime $|\mathcal A_\gamma|\ll1$. The exact drag integral~\eqref{eq:s56_pair_photon_eta_exact} uses $m_{\gamma,\infty}\to\omega_p$ and $T\to T_\gamma$. It is useful to write
\begin{equation}
 \eta_\gamma
 =\mathcal I_\gamma\frac{T_\gamma}{3\pi d^3}
 |\mathcal A_\gamma|^2\,,\qquad
 P_{{\rm drag},\gamma}=\eta_\gamma v+\mathcal O(v^3)\,.
 \label{eq:s56_asymmetric_drag}
\end{equation}
Here $\mathcal I_\gamma$ is defined by the exact integral and approaches unity when $|\mathcal A_\gamma|\ll1$, $\omega_pd\ll1$, and $T_\gamma d\gg1$. For $T_e=T_\gamma=T$, the separate leading estimates are
\begin{align}
 \eta_{\gamma,\rm local}^{(0)}
 &=\frac{2\sqrt{2}\,e^3}{3\pi}
 c_0^2 (n_e^{\rm asy})^{3/2}T^{3/2}
 \left(\frac{1}{m_e}+\frac{1}{m_p}\right)^2\,,
 \label{eq:s56_asymmetric_local_eta}\\
 \eta_{\gamma,\rm grad}^{(0)}
 &=\frac{2\sqrt{2}\,e^7}{3\pi}
 \frac{(n_e^{\rm asy})^{7/2}}{T^{5/2}}
 \left(\frac{C_e}{m_e^2}+\frac{C_p}{m_p^2}\right)^2\,.
 \label{eq:s56_asymmetric_gradient_eta}
\end{align}
These define two useful benchmarks: $(c_0,C_e,C_p)=(1,0,0)$ for a local response and $(0,1,0)$ for a suppressed gradient response. In Fig.~\ref{fig:tildeP_allT_summary}, we show the $\widetilde{P}_\gamma=\eta_{\gamma}^{(0)}/T^4$ for both local and gradient responses. The parameters $c_0=1$ and $C_e=1.9\times 10^{12}$ are selected to make $\widetilde{P}_\gamma$ is continuous with $T$ respectively, and $C_p=0$. If both amplitudes are present, their interference is retained in $|\mathcal A_\gamma|^2$. The $(n_e^{\rm asy})^{7/2}$ law is conditional on the specified gradient operator, its density-independent coefficients, and the adopted width. In a homogeneous ideal-gas calculation with no wall-dependent state or coupling, the subtracted response and the coherent reflection drag both vanish.

\paragraph{Recombination and residual ionization.}
In the same hydrogen-only approximation,
\begin{equation}
 n_e=n_p=x_en_H\,,\qquad
 n_H=\eta_Bn_\gamma(T_\gamma)\,,\qquad
 n_\gamma(T_\gamma)=\frac{2\zeta(3)}{\pi^2}T_\gamma^3\,.
 \label{eq:s56_residual_density}
\end{equation}
Here $n_H$ counts all hydrogen nuclei, including those in neutral atoms. Nonequilibrium recombination leaves a residual ionized fraction of order $x_e\sim10^{-4}$ rather than a continuing Saha exponential suppression~\cite{Ali-Haimoud:2010hou}. The free-carrier densities are treated as fixed during an individual wall-scattering event, while their cosmological evolution is supplied by the recombination history. Neutral-atom polarizability and collisional channels are not included in the charged-plasma reflection benchmarks above.

After matter thermally decouples from radiation, the response uses $T_e\simeq T_p\simeq T_m$, while the incident photons use $T_\gamma$. Equation~\eqref{eq:s56_asymmetric_drag} keeps these temperatures separate. There is then no common equilibrium Euclidean temperature. Continuing the wall profile with $d^{-2}=2e^2n/T_e$ is a quasistatic modeling assumption, conditional on the existence and relaxation of the wall state discussed in Section~\ref{sec:Z3_DW_property}. The microscopic drag coefficients calculated above provide inputs to the wall dynamics. Their relation to the surface inertia and the damping terms in the network model is discussed next.

\subsection{Dynamical coefficients and input to the network model}
\label{subsec:s56_dynamical_inputs}

The network evolution requires three surface coefficients: the stiffness $\mathcal T_X$, the inertia $\mu_X$, and the drag coefficient $\eta_X$. They characterize different responses and must be specified separately. 

Let $X(t,\bm x_\parallel)$ be the normal displacement of a locally planar wall. We use the Fourier convention $X\propto e^{-i\omega t+i\bm k_\parallel\cdot\bm x_\parallel}$ and introduce an external force per unit area through
\begin{equation}
 H_{\rm ext}
 =-\int\mathrm d^2x_\parallel\,f_{\rm ext}X ~.
\end{equation}
The retarded displacement susceptibility and its inverse are defined by
\begin{equation}
 X(\omega,\bm k_\parallel)
 =\chi_X^R(\omega,k_\parallel)
 f_{\rm ext}(\omega,\bm k_\parallel) \,,
 \qquad
 K_X^R\equiv(\chi_X^R)^{-1} ~.
\end{equation}
At fixed background temperature and density, a local description of slow, long-wavelength deformations takes the form
\begin{equation}
 K_X^R(\omega,k_\parallel)
 =\mathcal T_X k_\parallel^2
 -\mu_X\omega^2-i\eta_X\omega+\cdots \,,
 \qquad
 K_X^R(0,0)=0 ~.
 \label{eq:s56_wall_kernel}
\end{equation}
The retarded boundary prescription is understood. The vanishing constant term expresses translational invariance in the homogeneous environment. With this convention, the corresponding equation of motion is
\begin{equation}
 \mu_X\ddot X+\eta_X\dot X
 -\mathcal T_X\nabla_\parallel^2X=f_{\rm ext} \,,
 \label{eq:s56_local_wall_eom}
\end{equation}
where $\mu_X>0$ and $\mathcal T_X>0$ describe positive inertia and stiffness, and $\eta_X\geq0$ gives damping.

The coefficients in Eq.~\eqref{eq:s56_wall_kernel} are
properties of the physical displacement response. When the indicated limits exist, they can be extracted as 
\begin{equation}
 \begin{aligned}
 \mathcal T_X
 &=\lim_{k_\parallel\to0}
   \frac{\operatorname{Re}K_X^R(0,k_\parallel)}
        {k_\parallel^2} \,,
 \\
 \mu_X
 &=-\frac12
   \left.
   \frac{\partial^2}{\partial\omega^2}
   \operatorname{Re}K_X^R(\omega,0)
   \right|_{\omega=0} \,,
 \\
 \eta_X
 &=-\lim_{\omega\to0^+}
   \frac{\operatorname{Im}K_X^R(\omega,0)}{\omega} ~.
 \end{aligned}
 \label{eq:s56_coefficient_matching}
\end{equation}
These definitions assume that the response admits the stated expansion. Long-lived plasma modes or nonanalytic frequency dependence can instead require a memory kernel or additional dynamical variables.

\paragraph{Static stiffness and plasma drag.}
For an isotropic equilibrium interface, a slowly varying displacement changes the surface free energy by
\begin{equation}
 \Delta F
 =\sigma\,\int\mathrm d^2x_\parallel
 \left[\sqrt{1+|\bm\nabla_\parallel X|^2}-1\right]
 =\frac{\sigma}{2}
 \int\mathrm d^2x_\parallel
 |\bm\nabla_\parallel X|^2+\cdots.
\end{equation}
Consequently,
\begin{equation}
 \mathcal T_X=\sigma ~.
 \label{eq:s56_static_stiffness}
\end{equation}
This equality determines the static restoring force. It does not identify the surface free energy with the inertial mass, because the plasma selects a preferred rest frame.

The drag coefficient is determined by the force opposing motion relative to the plasma. If $P_{\rm drag}(v)$ denotes the positive drag pressure for $v>0$, then
\begin{equation}
 \eta_X
 =\lim_{v\to0^+}\frac{P_{\rm drag}(v)}{v} ~.
 \label{eq:s56_drag_slope}
\end{equation}
Thus the scattering calculations in the preceding subsections directly provide contributions to $\eta_X$, subject to their response and scattering assumptions, and for independently identifiable channels,
\begin{equation}
 \eta_X=\eta_q+\eta_{e^-e^+}+\eta_\gamma+\cdots.
 \label{eq:s56_total_drag}
\end{equation}

\paragraph{What the thermal transit calculation determines.} 
An ideal-electron calculation provides a microscopic time scale associated with a translated thermal twist. It does not calculate the physical displacement kernel in Eq.~\eqref{eq:s56_wall_kernel}. We summarize the calculation here to make its role in the inertia parameterization explicit.

Use $t$ for real time and $\tau_E$ for imaginary time, with $\beta_e=1/T_e$. The twisted electron boundary condition and the corresponding one-particle insertion are
\begin{equation}
 \psi(\beta_e,z)
 =-e^{i\theta_0(z-X)}\psi(0,z) \,,
 \qquad
 u_X(z)=e^{-i\theta_0(z-X)} ~.
\end{equation}
An infinitesimal displacement changes this insertion by $\delta u=\delta X\,b(z)$, where
\begin{equation}
 b(z)\equiv
 \left.\frac{\partial u_X(z)}{\partial X}\right|_{X=0}
 =-u_0'(z) \,,
 \qquad
 w(z)\equiv u_0(z)-1 ~.
 \label{eq:s56_twist_displacement_insertion}
\end{equation}
To track the decorrelation of this change under free electron motion, define the auxiliary operator $\widehat{\mathcal B}=\sum_j b(\widehat z_j)$ and its connected correlation per transverse area,
\begin{equation}
 \mathcal G_{\rm tw}^{>}(t)
 \equiv\frac{1}{A_\parallel}
 \left\langle
 \widehat{\mathcal B}^{\dagger}(t)
 \widehat{\mathcal B}(0)
 \right\rangle_{0,c} ~.
 \label{eq:s56_twist_correlation_definition}
\end{equation}
Here the average is taken in the homogeneous ideal electron gas, the subscript $c$ means that the product of the one-point averages is subtracted, and time evolution is generated by the free-electron Hamiltonian. The superscript $>$ denotes the displayed, non-time-ordered correlation, with no factor of $-i$.

At leading order in the dilute-gas expansion, free-fermion contractions, equivalently the trace identity of Ref.~\cite{Klich:2002xau}, give 
\begin{equation}
 \mathcal G_{\rm tw}^{>}(t)
 =n_e\int\frac{\mathrm dq}{2\pi}\,
 q^2|\widetilde w(q)|^2
 \exp\left[
 -\frac{q^2v_{Te}^2t^2}{2}
 -\frac{iq^2t}{2m_e}
 \right] \,,
 \qquad
 v_{Te}^2=\frac{T_e}{m_e} \,,
 \label{eq:s56_twist_correlation}
\end{equation}
where $\widetilde w(q)=\int\mathrm dz\,e^{-iqz}w(z)$ and $n_e$ includes both electron spin states. The factor $q^2|\widetilde w|^2$ follows from $\widetilde b=-iq\widetilde w$. The Gaussian is the Maxwell average of $e^{-iqv_z t}$, while the remaining phase follows from the recoil energy $q^2/(2m_e)$.

For the sine-Gordon profile in Eq.~\eqref{eqn:sineGordon_tprofile_Tasy_me}, $\int\mathrm dz\,|u_0'|^2=8/d$ and $\int\mathrm dz\,|u_0''|^2=24/d^3$. Neglecting recoil when $d\sqrt{m_eT_e}\gg1$ therefore gives the short-time expansion
\begin{equation}
 \frac{\mathcal G_{\rm tw}^{>}(t)}
      {\mathcal G_{\rm tw}^{>}(0)}
 =1-\frac32\left(\frac{v_{Te}t}{d}\right)^2
 +\mathcal O\!\left(\frac{v_{Te}^4t^4}{d^4}\right),
 \qquad
 t_{\rm tr}\sim\frac{d}{v_{Te}}.
 \label{eq:s56_twist_time}
\end{equation}
The result identifies the thermal transit time of the wall width. The decorrelation in this calculation arises from free-particle dephasing.

There is no direct identification between the coefficient $3/2$ in Eq.~\eqref{eq:s56_twist_time} and any coefficient in Eq.~\eqref{eq:s56_wall_kernel}. The former describes a short-time moment of an auxiliary correlation, whereas $\mu_X$ and $\eta_X$ describe the low-frequency physical displacement response. Moreover, normalization by $\mathcal G_{\rm tw}^{>}(0)$ removes the overall correlation amplitude: rescaling $\widehat{\mathcal B}\to g\widehat{\mathcal B}$ leaves this ratio unchanged.

The thermal transit time nevertheless provides useful guidance for modeling the wall dynamics. If electron transport controls the response to a wall displacement, $t_{\rm tr}$ suggests a characteristic microscopic response time, while the associated velocity scale $d/t_{\rm tr}=v_{Te}$ provides a benchmark for the propagation speed of wall deformations. These scales can guide a phenomenological description of the wall inertia and the response of the force conjugate to $X$.

This interpretation remains conditional: the auxiliary operator $\widehat{\mathcal B}$ has not been matched to the physical force, and the transit calculation does not determine its magnitude or its reactive and dissipative components. Establishing these requires the wall-dependent real-time coupling and its electromagnetic constraints. The result therefore motivates an electron-controlled dynamical scale without fixing $\mu_X$ or $\eta_X$, imposing a strict speed bound, or establishing the existence of a local propagating wall mode.

\paragraph{Surface inertia as a matching parameter.}
To proceed with the network evolution, we adopt the electron-controlled hypothesis that the ratio of the temporal and spatial response coefficients is set by the electron thermal velocity. This is an additional assumption motivated by the transit scale, rather than a consequence of the normalized correlation. We parameterize it by
\begin{equation}
 \mu_X=C_\mu\,\sigma\,\frac{m_e}{T_e}\,,
 \qquad C_\mu>0~.
 \label{eq:s56_inertia_ansatz}
\end{equation}
We take $C_\mu$ to be constant within each illustrative evolution benchmark and vary it independently of the specified drag input. Its value must ultimately be determined by matching the physical displacement response.

If a weakly damped range of wall modes exists, Eqs.~\eqref{eq:s56_static_stiffness} and \eqref{eq:s56_inertia_ansatz} imply
\begin{equation}
 c_X^2\equiv\frac{\mathcal T_X}{\mu_X}
 =\frac{T_e}{C_\mu m_e}
 =\frac{v_{Te}^2}{C_\mu} ~.
 \label{eq:s56_wall_mode_speed}
\end{equation}
This equation expresses the assumed electron-controlled scaling. The ideal-gas calculation does not prove $C_\mu=\mathcal O(1)$ or impose a strict upper bound $v\leq v_{Te}$ on the translational wall velocity. Whether $c_X$ describes a propagating wall mode depends on the damping at the wavelength of interest. For reference, the small-velocity limit of the conventional tension-based baseline model identifies
\begin{equation}
 \mu_X^{0}=\mathcal T_X^{0}=\sigma^0 ~.
 \label{eq:s56_baseline_inertia}
\end{equation}
For $\eta_X>0$, mobility length $\ell_{\rm mob}\equiv\mathcal T_X/\eta_X$ characterizes the transport in the collective coordinates, while $\tau_{\rm rel}\equiv\mu_X/\eta_X$ represents the velocity relaxation. With $\mathcal T_X=\sigma$ and the inertia ansatz, these become
\begin{equation}
 \ell_{\rm mob}=\ell_f \,,
 \qquad
 \tau_{\rm rel}
 =C_\mu\frac{m_e}{T_e}\,\ell_f \,,
 \qquad
 \ell_f\equiv\frac{\sigma}{\eta_X} ~.
 \label{eq:s56_mobility_inertia_relation}
\end{equation}
Thus the conventional friction length equals the mobility length, but it equals the velocity relaxation time only in baseline $\mu_X^0=\sigma_0$.

The ansatz is applied below $T_{\rm asy}\simeq20\,\mathrm{keV}$ and is not extrapolated into the relativistic electron or quark regimes. After matter and radiation cease to share a common temperature, its use also assumes that the adopted wall profile and local surface description remain applicable.

\paragraph{Photon contribution to network damping.}
The electromagnetic response benchmarks in the preceding subsection supply conditional estimates of $\eta_\gamma$. Their conversion into velocity damping requires the surface inertia. If the same sine-Gordon free-energy functional supports the entire wall, then
$\sigma=8T_e^2/(e^2d)$, and
\begin{equation}
 \ell_{f,\gamma}^{-1}
 \equiv\frac{\eta_\gamma}{\sigma}
 =\mathcal{I}_\gamma
 \frac{e^4\,n\,T_\gamma}{12\pi\,T_e^3}\,
 |\mathcal A_\gamma|^2 ~.
 \label{eq:s56_photon_vos_input}
\end{equation}
The corresponding contribution to velocity damping is
\begin{equation}
 \Gamma_\gamma
 =\frac{\eta_\gamma}{\mu_X}
 =\frac{\sigma}{\mu_X}\ell_{f,\gamma}^{-1}
 =\frac{T_e}{C_\mu m_e}\ell_{f,\gamma}^{-1},
 \label{eq:s56_photon_relaxation}
\end{equation}
where the last equality uses Eq.~\eqref{eq:s56_inertia_ansatz}. For the baseline $\mu_X^0=\sigma^0$, instead, $\Gamma_\gamma^{\rm base}=\ell_{f,\gamma}^{-1}$.

With the adopted width $d^{-2}=2e^2n/T_e$, fixed $T_e$ and $T_\gamma$, and fixed response coefficients, the leading local and gradient benchmarks give $\eta_\gamma\propto x_e^{3/2}$ and $x_e^{7/2}$, respectively. Dividing by the corresponding surface free-energy tension gives $\ell_{f,\gamma}^{-1}\propto x_e$ and $x_e^3$, up to the correction factor $\mathcal F_\gamma$. These scalings depend on the assumed wall profile and response operators.

For orientation, at $T_e=T_\gamma=T=10\,\mathrm{keV}$, $x_e=1$, and $\eta_B=6.1\times10^{-10}$, evaluation of the exact scalar integral gives
\begin{equation}
 \left(
 \frac{\eta_\gamma}{T^4},
 \ell_{f,\gamma}^{-1}
 \right)
 \simeq
 \begin{cases}
 (4.58\times10^{-21},\;
  1.53\times10^2\,\mathrm{s}^{-1}),
 & (c_0,C_e,C_p)=(1,0,0),\\[2pt]
 (3.44\times10^{-46} \,,\;
  1.15\times10^{-23}\,\mathrm{s}^{-1}),
 & (c_0,C_e,C_p)=(0,1,0) ~.
 \end{cases}
 \label{eq:s56_photon_drag_benchmarks}
\end{equation}
For the inertia ansatz, the corresponding damping rates are approximately $2.99\,C_\mu^{-1}\,\mathrm{s}^{-1}$ and $2.25\times10^{-25}\,C_\mu^{-1}\,\mathrm{s}^{-1}$.

For an illustrative post-recombination state with $T_e=T_\gamma=0.1\,\mathrm{eV}$ and $x_e=10^{-4}$, the same two benchmarks give $\ell_{f,\gamma}^{-1}\simeq
1.9\times10^{-17}\,\mathrm{s}^{-1}$ and $1.4\times10^{-60}\,\mathrm{s}^{-1}$, respectively. Their velocity damping rates follow from Eq.~\eqref{eq:s56_photon_relaxation}.

These estimates illustrate the sensitivity to the electromagnetic matching coefficient $c_0$. They do not establish photon dominance over other drag channels. We therefore specify the response benchmark and $C_\mu$ separately in the network evolution. Treating them as independent effective inputs does not imply that inertia and dissipation are fundamentally unrelated: a complete microscopic calculation must determine a common causal response. The available static free energy and low-velocity drag estimate, however, do not suffice to fix its inertial coefficient.

\section{Cosmological evolution of the domain-wall network}
\label{sec:s56_network_evolution}

A thermal holonomy wall differs from the scalar interfaces usually described by the velocity-dependent one-scale (VOS) model. For a relativistic scalar wall with fixed tension $\sigma_0$, one coefficient determines its rest energy, surface stiffness, and small-velocity inertia. A thermal wall exists in a medium that selects a rest frame, and its static properties change with the plasma. We denote its excess surface free energy by $\sigma$ and identify the restoring stiffness with this tension, $\mathcal T_X=\sigma$. The inertia $\mu_X$ and drag $\eta_X$ remain separate dynamical inputs. This tension $\mathcal T_X$ is assumed to equal the excess gravitational energy per area.

We evolve a conditional network from $T_{\rm form}=10^3,10^6,10^{10},10^{14}\,\mathrm{GeV}$ to $z_{\rm rei}=20$, using quark, electron--positron, and photon drag in their respective temperature regimes. Above $m_e$ we adopt $\mu_X=\sigma$. In the nonrelativistic regime the electron-transit argument motivates
\begin{equation}
 \mu_X=C_\mu\sigma\frac{m_e}{T_e} \,,\qquad
 c_X^2\equiv\frac{\sigma}{\mu_X}
       =\frac{T_e}{C_\mu m_e} \,,\qquad
 \Gamma_X\equiv\frac{\eta_X}{\mu_X}
       =\frac{T_e}{C_\mu m_e}\frac{\eta_X}{\sigma} ~.
 \label{eq:s6_inertia_benchmark}
\end{equation}
The scan $C_\mu=0.1,1,10$ changes the low-temperature inertia while holding the tension, drag, cosmology, and formation prescription fixed. In particular, $C_\mu=1$ gives $c_X=v_{Te}=\sqrt{T_e/m_e}$, rather than $\mu_X=\sigma$.

All four formation temperatures exceed the electroweak crossover, $T_{\rm EW}\simeq160\,\mathrm{GeV}$~\cite{DOnofrio:2015gop}. Above this crossover the electromagnetic wall description used here requires matching to the electroweak theory. We explicitly extrapolate the specified quark-wall benchmark into this region to study sensitivity to the inherited separation. These early trajectories are not a calculation of wall formation or survival in the full SM. The endpoint $z_{\rm rei}=20$ likewise specifies the end of the calculation; we do not model astrophysical reionization.

\subsection{Formation and the choice of network scale}
\label{subsec:s56_formation}

The existence of symmetry-related thermal configurations alone does not determine the network production rate or correlation length. One possible formation channel is reheating. If the maximum plasma temperature enters the regime in which the thermal holonomy potential admits multiple symmetry-related minima, causally disconnected regions may relax into different configurations. The interfaces between these regions then form thermal domain walls. If the system passes through this regime sufficiently rapidly, the resulting correlation length is controlled by the thermal relaxation and quench time, as in the Kibble--Zurek picture, and is bounded by the horizon scale~\cite{Kibble:1976sj,Zurek:1985qw}.

A second possibility is production during preheating. Resonant, nonthermal fluctuations of gauge and charged fields may drive the local holonomy into different basins of the effective potential~\cite{Kofman:1994rk,Kofman:1997yn}. In this case, the initial separation is set by the characteristic wavelength of the amplified modes and may be substantially smaller than the horizon. Subsequent thermalization preserves only those interfaces that relax into domain walls supported by the finite-temperature effective potential. The efficiency of either mechanism depends on the reheating history, relaxation rates, and couplings to the inflaton sector, and we do not attempt to calculate it here.

The number of vacua $N=6/p$ enters both the initial conditions and the network parameters. For randomly assigned domains the mean number of elementary walls per domain boundary is $1.5$, $0.67$ and $0.5$ for $N=6,3,2$. The $N=2$ network ($\Gamma=\Z_3$) has no junctions, whereas the $N=3$ and $N=6$ networks contain three-wall junctions.
We instead parameterize these formation uncertainties by $T_{\rm form}$ and $\xi_{\rm form}$ and specify
\begin{equation}
L_{\rm form}=\frac{\xi_{\rm form}}{H_{\rm form}} \,,\qquad
\xi_{\rm form}=1 \,,\qquad v_{\rm form}=0,
\label{eq:s56_initial_condition}
\end{equation}
for each numerical benchmark. The choice $\xi_{\rm form}=1$ corresponds to a horizon-scale initial separation and therefore to the largest separation expected from a causal post-inflationary formation process. It should be regarded as a benchmark rather than a prediction of either reheating or preheating. A subhorizon formation scale can be described by $\xi_{\rm form}<1$. We take $v_{\rm form}=0$ to represent an initially overdamped network.

This common horizon-scale initial condition replaces an independently assigned late-time separation. Wall motion and area removal are initially negligible for the reasons discussed in the following subsections, and entropy conservation gives
\begin{equation}
\dot L\simeq HL \,,\qquad
L(T)\simeq\frac{\xi_{\rm form}}{H_{\rm form}}
\frac{T_{\rm form}}{T}
\left[\frac{g_{*s}(T_{\rm form})}{g_{*s}(T)}\right]^{1/3} ~.
\label{eq:s56_stretching}
\end{equation}
For approximately constant relativistic degrees of freedom during radiation domination, $HL\propto T/T_{\rm form}$. Earlier formation therefore yields a smaller separation in Hubble units at a fixed later temperature.

We define the network geometric length $L$ and the energy length $L_E$ by
\begin{equation}
L\equiv\frac{V_{\rm phys}}{A}\,,\qquad L_E\equiv\frac{\sigma}{\rho} \,.
\end{equation}
Here $A$ is the physical area of the selected long-wall population contained in a fixed comoving volume. For a narrow velocity distribution, we use a representative $v$ for the rms velocity. The scalar network then satisfies $L=\gamma_vL_E$, where $\gamma_v=(1-v^2)^{-1/2}$. A surface description also requires the wall width to be much smaller than both $L$ and $H^{-1}$, together with sufficiently rapid relaxation of the wall profile.

The physical area in a fixed comoving volume evolves according to $\dot A/A=2H-\kappa\,v/a-\Gamma_A$, where $\Gamma_A$ is its fractional removal rate and $a$ is the scale factor. Under our curvature convention, $\kappa\,v>0$ corresponds to area shrinkage after averaging over the wall network. The curvature $\kappa$ is defined in Appendix~\ref{app:modified_VOS}. This implies that the evolution of $L$ takes the form
\begin{align}
\dot L=HL+ L\,\kappa\,v/a+\Gamma_A L~.
\label{eq:L_evolution_geometry}
\end{align}
The evolution of $L_E$ and $v$ is described by the VOS model in the following subsection.

\subsection{Thermal VOS equations for domain walls}
\label{subsec:s56_vos_model}

For a scalar wall with constant tension, the standard VOS equations use the energy length $L_E$ and rms velocity $v$:
\begin{align}
 \dot L_E&=HL_E+D_Ev^2L_E+c_wv \,,\nonumber\\
 \dot v&=(1-v^2)
       \left[\frac{k_w}{L_E}-D_Ev\right] \,,\qquad
 D_E=3H+\frac{\eta_0}{\sigma_0} ~.
 \label{eq:s56_minimal_vos}
\end{align}
Here $\sigma_0$ represents the constant tension, and $\eta_0$ denotes the small-velocity drag in the conventional relativistic friction prescription, while $k_w$ and $c_w$ parametrize curvature and loss from the long-wall population~\cite{Avelino:2005kn}. Transforming Eq.~\eqref{eq:s56_minimal_vos} for the energy length to the geometric length gives the same form for the evolution of $L$ as Eq.~\eqref{eq:L_evolution_geometry}:
\begin{align}
    \dot{L}=HL+\gamma_v(k_w+c_w)v,
\end{align}
because the explicit damping contribution to $\dot L_E$ cancels against the corresponding term from $\dot\gamma_v$. Here the parameter in the energy-length equation is $k_w=\kappa L_E/a$. The corresponding geometric curvature coefficient is $k_g\equiv\kappa L/a=\gamma_v k_w$, so $k_gv$ describes curvature-driven area shrinkage. The distinction is negligible for $v\ll1$. The term $c_{w}v\gamma_v=\Gamma_A L$ counts removal from the selected population, including the production of closed walls when they cease to contribute to the area $A$. Velocity-dependent scalar-network calibrations and scalar radiation can be included~\cite{Martins:2016ois}, but these do not provide a calibration for a thermal holonomy network.

For a thermal holonomy network, the speed of domain wall propagation is bounded by the thermal interaction speed, which differs from the speed of light at low temperatures. We describe it using $c_X^2=\sigma/\mu_X$ and a phenomenological parameter $C_\mu$. The TVOS equations are
\begin{align}
 &\dot{L}_E=HL_E+D_X  \frac{v^2}{c_X^2}L_E+c_w v \,,\qquad\dot v=\left(1-\frac{v^2}{c_X^2}\right)\left(c_X^2\frac{k_E}{L_E}-D_Xv\right)+\frac{\dot{c}_X}{c_X}v\,,\nonumber\\
 &D_X=3H+\frac{\dot\mu_X}{\mu_X}+\frac{\eta_X}{\mu_X}+\frac{\dot{c}_X}{c_X} ~,
 \label{eq:s56_modified_vos_cSound}
\end{align}
as derived in Appendix~\ref{app:modified_VOS}. The time-dependent inertia enters the damping in a manner analogous to the time-dependent tension in the VOS model of Ref.~\cite{Bai:2026ufp}. Setting $\mu_X=\sigma=\sigma_0$ and $\eta_X=\eta_0$ recovers the standard VOS equations in Eq.~\eqref{eq:s56_minimal_vos}. We should point out that the equation of $\dot{L}$ is no longer as simple as the standard VOS results with $c_X\ne1$.

\paragraph{Temperature-dependent coefficients.}
Away from crossover regions we use
\begin{equation}
 (\mu_X,\quad \eta_X)=
 \begin{cases}
  (\sigma,\quad T^4\widetilde P_q(T))\,,&T>T_{\rm QCD}\,,\\[2pt]
  (\sigma,\quad T^4(\widetilde P_e(T)+\widetilde P_\gamma(T)))\,,&m_e<T<T_{\rm QCD}\,,\\[2pt]
  (C_\mu\sigma m_e/T_e,\quad T^4\widetilde P_\gamma(T))\,,&T\ll m_e\,,
 \end{cases}
 \label{eq:s56_piecewise_inputs}
\end{equation}
The corresponding curvature and damping coefficients are
\begin{equation}
 c_X^2=
 \begin{cases}
  1\,,&T>m_e \,,\\[2pt]
  T_e/(C_\mu m_e)\,,&T\ll m_e \,,
 \end{cases}
 \label{eq:s56_piecewise_curvature}
\end{equation}
and
\begin{equation}
 D_X=
 \begin{cases}
  3H+\dot\sigma/\sigma+T^4\widetilde P_q/\sigma\,,
       &T>T_{\rm QCD}\,,\\[3pt]
  3H+\dot\sigma/\sigma+T^4(\widetilde P_e(T)+\widetilde P_\gamma(T))/\sigma\,,
       &m_e<T<T_{\rm QCD}\,,\\[3pt]
  3H+\dot\sigma/\sigma-\dot T_e/T_e+\dot{c}_X/c_X
       +T_e\eta_\gamma/(C_\mu m_e\sigma)\,,&T\ll m_e ~.
 \end{cases}
 \label{eq:s56_piecewise_damping}
\end{equation}
Around QCD transition, the nonrelativistic effect with non unity $c_X$ does not influence the evolution, because the drag pressure drops very quickly within a short temperature interval. Thus, we take $c_X=1$ around $T_{\rm QCD}$ as well. To smoothly connect the $\mu_X$ across the two regimes near $m_e$, we replace $c_X^{-2}$ by
\begin{align}
    c_X^{-2}\approx 1+\frac{C_\mu m_e/T_e}{1+(T/m_e)^4}\,,
\end{align}
and $\mu_X=\sigma/c_X^2$ . We also set $3H+\dot\sigma/\sigma=0$ around the recombination era, to avoid the a negative pulse in damping rate. Thus, all inputs to the TVOS equations are nearly continuous.

Below $m_e$, the inertia brings in the additional term
$-\dot T_e/T_e$. For the asymmetric-plasma tension
$\sigma\propto n_e^{1/2}T_{\sigma}^{3/2}$ and
$n_e\propto x_ea^{-3}$, where $T_{\sigma}$ denotes the temperature used in the tension calculation. This gives
\begin{equation}
 3H+\frac{\dot\mu_X}{\mu_X}+\frac{\dot{c}_X}{c_X}=\frac{3}{2}H+\frac{3}{2}\frac{\dot{T}_\sigma}{T_\sigma}-\frac{1}{2}\frac{\dot{T}_e}{T_e}+\frac{1}{2}\frac{\dot{x}_e}{x_e}.
 \label{eq:s56_expansion_damping}
\end{equation}
At fixed ionization, this term is positive while $T_e=T_\gamma\propto a^{-1}$. After thermal decoupling at $z\simeq150$, $T_e\propto a^{-2}$ and the notion of an equilibrium wall tension becomes ambiguous, as discussed in Section~\ref{subsubsec:tensionTlTrec}. We consider (i) $T_\sigma=T_\gamma$ and (ii) $T_\sigma=T_e$. The background damping is $H/2$ before decoupling; afterward it is $H$ in (i) and $-H/2$ in (ii), at fixed ionization. Negative damping increases $r=v/c_X$ in the model, but $r=1$ remains an invariant boundary. Since $c_X$ decreases, a larger $r$ does not require the physical velocity to increase with time. The calculations remain finite through the chosen endpoint $z_{\rm rei}=20$. This ambiguity introduces an order-of-magnitude difference in the final results for $HL$, $v$, and $\rho_{\rm DW}/\rho_{\rm tot}$. We use benchmark (i) in the following discussion.

\subsection{Thermal history of the domain wall network from TVOS solutions}

\begin{figure}[th!]
 \centering
 \includegraphics[width=0.49\linewidth]{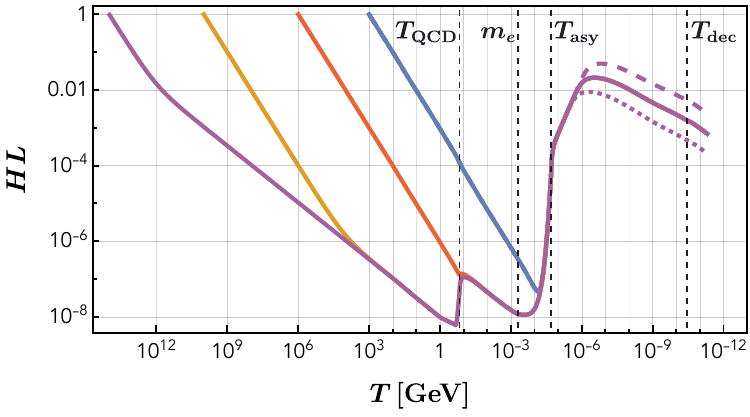}\hfill
 \includegraphics[width=0.49\linewidth]{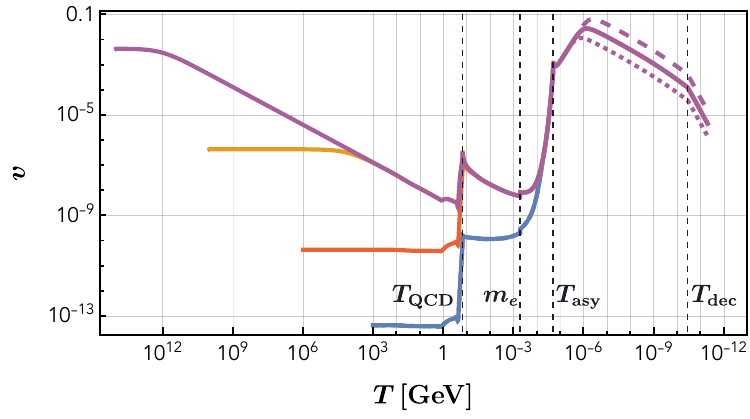}\\[3pt]
 \includegraphics[width=0.49\linewidth]{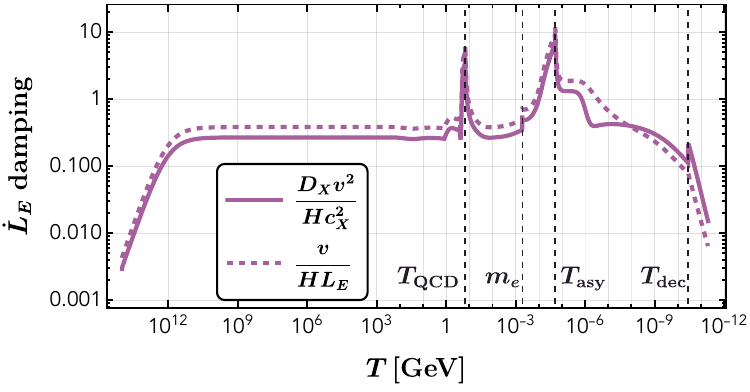}\hfill
 \includegraphics[width=0.49\linewidth]{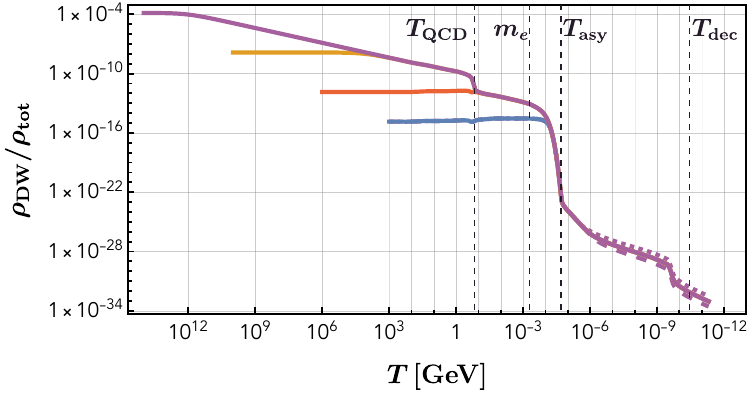}
 \caption{Assuming a gradient friction response for $T<T_{\rm asy}$, we show the evolution for $T_{\rm form}=10^3,10^6,10^{10},10^{14}\,\mathrm{GeV}$ (blue, red, orange, and purple, respectively) and $C_\mu=0.1,1,10$ (dashed, solid, and dotted, respectively), with $H_{\rm form}L_{\rm form}=1$ and $v_{\rm form}=0$. The parameters $k_w=0.7$ and $c_w=0.6$ are held fixed. To match the friction at $T_{\rm asy}$, we use $(c_0,C_e,C_p)=(0,1.9\times10^{12},0)$. The top row shows the evolution of $HL$ and $v$ as the Universe cools. The bottom-left panel shows the two damping terms in the $L_E$ equation, normalized to the Hubble rate, for $T_{\rm form}=10^{14}~{\rm GeV}$. The bottom-right panel shows the ratio of the domain-wall energy density to the total energy density. All trajectories end at the adopted reionization redshift, $z_{\rm rei}=20$.}
 \label{fig:vos_friction_results_grad}
\end{figure}

In this subsection, we discuss the solutions of the TVOS equations. We use the numerical Standard Model degrees of freedom from Ref.~\cite{Saikawa:2018rcs}: $g_{*s}$ determines entropy conservation, while $g_{*\rho}$ determines the radiation density. The results are shown in Fig.~\ref{fig:vos_friction_results_grad}. In this computation, we adopt $k_w=0.7$ and $c_w=0.6$~\footnote{The VOS parameters for domain walls can also vary with time and system properties, see Refs.~\cite{Avelino:2011ev, Martins:2016ois, Avelino:2020ubr, Avelino:2022zem}.}, as in the standard domain wall network, although simulations are needed to justify the assumption that these parameters are of order unity. The physical velocity is small, so the reporting convention gives $L\simeq L_E$, although $v/c_X$ can approach unity. The energy density ratio is
\begin{align}
 \Omega_\sigma\equiv\frac{\rho_{\rm DW}}{\rho_{\rm tot}}=\frac{\sigma/L_E}{\rho_{\rm tot}}
 =\frac{8\pi\sigma}{3M_{\rm pl}^2H(HL_E)}.
\end{align}
Here $M_{\rm pl}=G^{-1/2}=1.22\times10^{19}\,\mathrm{GeV}$ is the unreduced Planck mass. During radiation domination this reduces to
\begin{align}
 \Omega_\sigma=\frac{4\sqrt5\,\widetilde\sigma}{\sqrt{\pi g_{*\rho}}}\frac{T}{M_{\rm pl}}\frac{1}{HL_E},\qquad \widetilde\sigma\equiv\sigma/T^3.
\end{align}

\paragraph{Stretching and coarsening network.} When $T>m_e$ and $g_{*s}$ is constant, Hubble damping is exactly canceled by $\dot{\sigma}/\sigma$ (or slightly overcompensated during the QCD transition), while friction dominates the evolution. The network initially follows the stretching regime discussed in Ref.~\cite{Bai:2026ufp}, until the motion-dependent terms become appreciable. The velocity becomes independent of its initial value and reaches the terminal velocity 
\begin{align}
    v_{\rm term}\simeq\frac{k_w\sigma}{\eta_X L_E}, \label{eq:s56_terminal}
\end{align}
on a timescale $D_X^{-1}\ll H^{-1}$ away from rapid input transitions. Substituting the terminal velocity into the $L_E$ equation gives
\begin{align}
    -\frac{\dd HL_E}{\dd \log T}=-HL_E+\frac{\sigma k_w(k_w+c_w)}{\eta_X}\frac{1}{L_E}=-HL_E+\frac{ k_w(k_w+c_w)}{D_X/H}\frac{1}{L_EH},\label{eq:initial_HL_T}
\end{align}
with slow $g_{*s}$ variance gives $\dd/\dd t=-HT\dd/\dd T$. The $D_X/H=\widetilde{P}M_{\rm pl}/(1.66 g_{*\rho}^{1/2}\tilde{\sigma}T)$, where $\widetilde{P}$ is about 1 above $T_{\rm QCD}$ and $\mathcal{O}(10^{-5})$ below. The solution can be written as 
\begin{align}
    HL_E=\sqrt{\left(\frac{T}{T_{\rm form}}\right)^2\left(1-\frac{T_{\rm form}}{\widetilde{M}}\right)+\frac{T}{\widetilde{M}}}\,,\qquad \widetilde{M}=\frac{M_{\rm pl}\widetilde{P}}{2k_w(k_w+c_w)1.66 g_{*\rho}^{1/2}\tilde{\sigma}}\,.
\end{align}
There are two limits. When the first term dominates, we have
\begin{align}
    HL_E\approx \frac{T}{T_{\rm form}}\Rightarrow L_E\approx \frac{M_{\rm pl}}{1.66g_{*\rho}^{1/2}T_{\rm form}}\frac{1}{T}\,,
\end{align}
as the stretching regime. The characteristic number of wall crossings along a Hubble-length line then scales as $N_{\rm cross}\sim(HL_E)^{-1}\propto T^{-1}$. This also shows that the physical separation is automatically much larger than the domain wall width, $L\gg \ell$. The second regime is when the second term dominates, corresponding to
\begin{align}
    T<T_v\equiv T_{\rm form}\frac{T_{\rm form}/\widetilde{M}}{1-T_{\rm form}/\widetilde{M}} \,,
\end{align}
which gives the coarsening law $HL_E\propto T^{1/2}$. This happens around $10^3~{\rm GeV}$ and $10^{-1}~{\rm GeV}$ for $T_{\rm form}=10^{10}~{\rm GeV}$ and $T_{\rm form}=10^{6}~{\rm GeV}$ respectively as shown in Fig.~\ref{fig:vos_friction_results_grad}. When $T_v>T_{\rm QCD}$, this transition happens in the quark-friction-dominated era $\widetilde{P}=\mathcal{O}(1)$, and the critical $T_{\rm form}=10^{8}~{\rm GeV}$.

Another feature is that trajectories with $T_{\rm form}\gtrsim10^6~{\rm GeV}$ approximately converge after the QCD transition. This can be understood by Eq.~\eqref{eq:initial_HL_T}, because confinement sharply reduces $\widetilde P$, accelerating curvature-driven coarsening. If the $HL_E$ is too small at $T_{\rm QCD}$, the second term overwhelms the first one, so $\dd HL_E/\dd \log T<0$. The condition is 
\begin{align}
    (HL_E)_{\rm QCD}<\sqrt{\frac{k_w(k_w+c_w)}{(D_X/H)_{\widetilde{P}=10^{-5}}}}\approx \mathcal{O}(10^{-7})\sqrt{k_w(k_w+c_w)}\,,
\end{align}
If $k_w$, $c_w$ are order unity, this critical temperature $T_{\rm form, cr}=\mathcal{O}(10^{6})~{\rm GeV}$. Below this criteria, the peak value of $v$ is $(1.66g_{*\rho}^{1/2}\widetilde{\sigma}k_w/\widetilde{P})(T_{\rm form}/M_{\rm pl})$. Another relevant diagnostic is $\Omega_{\rm DW}$ at a representative Big Bang nucleosynthesis (BBN) temperature. Using the simple relation for $L_E$, we obtain
\begin{align}
    \Omega_{\sigma}(T_{\rm BBN})\approx \frac{5.05\widetilde{\sigma}_e}{\sqrt{g_{*\rho}}}\left(\frac{M_{\rm pl}^2}{T_{\rm form}^2}+\frac{2k_w(k_w+c_w)1.66g_{*\rho}^{1/2}\widetilde{\sigma}_eM_{\rm pl}}{\widetilde{P}_e}\left(\frac{1}{T_{\rm BBN}}-\frac{1}{T_{\rm QCD}}\right)\right)^{-1/2},
\end{align}
where we assume $g_{*\rho}=10$ and $\widetilde{\sigma}_e=16$ is constant after QCD transition. This ratio is smaller than $10^{-5}$ (the matter energy density ratio) as long as the $k_w,c_w$ is not too small, so this thermal domain wall cannot influence BBN through the energy density mechanism.

\paragraph{Nonrelativistic evolution and loss of formation memory.} When $T\ll m_e$, the speed $c_X$ is much smaller than unity, and we can use normalized TVOS equations to understand the evolution. Define $r\equiv v/c_X$, $h\equiv HL_E/c_X$, $d\equiv D_X/H$, $B\equiv\dot{c}_X/Hc_X$, and $1/\lambda=-\dot{H}/H^2$ with $\lambda=1/2$ in a radiation-dominated Universe and $\lambda=2/3$ in a matter-dominated Universe. Here, we use $'$ to denote $\dd/\dd\log a=-\dd/\dd \log T$, because $g_{*s}$ is constant in this era.
\begin{align}
    r'=(1-r^2)\left(\frac{k_w}{h}-dr\right),\qquad \frac{h'}{h}=1-\frac{1}{\lambda}-B+d r^2+\frac{c_w r}{h}.\label{eq:reduced_modified_VOS}
\end{align}
These equations separate the explicit $c_X$ dependence from the evolution of $h$ and $r$. We distinguish epochs dominated by microscopic friction, $\eta_X/(\mu_XH)$, from those dominated by the adopted background damping.

When $T_{\rm asy}<T\ll m_e$, terminal motion gives $r=k_w/hd$ and $h'/h=-1/2+k_w(k_w+c_w)/h^2d$. $d$ contains the Boltzmann factor $\exp(-m_e/T)$, and its rapid decrease drives the transition away from stretching. We use the approximation in Eq.~\eqref{eq:s56_pair_photon_eta_leading} for $\eta_X$, and find approximately
\begin{align}
    h(T)\approx \sqrt{C_1 \left(\frac{T}{m_e}\right)+\frac{4C_0}{3}\sqrt{\frac{m_e}{T}} \, _1F_1\left(\frac{3}{2};\frac{5}{2};\frac{m_e}{T}\right)},~ C_0=\frac{4\pi^4 k_w(c_w+k_w)C_\mu g_{*\rho}^{1/2}}{e^4\sqrt{10}}\frac{m_e}{M_{\rm pl}},
\end{align}
where $_1F_1$ is the confluent hypergeometric function, and $C_1$ is fixed by the initial condition, $C_1\approx h(T=m_e)$. The hypergeometric function grows exponentially, so the particular solution dominates and the benchmark trajectories approximately converge near $T\sim10^{-5}~{\rm GeV}$. The temperature at which $HL_E$ reaches its minimum has the leading large-$m_e/T$ estimate
\begin{align}
    T_{HL\,{\rm min}}=\frac{2m_e}{3 W[2/3 (C_1/C_0)^{2/3}]},
\end{align}
where $W(x)$ is the Lambert $W$ function, which can be expanded as $\log x-\log\log x+\dots$. Below $T_{\rm asy}$, formation-temperature dependence is negligible for the benchmarks shown.

When $T<T_{\rm asy}$, $\eta\propto T^6$ or $T^8$, depending on the local and gradient responses, respectively as discussed in Section~\ref{subsec:s56_photon_friction}, so during radiation domination we approximate $d=1/2+C_2T^p$, with $p\simeq2$ and $4$, respectively. The gradient response has the $C_2\approx1.3\times10^{-4}\eta_Q^3e^8C_e^2 M_{\rm pl}/(C_\mu m_e^5)$, for example. The terminal approximation holds when $1/2\ll C_2 T^{\,p}$. The solution is
\begin{align}
    h^2(T)\approx h_{\rm asy}^2\left(\frac{T}{T_{\rm asy}}\right)+\frac{2k_w(k_w+c_w)}{C_2(p+1)}\left(T^{-p}-\frac{T}{T_{\rm asy}^{p+1}}\right).\label{eqn:h_T_d0p5}
\end{align}
As $d$ approaches $1/2$, terminal motion ceases to be accurate. Defining $r_c=rhd/k_w$, the reduced equations give 
\begin{align}
    \frac{h'}{h}=1-\frac{1}{\lambda}-B+\frac{2r_ck_w(r_c k_w+c_w)}{h^2},
\end{align}
and one can check the maximum point of $HL$ from $\eval{(HL)'/HL}_{h=h_m}=h'/h+c_X'/c_X=0$,
because $r_c$ is not far away from 1 depending on the onset of friction negligible and $k_w,c_w$, this implies $h_m\approx 1$. Inverting the Eq.~\eqref{eqn:h_T_d0p5} can give the peak is around $T\approx[2k_w(k_w+c_w)/C_2(p+1)]^{1/p}\approx10^{-6}~{\rm GeV}$ for the example we show. However, the $v'=0$ is not easy to estimate from either Eq.~\eqref{eq:s56_modified_vos_cSound} or \eqref{eq:reduced_modified_VOS}, but it is expected closed to the peak of $HL$. After the peak, $h$ and $r$ approach to the attractor
\begin{align}
    r_*=\sqrt{\frac{k_w}{c_w+k_w}}\sqrt{\frac{1/\lambda -1+B}{d}},\qquad h_*=\sqrt{\frac{k_w(k_w+c_w)}{d(1/\lambda-1+B)}},\label{eq:reduced_modified_VOS_scaling}
\end{align}
but the changing cosmological background interrupts relaxation. In the ideal matter-dominated, pre-decoupling limit, there is no attractor and the $h^2$ evolves in logarithm. The physical quantities $HL_E=c_Xh$ and $v=c_Xr$ therefore have slowly varying corrections to the explicit $c_X\propto a^{-1/2}$ dependence. For the alternative $C_e=1$ benchmark, the friction becomes negligible immediately at $T_{\rm asy}$, giving $r\to 1$ quickly, because $dr$ becomes much smaller than the curvature term $k_w/h$. Consequently, $r$ is too close to 1 to relax within the available cosmological time, affecting the slope of the late-time evolution. 

\begin{table}[bh!]
\renewcommand{\arraystretch}{1.15}
    \centering
    \begin{tabular}{c|l||c|c}
    \hline\hline
        $T<T_{\rm dec}$ tension & $T<T_{\rm asy}$ response & $HL$ & $v$  \\ \hline
        \multirow{3}{*}{$\sigma\propto T_\gamma^3$}& local  & $7.2\times 10^{-4}$ & $5.5\times 10^{-6}$ \\
        & gradient matched $C_e$ & $7.0\times 10^{-4}$ & $4.5\times 10^{-6}$ \\
        & gradient $C_e=1$ & $2.3\times 10^{-2}$ & $3.7\times 10^{-5}$ \\ \hline
        \multirow{3}{*}{$\sigma\propto T_e^3$}& local & $3.2\times 10^{-4}$ & $3.4\times 10^{-5}$ \\
        & gradient matched $C_e$ & $3.9\times 10^{-4}$ & $3.2\times 10^{-5}$ \\
        & gradient $C_e=1$ & $1.2\times 10^{-3}$ & $3.7\times 10^{-5}$ \\ \hline\hline
    \end{tabular}
    \caption{The $HL$ and $v$ value at the onset of reionization, $z_{\rm rei}=20$, for different cases with $C_\mu=1$, $k_w=0.7$, and $c_w=0.6$.}
    \label{tab:results_HLv_z20}
\end{table}

The effects of the coefficient $C_\mu$ in the TVOS model become apparent when $d=1/2$. The dependence on $C_\mu$ is contained in $c_X$, so we expect it to provide only an overall factor in $HL_E$ and $v$. This is confirmed by the solutions in Fig.~\ref{fig:vos_friction_results_grad}, with approximately $HL\propto C_\mu^{-1/2}$ and $v\propto C_\mu^{-1/2}$. If the TVOS parameters $k_w$ and $c_w$ are both held fixed at values of order unity, their influence on the evolution are illustrated by the $h$ solutions for instance. In the examples we present, we choose $k_w=0.7$, $c_w=0.6$. The results at the onset of reionization are shown in Table~\ref{tab:results_HLv_z20}. Within the assumptions of our phenomenological model, $HL\sim \mathcal{O}(10^{-3})$ and $v\sim\mathcal{O}(10^{-5})$ when the Universe enters the reionization era.

\subsection{Thermal walls in the present-day universe}
\label{eq:current-universe}

After reionization, the notion of a single cosmological temperature is no longer appropriate for the wall microphysics. The photon temperature, the kinetic temperature of the ionized gas, and the temperatures of different astrophysical environments can be very different. We therefore interpret the low-temperature wall solution locally and quasistatically, using the free-electron density $n_e^{\rm free}$ and the matter temperature $T_m$. In a dilute, nonrelativistic, approximately neutral electron--proton plasma, the plasma-induced width and tension are
\begin{equation}
 d_{\rm DW}=\sqrt{\frac{T_m}{2e^2 n_e^{\rm free}}}\,,
 \qquad
 \sigma_{\rm pl}=\frac{8T_m}{e}\sqrt{2T_m n_e^{\rm free}}\,.
 \label{eq:current_wall_microphysics}
\end{equation}
Thus, $d_{\rm DW}\propto T_m^{1/2}(n_e^{\rm free})^{-1/2}$, whereas $\sigma_{\rm pl}\propto T_m^{3/2}(n_e^{\rm free})^{1/2}$. Numerically,
\begin{align}
 d_{\rm DW} &\simeq 5.26~{\rm m}
 \left(\frac{T_m}{\rm eV}\right)^{1/2}
 \left(\frac{n_e^{\rm free}}{{\rm cm}^{-3}}\right)^{-1/2},
 \label{eq:current_wall_width_numeric}\\
 \sigma_{\rm pl} &\simeq 3.28\times10^{-6}~{\rm eV}^{3}
 \left(\frac{T_m}{\rm eV}\right)^{3/2}
 \left(\frac{n_e^{\rm free}}{{\rm cm}^{-3}}\right)^{1/2}.
 \label{eq:current_wall_tension_numeric}
\end{align}
These expressions show that a surviving wall would not have uniform microscopic properties in the current universe. It would broaden in a dilute ionized region, become narrower in a denser ionized region at fixed temperature, and delocalize as $n_e^{\rm free}\rightarrow0$ in a neutral region unless an additional microscopic core persists.

\begin{table}[hb!]
\centering

\renewcommand{\arraystretch}{1.15}
\begin{tabular}{l||c|c|c|c}
\hline\hline
Environment & $T_m$~[K] & $n_e^{\rm free}~[\mathrm{cm}^{-3}]$
& $d_{\rm DW}$~[m] & $\sigma_{\rm pl}~[\mathrm{eV}^{3}]$ \\
\hline
Mean ionized IGM & $10^4$ & $2.2\times10^{-7}$
& $1.0\times10^4$ & $1.2\times10^{-9}$ \\
Warm ionized medium & $8\times10^3$ & $1.4\times10^{-2}$
& $3.7\times10^1$ & $2.2\times10^{-7}$ \\
Intracluster medium & $5.8\times10^7$ & $10^{-3}$
& $1.2\times10^4$ & $3.7\times10^{-2}$ \\
\hline\hline
\end{tabular}
\caption{Illustrative present-day values of the plasma-induced wall width and tension obtained from Eqs.~\eqref{eq:current_wall_width_numeric} and \eqref{eq:current_wall_tension_numeric}. For the intracluster benchmark, $5.8\times10^7~{\rm K}$ corresponds to $T_m\simeq5~{\rm keV}$, the convention commonly used for X-ray plasma temperatures. These values describe the local electron--proton approximation and should be regarded as order-of-magnitude estimates.}
\label{tab:current_wall_estimates}
\end{table}

Some representative estimates are given in Table~\ref{tab:current_wall_estimates}. For the mean reionized intergalactic medium (IGM), we use $\bar n_{e,0}\simeq2.2\times10^{-7}~{\rm cm}^{-3}$, inferred from the present baryon density measured by Planck and a primordial hydrogen--helium composition~\cite{Planck:2018vyg}. The remaining entries are illustrative local plasma benchmarks; the warm-ionized-medium density is guided by pulsar dispersion and emission-measure measurements~\cite{Gaensler:2008ec}, while the intracluster value is representative of X-ray-inferred cluster atmospheres~\cite{Vikhlinin:2005mp}.

The abundance of walls is much less certain than these local microscopic quantities. Since a wall is an extended object, the useful measure is its surface area per physical volume, $A/V_{\rm phys}=L^{-1}$, rather than a particle-like number density. As a reference estimate, suppose that the preceding network evolution gives $\xi_{\rm rei}\equiv(HL)_{\rm rei}\simeq10^{-3}$ at $z_{\rm rei}=20$ and that the network is subsequently only conformally stretched, $L\propto a$. During matter domination this gives
\begin{align}
 L_0^{\rm stretch}
 \simeq \frac{\xi_{\rm rei}c}
 {H_0\sqrt{\Omega_m}\sqrt{1+z_{\rm rei}}}
 \simeq 1.7~{\rm Mpc}
 \left(\frac{\xi_{\rm rei}}{10^{-3}}\right)
 \left(\frac{21}{1+z_{\rm rei}}\right)^{1/2},
 \label{eq:current_stretched_scale}
\end{align}
where $H_0=67.4~{\rm km\,s}^{-1}{\rm Mpc}^{-1}$ and $\Omega_m=0.315$ have been used \cite{Planck:2018vyg}. The corresponding mean area density is $(A/V_{\rm phys})_0\sim0.6~{\rm Mpc}^{-1}$. Combining this scale with the mean-IGM tension gives
\begin{equation}
 \frac{\rho_{\rm DW}^{\rm pl}}{\rho_{c,0}}
 \sim \frac{\sigma_{\rm pl}}{L_0^{\rm stretch}\rho_{c,0}}
 \sim 10^{-28},
 \label{eq:current_wall_energy_fraction}
\end{equation}
so the plasma-induced part of the wall energy is cosmologically negligible for this benchmark. Equation~\eqref{eq:current_wall_energy_fraction} does not include any additional core contribution that may survive from the earlier evolution.

It is nevertheless possible that the local wall area density is correlated with the matter distribution. If walls remain sufficiently coupled to the ionized plasma and are carried into a region that contracts approximately isotropically by an electron overdensity $\Delta_e=n_e^{\rm free}/\bar n_{e,0}$, simple geometric compression would give
\begin{equation}
 L_{\rm local}\sim L_0^{\rm stretch}\Delta_e^{-1/3},
 \qquad
 \left(\frac{A}{V_{\rm phys}}\right)_{\rm local}
 \sim \left(\frac{A}{V_{\rm phys}}\right)_0\Delta_e^{1/3}.
 \label{eq:current_wall_clustering}
\end{equation}
For the warm ionized medium and intracluster benchmarks in Table~\ref{tab:current_wall_estimates}, this conditional estimate gives $L_{\rm local}\sim40~{\rm kpc}$ and $0.1~{\rm Mpc}$, respectively. It therefore illustrates how the wall area density could be enhanced in galaxies or galaxy clusters.

These last estimates should not be interpreted as a prediction of the present wall abundance. Reionization fronts, spatially varying ionization, plasma friction, curvature-driven collapse, reconnection, and the nonlinear assembly of halos can all change the wall geometry or remove walls. Moreover, a multicomponent ionized plasma modifies the simple electron--proton coefficients, and a fully time-dependent treatment should use the local distribution functions rather than an equilibrium effective potential. Determining whether walls survive, become concentrated in ionized structures, or are regenerated during reionization is therefore left for future work.  The robust conclusion at present is only that, if such walls are present, their width and plasma-induced tension are highly environment dependent and are set primarily by the local ionized-electron density and temperature.

\section{Discussion and conclusions}
\label{sec:conclusions}
Thermal domain walls associated with electric one-form symmetries offer a complementary probe of the global structure of gauge theories and their unified embeddings. For standard embeddings, $SU(5)$, $\mathrm{Spin}(10)$, and $E_6$ give the $\mathbb{Z}_6$ quotient, while the direct-product Pati--Salam and trinification groups give the $\mathbb{Z}_3$ and $\mathbb{Z}_2$ quotients, respectively~\cite{Alonso:2025rkk}. An unambiguous identification of such walls would exclude $\Gamma=\Z_6$, and hence the standard $SU(5)$, Spin$(10)$ and $E_6$ embeddings, while direct-product Pati--Salam ($\Gamma=\Z_3$, two thermal vacua) and trinification ($\Gamma=\Z_2$, three vacua) would remain compatible. The three surviving cases are distinguished by the number of vacua, i.e.\ by the topology of the network, not by the properties of an individual wall. This would complement searches for fractionally charged particles~\cite{Alonso:2024pmq,Koren:2024xof}. For example, a representation with charge $2\pmod 6$ under the diagonal $\mathbb{Z}_6$ center is incompatible with a $\mathbb{Z}_3$ quotient, since allowed representations must be invariant under the gauged subgroup. Fractionally charged particles have been investigated through terrestrial searches~\cite{Perl:2009zz} and cosmological constraints~\cite{Langacker:2011db}, whereas thermal topological defects remain less explored. Exotic matter charged under the relevant center also explicitly breaks the one-form symmetry and can bias the holonomy vacua, linking its particle properties to wall evolution.

Several aspects of wall formation require further study. Our two-flavor PNJL description of the QCD crossover still requires controlled matching to high-temperature perturbation theory. Moreover, the thermal $\mathbb{Z}_3$ symmetry is broken in the high-temperature phase, so formation differs from the usual cooling-induced symmetry-breaking scenario. Reheating may populate distinct holonomy vacua, while nonthermal fluctuations during preheating could seed domains that survive thermalization. The horizon-scale initial separation adopted here remains a benchmark pending a dynamical calculation of these processes. Unified gauge transitions provide another possible setting. In direct-product trinification with the usual bifundamental matter and Higgs fields, the diagonal ultraviolet $\mathbb{Z}_3$ center defines an electric one-form symmetry that descends to the low-energy $\mathbb{Z}_3^{\mathfrak e}$ symmetry~\cite{Alonso:2025rkk}. To determine whether gauge breaking produces walls, one must follow how distinct thermal holonomy vacua are populated during the transition.

The TVOS description assumes that the wall admits a collective coordinate with a well-defined inertial response. For nonrelativistic electrons, we adopt $c_X^2=\sigma/\mu_X=T_e/(C_\mu m_e)$ as an inertia benchmark motivated by the electron thermal speed. The coefficients $c_0$, $C_e$ and $C_p$ governing the response below $T_{\rm asy}$, the inertia coefficient $C_\mu$, and the network parameters $k_w$ and $c_w$ remain inputs to be determined microscopically or through simulations. Plasma friction can control the benchmark evolution, while inelastic scattering, radiation, and mode conversion may contribute additional dissipation. Recombination, electron--photon thermal decoupling at $z\lesssim150$, and reionization also change the medium supporting the wall. These effects should be incorporated into a common description of the coupled wall--plasma dynamics. The real-time formalism of Ref.~\cite{Hidaka:2009hs} and the Hamiltonian treatment of Wilson loops in Ref.~\cite{Pisarski:2022cuo} offer starting points for such a calculation.

Heating during reionization and galaxy formation may modify existing walls. Producing new walls, however, requires the plasma to populate distinct holonomy vacua coherently, even when the heated region is much larger than the wall thickness. Spontaneous thermal production is additionally suppressed by the surface free-energy cost in our equilibrium benchmark. Heavy-ion collisions provide another setting for investigating related structures. Ref.~\cite{Asakawa:2012yv} proposed that QCD center domains could contribute to the opacity and nearly ideal-fluid behavior of the quark-gluon plasma. Assessing production of the SM walls considered here requires the combined color--electromagnetic holonomy dynamics within the finite size and lifetime of the collision plasma.

Observational detection remains challenging. Under the adopted wall-energy prescription, our benchmark networks remain energetically subdominant as the tension decreases. The Zel'dovich bound and subsequent CMB constraints provide a reference scale $\sigma^{1/3}\lesssim\mathcal O(1\,\mathrm{MeV})$~\cite{Zeldovich:1974uw,Sousa:2015cqa,Caloni:2026dyu}, although their fixed-tension network assumptions do not directly apply to plasma-supported walls. Nongravitational probes may offer complementary opportunities, especially because late-time wall properties depend on the local plasma density and temperature. For the response in Eq.~\eqref{eq:s56_asymmetric_response}, the refractive-index contrast is $|\delta n_{\rm refr}|\simeq|\operatorname{Re}\delta\Pi_T|/(2\omega^2)\sim10^{-22}$ for a $100\,\mathrm{GHz}$ photon and $n_e\sim10^{-7}\,\mathrm{cm}^{-3}$. Lower-frequency pulsar and fast-radio-burst observations could probe dispersive delays or lensing, while spacecraft could search for changes in local plasma-wave spectra during a wall encounter. Their sensitivity depends on the integrated response, wall geometry, crossing time, local abundance, and distinguishability from ordinary plasma backgrounds.

Our benchmarks also suggest very weak gravitational-wave (GW) emission because of the small inferred wall energy fraction and substantial plasma damping. Heavy fractionally charged particles can induce a bias that drives wall annihilation. Because the holonomy free energy depends only on densities and charges, the bias from a frozen-out relic, $\Delta V = T\,(n_X+n_{\bar X})[1-\cos(2\pi Q_X)]$, is not Boltzmann suppressed and scales as $T^4$; it overtakes the curvature and friction pressures once $e^\pm$ annihilation depletes the wall tension, so that for $10^{-14}\lesssim n_X/n_B\lesssim10^{-4}$ the network collapses at $T\simeq10$--$100$~keV, while survival to recombination requires $n_X/n_B\lesssim10^{-20}$. The collapse remains friction limited and releases at most $\sim10^{-10}\,n_X/n_B$ of the energy density, giving $\Omega_{\rm GW}h^2\lesssim10^{-19}$ for any abundance allowed by the CMB. Without a bias, closed walls can still contract under their tension and emit GWs through nonspherical motion. Friction can dissipate much of the energy of large closed walls before their collapse becomes rapid. Early, small walls could contribute at higher frequencies, with the amplitude depending on their production rate, size and shape distributions, and the nonlinear dynamics of collapse.

New physics may modify both the plasma drag and the holonomy potential. Additional light charged species could supply friction after recombination if their abundance persists and their interactions preserve the relevant center symmetry. Higher-dimensional interactions, including $\phi F_{\mu\nu}F^{\mu\nu}$ and $aF_{\mu\nu}\widetilde F^{\mu\nu}$, provide further possibilities. The axion coupling could introduce dissipation through plasma fluctuations, photon--axion conversion in physical magnetic fields, or an axion profile coupled to the wall. A consistent calculation must distinguish the Euclidean holonomy from physical electromagnetic fields and determine the momentum transferred to the interface through its real-time response. These interactions could slow coarsening and enhance wall survival, connecting thermal-wall cosmology with axion physics and cosmic magnetic fields.

In this work, we have constructed the effective potential for the coupled $SU(3)_c\times U(1)_{\rm EM}$ thermal holonomies and studied the resulting $\mathbb{Z}_{6/p}$ walls for compatible global forms of the SM gauge group. Our benchmark profiles give $\sigma\propto T^3$ and $\ell\propto T^{-1}$ in the high-temperature scaling regime, while the dilute electron--proton benchmark gives $\sigma\propto n_e^{1/2}T^{3/2}$ and $\ell\propto(T/n_e)^{1/2}$. These relations connect the tension and width to the plasma density, temperature, and ionization history, allowing kilometer-scale thicknesses around reionization. We have also developed a phenomenological thermal velocity-dependent one-scale model with temperature-dependent tension, independent surface inertia, and plasma drag. Its terminal coarsening rate is governed by the mobility $\sigma/\eta_X$, making the plasma response central to the surviving wall abundance. The existence of these walls today requires cosmological production and survival as well as a compatible global gauge structure. If realized, these macroscopic thermal interfaces could provide a cosmological probe of the global structure and higher-form symmetries of the SM and its extensions.

\section*{Acknowledgment}

We would like to thank Ting-Kuo Chen and Yun Jiang for helpful discussion. The work is supported by the U.S. Department of Energy under the contract DE-
SC0017647. The authors used ChatGPT (OpenAI) and Claude Fable (Anthropic) for theoretical discussions, calculation checks, and manuscript preparation. They independently verified the scientific content and take full responsibility for the work.

\newpage


\newpage
\appendix

\section{The holonomy contribution to self-energy at one-loop}
\label{app:1Loop-selfE}

In this appendix, we show the detailed 1-loop self-energy for fermions and gauge bosons, with the gauge holonomy. Notice that this effect does not occur at tree-level, because the nontrivial holonomy $Q_a\equiv2\pi\phi_a$ will be Wick rotated together with the Matsubara frequency to the real time formalism. This can be understood, e.g., from the pole structure of the two-point Green function. However, its effect remains in the thermal loop calculation. From the thermal field calculation, one can get the self-energy correction for a quark with energy $\omega=i\omega_n+iQ_a\to E$ is 
\begin{align}
    \Xi_a(z;\omega,\bm p)=g_s^2\sum_c \mathcal{C}_{ac}\int\frac{\dd^3k}{(2\pi)^3}\sum_{s,t=\pm 1}\frac{t\,\mathcal{M}_s}{2q}\frac{1+f_B(t\,q-i Q_{ac})-f_F(s E_k-iQ_c)}{\omega-s\,E_k-t\,q},
\end{align}
in Feynman gauge, where $\bm{k}$ is the internal quark momentum, and $\bm{q}$ is the internal gluon momentum (not to be confused with the holonomy $q_i$), $E_k=\sqrt{k^2+m^2}$, and $q=|\bm{q}|=\sqrt{k^2+p^2-2 kp\cos\theta}$ with $p=|\bm{p}|$ and $x=\cos\theta$ the angle between $\bm{k}$ and $\bm{p}$. The spinor structure $\mathcal{M}_s$ is defined as
\begin{align}
    \mathcal{M}_\pm=\gamma^0\mp \frac{k}{E_k}\bm{\gamma}\cdot\hat{\bm{k}}\mp 2\frac{m}{E_k}\,, 
\end{align}
and the color projection operator
\begin{align}
    \mathcal{C}_{ac}=\frac{1}{2}\left(1-\frac{1}{3}\delta_{ac}\right).
\end{align}
The $Q_{a=1,2,3}$ is the holonomy for the three colors and $Q_{ac}=Q_{a}-Q_{c}$ comes from the gluon contribution so it is in the adjoint way. In the limit $k\gg p$ one can reproduce the HTL results~\cite{Hidaka:2009hs}. Notice that the $Q_{a}$ here will not only contain the $q_{1,2}$ but also the $t$ contribution, because it originates from the fermion in the loop, carrying the full information of $SU(3)_c\times U(1)_{\rm EM}$. This demonstrates that the $\mathbb{Z}_{6/p}$ symmetry is also preserved at 1-loop level. One can also write the self-energy from photon loop for the electron and quarks. Collect the self-energy into
\begin{align}
    \Xi_a(z;\omega,\bm p)=A_a\gamma^0+B_a\bm{\gamma}\cdot\hat{\bm{k}}+C_a,
\end{align}
where
\begin{align}
    &A_a(z;\omega,\bm p)=\frac{g_s^2}{8\pi^2}\sum_c \mathcal{C}_{ac}\int_0^\infty\dd k k^2\int_{-1}^1\dd x\frac{\mathcal{F}_{+}^{ac}+\mathcal{F}_{-}^{ac}}{q},\nonumber\\
    &B_a(z;\omega,\bm p)=\frac{g_s^2}{8\pi^2}\sum_c \mathcal{C}_{ac}\int_0^\infty\dd k k^2\int_{-1}^1\dd x\frac{k x}{E_k}\frac{\mathcal{F}_{-}^{ac}-\mathcal{F}_{+}^{ac}}{q},\\
    &C_a(z;\omega,\bm p)=\frac{g_s^2}{8\pi^2}\sum_c \mathcal{C}_{ac}\int_0^\infty\dd k k^2\int_{-1}^1\dd x\frac{2m}{E_k}\frac{\mathcal{F}_{-}^{ac}-\mathcal{F}_{+}^{ac}}{q},\nonumber
\end{align}
and
\begin{align}
    &\mathcal{F}_{+}^{ac}=\frac{\mathcal{N}_{++}^{ac}}{\omega-E_k-q}+\frac{\mathcal{N}_{+-}^{ac}}{\omega-E_k+q},~~\mathcal{F}_{-}^{ac}=\frac{\mathcal{N}_{-+}^{ac}}{\omega+E_k-q}+\frac{\mathcal{N}_{--}^{ac}}{\omega+E_k+q},\\
    &\mathcal{N}_{++}^{ac}=1+f_B(q-i Q_{ac})-f_F(E_k-iQ_c),~~\mathcal{N}_{--}^{ac}=\left(\mathcal{N}_{++}^{ac}\right)^*,\nonumber\\
    &\mathcal{N}_{+-}^{ac}=f_B(q+i Q_{ac})+f_F(E_k-iQ_c),~\qquad\mathcal{N}_{-+}^{ac}=\left(\mathcal{N}_{+-}^{ac}\right)^*.
\end{align}
It is straightforward to write these formulas for electron in the EM holonomy as well.

The gauge holonomy also gives thermal correction to the gauge boson itself. Notice that the physical meaning of nontrivial $\phi_i$ is the shift of Matsubara frequency, and we should not treat it as a tree-level electric field $A_0$ in Minkowski space. The thermal fermion loop will induce the color $a,b$ gluon self-energy,\footnote{In principle, one should also include the gluon and ghost fields contribution for completeness~\cite{Hidaka:2009hs}, but that is not the aim of this paper}
\begin{align}
    \Pi^{\mu\nu}_{ab}(z;\omega,\bm p)&=\,2g_s^2\sum_f\int \frac{\dd^3k}{(2\pi)^3}\frac{\mathcal{G}_s^{ab,f}}{2E_k}\frac{K\cdot P (K^\mu P^\nu+K^\nu P^\mu)-P^2K^\mu K^\nu-\eta^{\mu\nu}(K\cdot P)^2}{(K\cdot P)^2-(P^2)^2/4}\nonumber\\
    &+g_s^2\sum_f\int \frac{\dd^3k}{(2\pi)^3}\frac{\mathcal{G}_a^{ab,f}}{2E_k}\frac{4K\cdot P K^\mu K^\nu-P^2(K^\mu P^\nu+K^\nu P^\mu)+(K\cdot P)P^2\eta^{\mu\nu}}{(K\cdot P)^2-(P^2)^2/4},
\end{align}
where we denote $P=(\omega,\bm{p})$ and $K=(E_k,\bm{k})$ the momentum of fermion, the antifermion has momentum $P-K$. $f$ is the flavor of quark. The color symmetric $\mathcal{G}_s^{ab}$ and color antisymmetric $\mathcal{G}_a^{ab}$ are
\begin{align}
    \mathcal{G}_s^{ab,f}=\frac{1}{2}\left[f_F(E_k-i Q_a^f)+f_F(E_k-i Q_b^f)+\text{c.c.}\right],\mathcal{G}_a^{ab,f}=-\frac{1}{2}\left[f_F(E_k-i Q_a^f)-f_F(E_k-i Q_b^f)-\text{c.c.}\right].
\end{align}
The color antisymmetric part $\mathcal{G}_a^{ab,f}$ is purely imaginary, and it is zero at the high-temperature vacuum $(q_1,q_2,t)=(0,0,0)$. Moreover, the Ward identity is readily satisfied in the color symmetric part, but the antisymmetric part gives
\begin{align}
    P_\mu \Pi^{\mu\nu}_{ab}\supset g^2\sum_f\int \frac{\dd^3k}{(2\pi)^3}\frac{\mathcal{G}_a^{ab,f}}{2E_k}K^\nu.
\end{align}
For $\nu=1,2,3$ the identity holds, whereas for $\nu=0$ it holds only when $f_F=0$. This provides one way to understand how the Polyakov loop characterizes confinement. Because in the low-temperature vacuum $(q_1,q_2,t)\to(1/3,1/2,0)$, the contribution from $\mathcal{G}_a^{ab,f}$ does not cancel. Only when the thermal quark population vanishes, such that $f_F$ no longer has physical meaning but can only vanish with $f_F=0$, the Ward identity holds. The confinement transition can be very complicated, so we will ignore this issue. The transverse and longitudinal mode are
\begin{align}
    &\Pi_T^{ab}(z;\omega,\bm p)=\frac{g_s^2}{4\pi^2}\sum_f\int_0^\infty \dd k\frac{k^2}{E_k}\int_{-1}^1\dd x\left[\mathcal{G}_s^{ab,f}\frac{\iota^2-s k^2(1-x^2)/2}{\iota^2-s^2/4}+\mathcal{G}_a^{ab,f}\frac{\iota(k^2(1-x^2)-s/2)}{\iota^2-s^2/4}\right],\\
    &\Pi_L^{ab}(z;\omega,\bm p)=\frac{g_s^2}{4\pi^2}\sum_f\int_0^\infty \dd k\frac{k^2}{E_k}\int_{-1}^1\dd x\left[\mathcal{G}_s^{ab,f}\frac{s(E_k^2-k^2x^2)}{\iota^2-s^2/4}+\mathcal{G}_a^{ab,f}\frac{\iota(4\xi^2-s^2)}{\iota^2-s^2/4}\right],
\end{align}
where $s=P^2=\omega^2-p^2$, $\iota=K\cdot P=E_k \omega-kp x$, $\xi=K\cdot \epsilon_L\sqrt{s}=p E_k-\omega kx$. One can perform the same calculation for photon self-energy. 

\section{The fermion Weyl expansion for reflection}
\label{app:WKB_reflection}

The derivation in Appendix~\ref{app:1Loop-selfE} indicates the nontrivial holonomies modify the thermal properties of the particles at loop level. The self-energy varies along the domain wall direction as the interaction between thermal domain wall and plasma. We consider the reflection of a massless electron. The differential operator in Dirac equation is
\begin{align}
    \mathcal{D}_{\rm Dirac}=\begin{pmatrix}
    0& \Omega-P \bm{\sigma}\cdot \hat{\bm{p}}\\
    \Omega+P\bm{\sigma}\cdot \hat{\bm{p}} & 0
    \end{pmatrix},
\end{align}
with $\Omega=\omega-A(z;\omega,\bm p), ~~P=p+B(z;\omega,\bm p)$. Assuming $p_x\ne 0$ and $p_y=0$~\footnote{The longitudinal momentum is essential for nonzero reflection, because the chirality of fermion is preserved in this interaction.}, the left and right-handed fermions decouple on Weyl fermion $\xi_R$
\begin{align}
    \mathcal{D}\equiv\begin{pmatrix}
    \Omega-P_z & -P_x\\
    -P_x & \Omega+P_z
    \end{pmatrix},~~\mathcal{D}\xi_R=0,
\end{align}
with $P_i=P(z;\omega,\bm p)p_i/p$ the $P$ contain the 1-loop correction. This requires $\Omega^2-P^2=0$ and the physical mode corresponds to $\Omega-P=0$ to match the tree-level positive-energy. We can solve $p=p_{\rm WKB}$ from $\Omega(z;\omega,\bm p)-P(z;\omega,\bm p)=0$ so the WKB approximation for the classical momentum is 
\begin{align}
    p_\pm=\pm\sqrt{p_{\rm WKB}^2-p_x^2}\,,
\end{align}
The eigenvector $v_\lambda$, with $\lambda=\pm$, is determined from
\begin{align}
    \eval{\mathcal{D}}_{p_z=p_\pm}v_{\pm}=0,
\end{align}
with the normalization chosen such that the current $j^z_{\pm}=v_{\pm}^\dagger \sigma_3 v_{\pm}$ satisfies $j^z_+=1,~j^z_-=-1$ at tree-level. Assume the propagating wave function is
\begin{align}
    \psi_R(z)=a_+(z)v_+(z) e^{iS_+(z)}+a_-(z)v_-(z) e^{iS_-(z)},~~S_\pm(z)=\int^z\dd \tilde{z}\, p_{\pm}( \tilde{z}),
\end{align}
in WKB approximation. The Weyl expansion~\cite{Emmrich:1994rr,Bolte:1998ia} of the operator $\mathcal{D}$ on the $a_\lambda v_\lambda \exp(iS_\lambda)$ gives
\begin{align}
    &\mathrm{Op}^{W}(\mathcal{D})[a_\lambda v_\lambda e^{iS_\lambda/\hbar}]\nonumber\\
    =&e^{iS_\lambda/\hbar}\left\{\mathcal{D}a_\lambda v_\lambda+\frac{\hbar}{i}\left[\left(\frac{\partial \mathcal{D}}{\partial p_z}v_\lambda'+\frac{1}{2}\left(\frac{\partial \mathcal{D}}{\partial p_z}\right)'v_\lambda\right)a_\lambda +\left(\frac{\partial \mathcal{D}}{\partial p_z}\right)a'_\lambda v_\lambda\right]+\mathcal{O}(\hbar^2)\right\}\eval{}_{p_z=p_\lambda},
\end{align}
where $'$ denotes the spatial derivative $\dd/\dd z$ here. Requiring this expression to vanish order by order, we can get the WKB approximation and the first order of $\hbar$ gives 
\begin{align}
    \sum_\lambda(a_\lambda'w_\lambda+a_\lambda g_\lambda)e^{iS_\lambda(z)}=0,
\end{align}
where
\begin{align}
    w_\lambda\equiv\left(\frac{\partial \mathcal{D}}{\partial p_z}\right) v_\lambda,~~g_\lambda\equiv\left(\frac{\partial \mathcal{D}}{\partial p_z}v_\lambda'+\frac{1}{2}\left(\frac{\partial \mathcal{D}}{\partial p_z}\right)'v_\lambda\right)\,,
\end{align}
and we define the dual of $w_\lambda$ by $\widetilde{w}_j^{\dagger}w_i=\delta_{ij}$. The above equation can be simplified to
\begin{align}
    \frac{\dd}{\dd z}\begin{pmatrix}
        a_+(z)\\ a_-(z)
    \end{pmatrix}=-\begin{pmatrix}
        \widetilde{w}_{+}^{\dagger}g_{+} & \widetilde{w}_{+}^{\dagger}g_{-} e^{i(S_+ -S_-)}\\
        \widetilde{w}_{-}^{\dagger}g_{+}e^{i(S_- -S_+)} & \widetilde{w}_{-}^{\dagger}g_{-}
    \end{pmatrix}\begin{pmatrix}
        a_+(z)\\ a_-(z)
    \end{pmatrix}.
\end{align}
Integrating this equation determines the transfer matrix relating $a_\pm(-\infty)$ and $a_\pm(+\infty)$. We impose the boundary conditions $(a_+(-\infty),a_-(-\infty))=(1,r)$, and $(a_+(+\infty),a_-(+\infty))=(t,0)$, to read the reflection rate from transfer matrix. Integrating over the $(p_x,p_z)$ disk gives 
\begin{align}
    \widetilde{P}_e\approx1.0\times 10^{-5},
\end{align}
which is independent of temperature as stated in the main text. 

\section{TVOS equations derivation}
\label{app:modified_VOS}

The metric in conformal time $\eta$ is $\dd s^2=a^2(\eta)(\dd\eta^2-\dd \bm{x}^2)$, and the wall is parametrized as $\bm{x}=\bm{x}(\eta,\xi^1,\xi^2)$. Impose the transverse gauge $\bm{v}\cdot\partial_i\bm{x}=0$, $\bm{v}=\partial_0\bm{x}$, $v=|\bm{v}|$, $\hat{n}=\bm{v}/v$. The induced spatial metric on the wall is $h_{ij}=\partial_i\bm{x}\cdot \partial_j\bm{x}$, and $h=\det(h_{ij})$. The curvature is $\kappa\equiv h^{ij}\partial_i\partial_j\bm{x}\cdot\hat{n}$, so
\begin{align}
    \frac{(\sqrt{h})'}{\sqrt{h}}=-v\kappa,~~~\delta \sqrt{h}=\sqrt{h}h^{ij}\partial_i\bm{x}\cdot\partial_j\delta\bm{x}.
\end{align}

A system of frictionless domain wall with $c_X\ne 1$ can be described by the action
\begin{align}
    S=-\int \dd\eta\dd^2\xi \sigma(\eta) a^3(\eta)\sqrt{h}\sqrt{1-\frac{v^2}{c_X^2}}\,,
\end{align}
and it can be matched to the convention as follows. For a small normal displacement $X=\pi(\eta,\xi^1,\xi^2)$, $v=\partial_0\pi$, at quadratic order $h_{ij}=\delta_{ij}+\partial_i\pi\partial_j\pi$ and $\sqrt{h}=1+(\nabla\pi)^2/2$. One has
\begin{align}
    \sigma\sqrt{h}\sqrt{1-\frac{v^2}{c_X^2}}=\sigma+\frac{\sigma}{2}(\nabla\pi)^2-\frac{\sigma}{2c_X^2}(\partial_0\pi)^2\,,
\end{align}
with $\mu_X=\sigma/c_X^2$. Defining $F(\eta,y)=\sqrt{1-y/c_X(\eta)^2}$, variation of the action gives
\begin{align}
    \partial_\eta[\sigma  a^3\sqrt{h}(-2F_y(\eta,y)|_{y=v^2})\bm{v}]=\sigma a^3\partial_i[\sqrt{h}F(\eta,v^2)h^{ij}\partial_j\bm{x}]\,.
\end{align}
Defining $P=-2F_y=-2\partial F/\partial y$, and projecting this equation onto $\hat{n}$ gives
\begin{align}
    \partial_\eta[\sigma  a^3\sqrt{h}P(\eta,y)v]=\sigma a^3\sqrt{h}F(\eta,v^2)\kappa\,,
\end{align}
and therefore
\begin{align}
    \frac{\sigma'}{\sigma}+3\frac{a'}{a}+\frac{(\sqrt{h})'}{\sqrt{h}}+\frac{P_{\eta}+P_{y}y'}{P}+\frac{v'}{v}=\kappa \frac{1}{v}\frac{F}{P}\,,
\end{align}
where $'$ denotes a derivative with respect to $\eta$. This gives the $v$ equation in the TVOS equation,
\begin{align}
    v'=(1-\frac{v^2}{c_X^2})\left[c_X^2\kappa-(3\mathcal{H}+\frac{\mu'}{\mu})v\right]+\frac{v^3}{c_X^2}\frac{c_X'}{c_X}\,,
\end{align}
where we use $c_X'/c_X=(\sigma'/\sigma-\mu'/\mu)/2$. Collecting a factor of $1-v^2/c_X^2\equiv\gamma_X^{-2}$ from the last term and adding the velocity relaxation term $a/\tau_{\rm rel}=a\eta_X/\mu_X$ gives
\begin{align}
    v'=(1-\frac{v^2}{c_X^2})\left[c_X^2\kappa-\left(3\mathcal{H}+\frac{\mu'}{\mu}+\frac{a\eta_X}{\mu_X}+\frac{c_X'}{c_X}\right)v\right]+v\frac{c_X'}{c_X}\,.
\end{align}

The equation for $L_E$ follows from energy evolution. On the unit area $\dd A=a^2\sqrt{h}\dd^2\xi$, the physical energy of the domain wall $E_{\rm phys}=\sigma \gamma_X \dd A$, because
\begin{align}
    E_{\rm phys}=\dd A\left(\frac{\sigma}{2}(\nabla\pi)^2+\frac{\sigma}{2c_X^2}(\partial_0\pi)^2\right)\rightarrow\dd A\frac{\sigma}{\sqrt{1-v^2/c_X^2}}\,,
\end{align}
its evolution is then
\begin{align}
    \frac{E'_{\rm phys}}{E_{\rm phys}}=\frac{\sigma'}{\sigma}+\frac{\gamma'_X}{\gamma_X}+\frac{\dd A'}{\dd A}=\frac{\sigma'}{\sigma}+\frac{\gamma'_X}{\gamma_X}+2\mathcal{H}-v\kappa=\frac{\sigma'}{\sigma}+\frac{v^2}{c_X^2-v^2}\left(\frac{v'}{v}-\frac{c_X'}{c_X}\right)+2\mathcal{H}-v\kappa\,.
\end{align}
The time derivative of length scale $L_E\equiv \sigma/\rho=\sigma V_{\rm phys}/E_{\rm phys}$ is
\begin{align}
    \frac{L'_E}{L_E}=\frac{\sigma'}{\sigma}+3\mathcal{H}-\frac{E'_{\rm phys}}{E_{\rm phys}}=\mathcal{H}+\frac{v^2}{c_X^2}\left(3\mathcal{H}+\frac{\mu'}{\mu}+\frac{a}{l_f}+\frac{c_X'}{c_X}\right).
\end{align}
Adding the energy-loss term $+c_E va/L_E$ implies the $L_E$ equation in the TVOS.


\providecommand{\href}[2]{#2}\begingroup\raggedright\endgroup

\end{document}